\documentclass[12pt,bibliography]{article}
\usepackage{hyperref}
\usepackage{epsfig}
\usepackage{amsfonts} 
\usepackage{amsmath} 
\usepackage{color} 

\usepackage[utf8]{inputenc}
\usepackage[T1]{fontenc}

\def\tr{{\rm Tr}}

\def\a{\alpha}

\def\Or[#1]{{\text{O}}\left({#1}\right)}
\def\dotl[#1,#2]{\left\langle #1, #2 \right\rangle}
\def\dotlb[#1,#2]{[ #1, #2 ]}
\def\dotp[#1,#2]{(#1) \cdot (#2)}
\def\aff[#1,#2]{\hat{#1}(#2)}
\def\n4sym{{\cal N}=4 SYM}
\def\>{\rangle}
\def\<{\langle}
\def\weight[#1,#2,#3]{\{(#1),#2,#3\}}
\def\ads[#1]{$\text{AdS}_{#1}$}

\newcommand{\ba}{\begin{eqnarray}}
\newcommand{\ea}{\end{eqnarray}}

\title{Conformal Blocks in Mellin Space}

\newcommand{\be}{\begin{equation}}
\newcommand{\ee}{\end{equation}}  
\newcommand{\bi}{\begin{itemize}}
\newcommand{\ei}{\end{itemize}}

\newcommand{\Ocal}{{\mathcal O}}

\newcommand{\aslash}[1]{\,\,{\raise.15ex\hbox{/}\mkern-12mu #1}}
\newcommand{\bslash}[1]{\,\,{\raise.15ex\hbox{/}\mkern-9mu #1}}

\renewcommand{\bar}{\overline}
\renewcommand{\tilde}{\widetilde}
\renewcommand{\hat}{\widehat}
\renewcommand{\Im}{{\rm Im\,}}
\renewcommand{\Re}{{\rm Re\,}}

\newcommand\lrpar{\raise .8ex\hbox{$^\leftrightarrow$} \hspace{-9pt}
\partial}
\newcommand\lpar{\raise .8ex\hbox{$^\leftarrow$} \hspace{-9pt}
\partial}
\newcommand\rpar{\raise .8ex\hbox{$^\rightarrow$} \hspace{-9pt}
\partial}
\newcommand\lrd{\raise .8ex\hbox{$^\leftrightarrow$} \hspace{-9pt}
\nabla}

\newcommand{\gsim}{\lower.7ex\hbox{$\;\stackrel{\textstyle>}{\sim}\;$}}
\newcommand{\lsim}{\lower.7ex\hbox{$\;\stackrel{\textstyle<}{\sim}\;$}}

\renewcommand{\Im}{\text{Im}}
\renewcommand{\Re}{\text{Re}}

\let\a=\alpha  \let\g=\gamma \let\d=\delta 
    
  \let\n=\nu  
     
  \let\D=\Delta  
    
    \let\G=\Gamma

\renewcommand{\ba}{\begin{eqnarray}}
\renewcommand{\ea}{\end{eqnarray}}
\newcommand{\bea}{\begin{eqnarray}}
\newcommand{\eea}{\end{eqnarray}}

\newcommand\T{\rule{0pt}{2.6ex}}
\newcommand\B{\rule[-1.2ex]{0pt}{0pt}}

\begin{document}

\begin{titlepage}

\begin{center}

\ 
\vspace{1cm}

{\Large \bf  Conformal Regge theory}

\vspace{1cm}

{\bf Miguel S. Costa, Vasco Goncalves, Jo\~ao Penedones}

\vspace{1cm}

{\it  Centro de F\'{i}sica do Porto \\
Departamento de F\'{i}sica e Astronomia\\
Faculdade de Ci\^encias da Universidade do Porto\\
Rua do Campo Alegre 687,
4169--007 Porto, Portugal}
\\
\vspace{.3cm}
{\it  Perimeter Institute for Theoretical Physics \\ Waterloo, Ontario N2J 2W9, Canada}

\end{center}
\vspace{1cm}

\begin{abstract}
We generalize Regge theory  to correlation functions in conformal field theories.
This is done by exploring the analogy between Mellin amplitudes in AdS/CFT and 
S-matrix elements.
In the process, we develop the conformal partial wave expansion in Mellin space, elucidating the analytic structure of the partial amplitudes.
We apply the new formalism to the case of  four point correlation functions between protected scalar operators in ${\cal N}=4$ Super
Yang Mills, in cases where the Regge limit is controlled by the leading twist operators associated to the
pomeron-graviton Regge trajectory.
At weak coupling, we are able to predict to arbitrary high order in the 't Hooft coupling the behaviour near  $J=1$ of the OPE coefficients  
$C_{{\cal OO}J}$ between the external scalars 
and the spin $J$ leading
twist  operators.  At strong coupling, we use recent results for the anomalous dimension of  the leading 
twist  operators to improve current knowledge of the AdS graviton Regge trajectory - in particular, determining  the next and next to next leading 
order corrections to the intercept. Finally, by taking the flat space limit and considering the Virasoro-Shapiro S-matrix element,
we compute the strong coupling limit of the OPE coefficient $C_{{\cal LL}J}$ between two Lagrangians and the leading twist operators of spin $J$.

\end{abstract}

\bigskip
\bigskip

\end{titlepage}

\tableofcontents

\section{Introduction}

Regge theory \cite{ReggeOriginal, Gribov} is a technique to study high energy scattering.
It is particularly useful when an infinite number of resonances participate in a scattering process, like it happens in QCD or in  string theory.
In the 60s, Regge theory was very important in organizing the phenomenology of hadrons, which suggested a stringy description.
String theory was later abandoned as a theory of hadrons in favour of gauge theory (more precisely, QCD).
However, we now know that conformal gauge theories are equivalent to string theory in Anti-de Sitter (AdS) space \cite{Maldacena,GKP,Witten}.
It is then natural to look for a generalization of Regge theory to scattering in AdS.
We believe this is a good starting point to understand AdS/CFT correlation functions  in the stringy regime of finite string tension (or finite 't Hooft coupling),  beyond the gravity approximation.

The goal of this paper is to generalize Regge theory of scattering amplitudes to correlation functions of conformal field theory (CFT), building
on the previous work \cite{BPST,CornalbaRegge,ourBFKL}.
The main tool that we shall use  is the Mellin representation of conformal correlation functions introduced in \cite{Mack}.
As emphasized in  \cite{Mack, JPMellin,NaturalMellin,PaulosMellin, JLAnalyticMellin,VolovichMellin,PaulosMellinweak,JLBoundedMellin},
Mellin amplitudes are the natural analogue of scattering amplitudes for AdS/CFT.
In particular, for a four-point function the Mellin amplitude depends on two variables $s$ and $t$ that are similar to the Mandelstam invariants of a scattering process in flat spacetime.
Following this analogy, we confirm that Mellin amplitudes are the appropriate observable to generalize Regge theory to AdS/CFT.
The main ingredients involved in our construction are summarized in table \ref{tabela}, which 
also serves as a partial outline of the paper.

We dedicate the next section to a review of Regge theory applied to string theory in flat spacetime. This will be very useful to set up the language and to guide us in the generalization for CFTs. In section \ref{secMellin}, we review some key properties of Mellin amplitudes and introduce the conformal partial wave expansion necessary for performing the Regge theory re-summation of section \ref{secReggeMellin}. This gives the main result of this paper, which is summarized in the last two rows of table \ref{tabela}. The Mellin amplitude associated to a conformal four-point function 
has  Regge behavior controlled by the Reggeon spin $j(\nu)$ and residue $\beta(\nu)$. Moreover, the Reggeon spin $j(\nu)$ is the inverse function of the dimension $\D(J)$ of the operators in the leading Regge trajectory, and the residue $\beta(\nu)$ is determined by the OPE coefficients $C_{{\cal OO}J}$ of the same operators in the OPE ${\cal OO}$ of the external operators.

In section \ref{secSYM}, we apply the above results to the study of the Pomeron-graviton Regge trajectory in maximally supersymmetric Yang-Mills  theory (SYM).
One of the great values of Regge theory is that the relation between the Reggeon spin  $j(\nu)$ and dimension $\D(J)$ of the exchanged operators 
does not commute with perturbation theory
\cite{DressingWrapping}. 
We shall see that the same statement applies for the relation between  Regge residue $\beta(\nu)$ and OPE coefficient $C_{{\cal OO}J}$. In particular, 
from the two-loop calculation of the four point function of dimension two chiral primaries $\mathcal{O}$, we obtain the Regge residue  at leading order in perturbation theory, 
and are able to compute the leading behaviour of the OPE coefficient $C_{{\cal OO}J}$ near  $J=1$ to
any order in the 't Hooft coupling $\lambda=g_{\rm YM}^2 N$. More concretely,  we show that 
\be
C_{{\cal OO}J}^2= \frac{1}{N^2}\,\sum_{k=0}^\infty \lambda^k  \left[ \frac{a_k}{(J-1)^{k-1}}+
O\left( \frac{1}{(J-1)^{k-2}} \right) \right]  ,
\ee
and  compute the coefficients $a_k$. Actually, since the Regge residue is known at next to leading order, we are also able to compute the 
next to leading order behaviour around $J=1$ of $C_{{\cal OO}J}$.
The relation between  Reggeon spin  $j(\nu)$ and dimension $\D(J)$ also has implications to the strong coupling expansion. In particular, recent results for
the anomalous dimension of  the leading twist  operators imply  the following expansion of the  AdS graviton intercept  
\be
j(0) = 2 - \frac{2}{\sqrt{\lambda}} - \frac{1}{\lambda} + \frac{1}{4\lambda^{3/2}} +\dots\,.
\ee
The intercept of the Pomeron-graviton Regge trajectory should interpolate between this strong coupling expansion and the BFKL weak coupling expansion.
Finally, we show how the flat space limit of the correlation function in the Regge limit,  together with known type IIB S-matrix elements,  gives predictions
for the OPE coefficients  $C_{{\cal OO}J}$ involving the operators  dual to the short string states in the leading Regge trajectory. 

In section \ref{secConclusion}, we
summarize our main findings and discuss some open avenues.

\begin{table}[b!]
\begin{center}
  \begin{tabular}{  | c | c | }
    \hline
   {\bf  Strings in flat spacetime} \T \B & {\bf CFT$_d$ or Strings in AdS$_{d+1}$} \\ \hline
    Scattering amplitude \T &  Correlation function or Mellin amplitude \\ 
    $\mathcal{T}(s,t)$  \B &  $M(s,t)$  \\ \hline
    Tree-level:  $g_s \to 0$ \T  \B & Planar level: $N \to \infty$ \\ \hline
       Finite string length    \T    &        Finite 't Hooft coupling\footnote{}  \\ 
         $l_s=\sqrt{\alpha'} $  \B  &
      $\displaystyle{g^2 \sim  g_{YM}^2 N  =\frac{R^4}{\alpha'^2_{\phantom{1}} }}$ \\  \hline
   Partial wave expansion   \T \B & 
   Conformal partial wave expansion   \\  
    $\displaystyle{ \mathcal{T}(s,t)=\sum_{J} a_J(t)\, \underbrace{P_J(\cos \theta)}_{\rm partial\  wave}} $ \T \B & 
    $ \displaystyle{M(s,t) =\sum_{J_{\phantom{1}}} \int d\nu \,b_J(\nu^2) \,\underbrace{M_{\nu,J}(s,t)}_{\rm partial\  wave}  }$  \\ 
    \hline
   On-shell poles   \T \B &  On-shell poles   \\  
    $ \displaystyle{ a_J(t) \sim \frac{C^2(J)}{t-m^2(J)} }$ \T \B & 
    $  \displaystyle{b_J(\nu^2) \sim   \frac{C^2(J)}{\nu^2+ \big(\Delta(J)-\frac{d}{2} \big)^2_{\phantom{\frac{1}{1}}}} }$  \\ 
      \hline
   Leading Regge trajectory   \T \B & 
    Leading twist operators \\  
    $\displaystyle{      \phantom{\frac{1}{1_{1_1}}}    m^2(J) =  \frac{2}{\alpha'}(J-2)    \phantom{\frac{1}{1_{1_1}}}   }$  \T \B & 
    $ \Delta(J)=d-2+J+ \underbrace{\gamma(J,g^2)}_{\rm anomalous \atop dimension} $  \\ \hline
    Cubic couplings  \T \B & 
   3-pt functions or OPE coefficients \\  
   $C(J)\sim $\!
    \begin{tabular}{r}
     \includegraphics[scale=0.11]{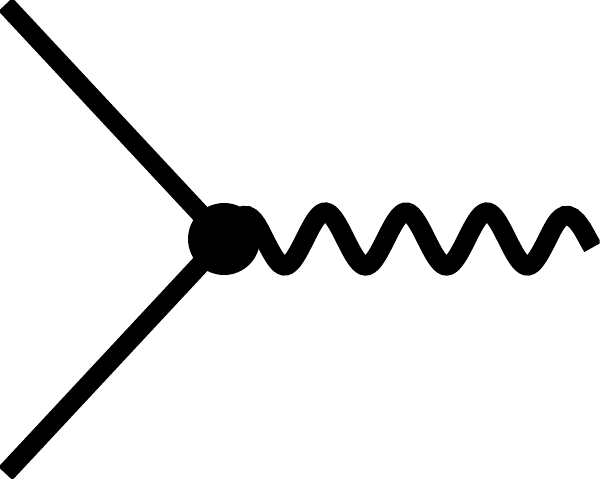}\\
    \end{tabular}
     \T \B & 
     $C(J)\sim $\!
    \begin{tabular}{r}
     \includegraphics[scale=0.11]{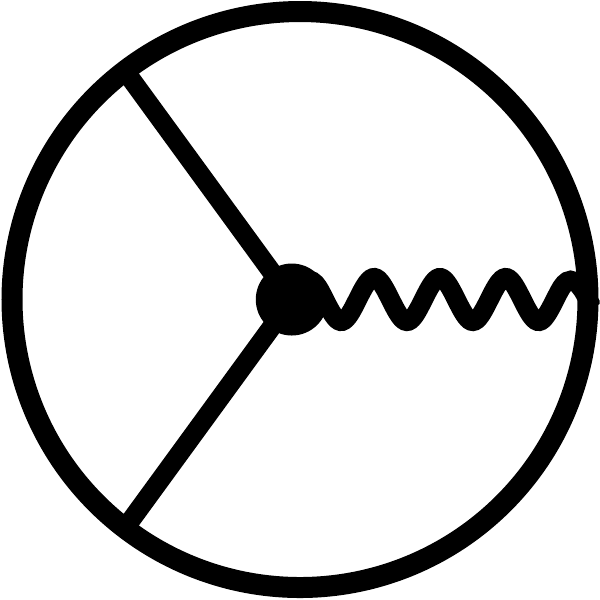} \\
    \end{tabular}
    \\ 
    \hline
    Regge limit:  $s \to \infty$ with fixed $t$   \T \B & 
    Regge limit:  $s \to \infty$ with fixed $t$  
    \\  
    $ P_J(\cos \theta) \approx
    \left(\frac{2s}{t}\right)^{J} $  \T \B  & 
    $ \displaystyle{M_{\nu,J} (s,t) \approx  \omega_{\nu,J}(t)  \,s^{J} }$  \\ 
     $ T(s,t) \approx \beta(t)\, s^{j(t)} $  \T \B & 
    $ \displaystyle{M(s,t) \approx \int_{\phantom{\frac{1}{1}}} d\nu \,\omega_{\nu,j(\nu)}(t) \,\beta(\nu) \,s^{j(\nu)} }$  \\ 
     \hline
    Regge pole and residue \T \B & 
 Regge pole and residue\\  
       $t-m^2(J) = 0$ $\ \Rightarrow\ $ $J=j(t)$  \T \B & 
    $ \left(\Delta(J)-\frac{d}{2}\right)^2+\nu^2=0$ $\ \Rightarrow\ $ $J=j(\nu)$  \\  
      $\beta(t) \sim C^2\!\big(j(t)\big) $  \T \B & 
    $\beta(\nu) \sim C^2\!\big(j(\nu)\big)$ \\ \hline
  \end{tabular}
  \caption{\label{tabela} Analogy between standard Regge theory for scattering amplitudes in flat spacetime and  conformal Regge theory.
  The notation will be explained later, but the analogy should already be clear for readers familiarized with Regge theory (and AdS/CFT).}
\end{center}
\end{table}

\footnotetext[1]{For concreteness we write the formula for a four-dimensional conformal gauge theory. In other  dimensions the 't Hooft coupling can be expressed in terms of the dimensionless ratio $R^2/\alpha'$, where $R$ is the AdS radius.}

\section{Regge theory review}
\label{ReviewRegge}

This section can be safely skipped by the knowledgeable reader. We include it mostly for pedagogical purposes and to emphasize the similarity between standard Regge theory in flat space and the conformal Regge theory we introduce in section \ref{secReggeMellin}. 

\begin{figure}
\begin{centering}
\includegraphics[scale=.35]{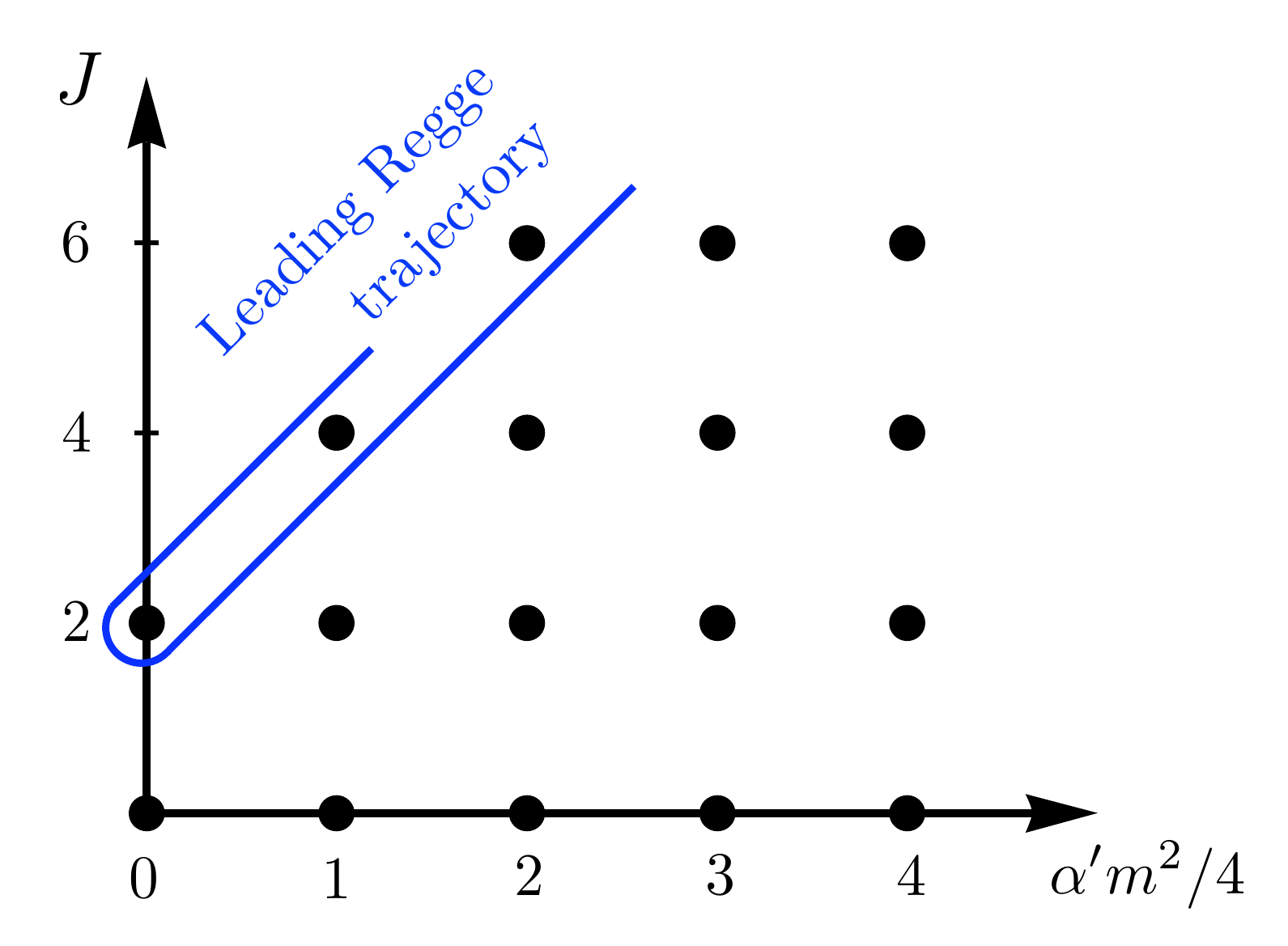}
\par\end{centering}
\caption{\label{ChewFrautschi}
Chew-Frautschi plot of the spectrum of exchanged particles in the Virasoro-Shapiro amplitude. }
\end{figure}

In type II superstring theory, the scattering amplitude of 4 dilatons 
is given by the Virasoro-Shapiro amplitude \cite{VirasoroShapiro}, 
\begin{equation}
\mathcal{T}(s,t)=8\pi G_N\left(\frac{tu}{s}+\frac{su}{t}+\frac{st}{u}\right)
\frac{\Gamma\big(1-\frac{\alpha's}{4}\big)\,\Gamma\big(1-\frac{\alpha'u}{4}\big)\,\Gamma\big(1-\frac{\alpha't}{4}\big)}
{\Gamma\big(1+\frac{\alpha's}{4}\big)\,\Gamma\big(1+\frac{\alpha'u}{4}\big)\,\Gamma\big(1+\frac{\alpha't}{4}\big)} \,, 
\label{VS}
\end{equation}
where $G_N$ is the 10-dimensional Newton constant, $\alpha'$ is the square of the string length and 
\be
s=-(p_1+p_3)^2\ ,\ \ \ \ \ \ 
t=-(p_1+p_2)^2\ ,\ \ \ \ \ \ 
u=-(p_1+p_4)^2\ ,
\label{Mandelstam}
\ee
are the Mandelstam invariants.
\footnote{We  interchanged 
$2\leftrightarrow 3$ relative to the standard definitions. As it will become clear later, the reason for this is to make the t-channel partial wave expansion the analogue of the conformal OPE expansion in the standard channel (12)(34).}
As the dilaton is massless the Mandelstam invariants satisfy
$s+t+u=0.$ This amplitude has an infinite number of poles that correspond to the exchange of an infinite number of particles, 
which can be organized in Regge trajectories as shown in figure \ref{ChewFrautschi}.
This follows from the partial wave expansion
\be
\mathcal{T}(s,t)=\sum_{J=0}^\infty a_J(t)\, P_J(z) \,, 
\label{pwexpansion}
\ee
where $P_J(z)$ are partial waves for 10-dimensional spacetime and 
\be
z=\cos \theta=1+2\, \frac{s}{t}
\ee
encodes the scattering angle $\theta$. 
\footnote{More precisely, the partial waves are just Gegenbauer polynomials (with $D=10$ in our case), 
\be
P_J(z)=\frac{J! \,\Gamma\!\left(\frac{D-3}{2}\right)}{2^J
\Gamma\!\left(J+\frac{D-3}{2}\right)}\, C_J^{\left(\frac{D-3}{2}\right)}(z)\ ,
\ee
which we normalized such that the highest degree term has unit coefficient, 
$P_J(z)=z^J+O(z^{J-1})$.}
In the present 
example only even spins contribute because the initial particles are identical scalars.
To determine the spectrum of exchanged particles we use the fact that each exchanged particle of mass $m$ and spin $J$ gives rise to a pole of $a_J(t)$ at $t=m^2$.
The full scattering amplitude has poles at $t=4 n/\alpha' $, for $n=0,1,2,\dots$, as can be seen in  figure \ref{ChewFrautschi}.
Computing the residues of equations (\ref{VS}) and (\ref{pwexpansion}), we obtain
\be
\sum_{J=0}^\infty  P_J(z) \,
{\rm Res}_{t=\frac{4n}{\alpha'}}a_J(t) ={\rm Res}_{t=\frac{4n}{\alpha'}}\mathcal{T}(s,t)=
- \frac{128\pi G_N}{(\alpha' n!)^2} \left(\frac{n z}{2}\right)^{2+2n} + O(z^{2n})\,,
\ee
where the RHS is a polynomial of degree $2n+2$ in $z$, whose leading term we wrote explicitly. 
In fact, this equation is satisfied with a finite sum over $J$ because this is just an equality between polynomials of $z$. 
More precisely, it tells us that $a_J(t)$ has poles at $t=4n/\alpha'$ for $n=\frac{J}{2}-1,\frac{J}{2},\frac{J}{2}+1,\dots$.
The first pole in this series gives 
\be
a_J(t) \approx 
\frac{r(J)}{t- m^2(J)}\,,
\label{leadingtpole}
\ee
where
\be
m^2(J)=\frac{2}{\alpha'}(J-2)\,,\ \ \ \ \ \ \ \ \
r(J)=  - \frac{128\pi G_N}{\alpha'^2 \Gamma^2(J/2)} 
\left(\frac{J-2}{4}\right)^J\,.
\ee
This pole describes the leading Regge trajectory, i.e. the lightest exchanged particle for each spin $J$. The residue of the pole encodes the cubic couplings between the external particles and the exchanged particles in the leading Regge trajectory.

\begin{figure}
\begin{centering}
\includegraphics[scale=0.3]{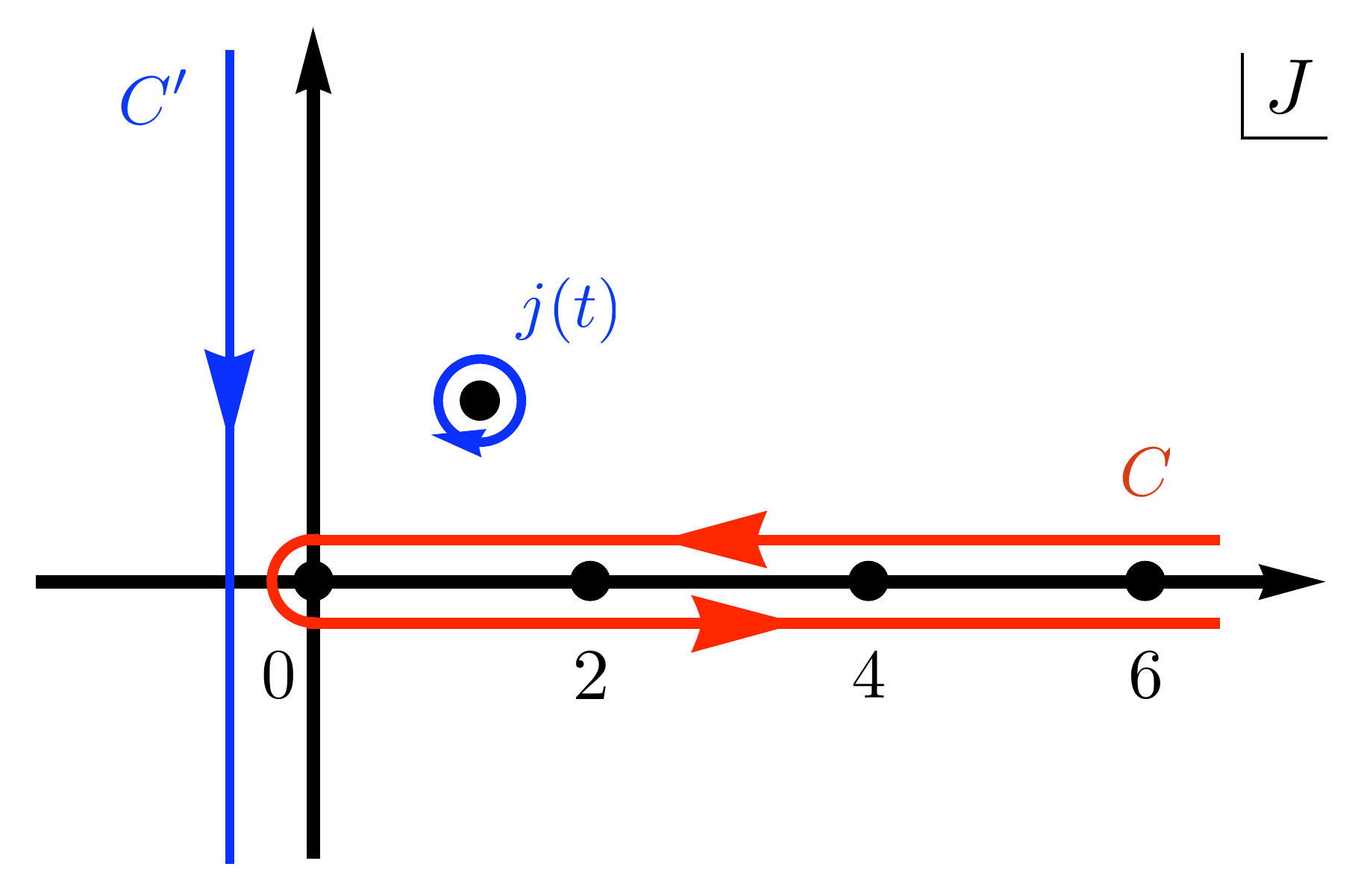}
\par\end{centering}
\caption{\label{fig:Sommerfeld-Watson-transformation}
Integration contours in the $J$-plane involved in the Sommerfeld-Watson transform. Initially, the sum is written as an integral over the contour $C$ encircling all positive integers. Then, the contour is deformed to the contour $C'$ plus the contribution of the Regge poles. }
\end{figure}

The goal of Regge theory is to describe the high energy limit of scattering processes. 
We shall think of the amplitude (\ref{VS}) describing elastic scattering of the initial particles 1 and 3 to the final particles 2 and 4, respectively.
Thus, the Regge regime is defined by large $s$ and fixed $t$ given by (\ref{Mandelstam}).
In this limit the amplitude  (\ref{VS})  simplifies to 
\footnote{This expression is valid for $|s|$ large in any direction of the complex plane, 
except along the real axis where the amplitude has an infinite series of poles in both directions. 
If we take $s$ large and almost real, the amplitude has a different phase depending if we go slightly above or below the real axis.
This is encoded in the $i\epsilon$ prescription.}
\begin{equation}
\mathcal{T}(s\pm i\epsilon,t) \approx \frac{32\pi G_{N}}{\alpha'}\,e^{\mp\frac{i\pi\alpha't}{4}}
\frac{\Gamma\big(-\frac{\alpha't}{4}\big)}
{\Gamma\big(1+\frac{\alpha't}{4}\big)}
\left(\frac{\alpha's}{4}\right)^{2+\frac{\alpha't}{2}} \,.
\label{eq:amplitude V-S regge}
\end{equation}
In this example, it was trivial to obtain the Regge limit of the scattering amplitude because we knew the exact amplitude (\ref{VS}).
However, the achievement of Regge theory is to derive the behaviour of the amplitude in the Regge limit without knowing the full result.
To understand how this works, it is instructive to stick to this example and ask the question: what is the minimal amount of information that we need to fix the amplitude in the Regge limit?
The answer is the spectrum of particles in the leading Regge trajectory and their cubic couplings to the external particles.
Let us review how this works.

The first step is to analytically continue the partial waves as a function of the spin $J$, and then transform the sum (\ref{pwexpansion}) into a contour integral in the $J$-plane, \footnote{Usually, this step requires more care because even and odd spin partial waves must be analytically continued separately \cite{Gribov}.
In our case, there are only even spins and one analytic continuation is sufficient. }
\be
\mathcal{T}(s,t)=\int_{\rm C} \frac{dJ}{2\pi i }
  \frac{\pi}{2 \sin(\pi J)}\,a_J(t)\big( P_J(z) +P_J(-z) \big) \,,
\ee
where the contour $C$ is shown in figure \ref{fig:Sommerfeld-Watson-transformation}.
The symmetry property $P_J(-z)=(-1)^J P_J(z)$ ensures that we are only summing over even spins. The final step is to continuously deform the integration contour $C$ to the contour $C'$ also
shown in figure \ref{fig:Sommerfeld-Watson-transformation}.
This is possible because of the large $J$ behaviour of the partial wave
\be
P_J(\cos\theta) \approx 
\frac{ \cos\!\left(J\theta+\frac{7}{2} \theta +\frac{\pi}{4} \right)}{
   2^{J+\frac{5}{2}}\sin^{\frac{7}{2}}(\theta )}\,,
\ee
and because the analytically continued partial amplitude $a_J(t)$ does not increase exponentially in any direction in the right half of the complex $J$--plane \cite{Gribov}.
In the contour deformation process, one picks up contributions from poles of $a_J(t)$ with $\Re (J)>0$. These are Regge poles and are directly related to the physical poles of the scattering amplitude.
In particular, the pole with largest $\Re (J)$ follows from the leading Regge trajectory (\ref{leadingtpole}) and can be written as 
\be
a_J(t) \approx   -\frac{j'(t)\,r\big(j(t)\big)}{J-j(t)}\,,
\ee
where $j(t)=2+\alpha' t/2$.
The contribution of this Regge pole for the scattering amplitude reads
\be
  j'(t)\,r\big(j(t)\big) \,\frac{\pi}{ 2 \sin\!\big(\pi j(t)\big) } 
\left( P_{j(t)}(z) +P_{j(t)}(-z) \right) \,,
\ee
and therefore in the Regge limit of large $s$ we obtain
\be
 \mathcal{T}(s ,t) \approx - \frac{32\pi^2 G_N }{\alpha'  \Gamma^2\big(j(t)/2\big)} 
  \left(\frac{\alpha' }{4}\right)^{j(t)}
 \frac{s^{j(t)}+(-s)^{j(t)}}{\sin\!\big(\pi j(t)\big)}\,.
\ee
This gives exactly the Regge limit (\ref{eq:amplitude V-S regge}) of the Virasoro-Shapiro amplitude.
This precise matching follows from the fact that the other Regge poles have
smaller $\Re (J)$ and therefore are subdominant in the Regge limit.
\section{Mellin amplitudes \label{secMellin}}

Mellin amplitudes were introduced by Mack in \cite{Mack}.
They are very convenient because they make manifest  the analogy between scattering amplitudes and conformal correlation functions \cite{JPMellin,NaturalMellin,PaulosMellin, JLAnalyticMellin,VolovichMellin,PaulosMellinweak,JLBoundedMellin}.
In this section, we will review the definition and discuss the  main properties of Mellin amplitudes.
In addition, we shall build the necessary tools for the conformal Regge theory developed in section \ref{secReggeMellin}.

The Mellin amplitude $M(\delta_{ij})$ associated to the connected part of the four-point function of scalar primary operators is defined by
\be
A(x_i)=\langle \mathcal{O}_1(x_1)\dots \mathcal{O}_4(x_4) \rangle_{\rm c}
= \int [d\delta]\, M(\delta_{ij}) \prod_{1\le i<j\le 4} \Gamma(\delta_{ij}) (x_{ij}^2)^{-\delta_{ij}}\,,
\ee 
where the integrals run parallel to the imaginary axis.
Conformal invariance constraints the integration variables to satisfy
\be
\sum_{j=1}^4 \delta_{ij} =0 \,,
\label{deltaconstraints}
\ee
where $-\delta_{ii}\equiv \Delta_i$ is the dimension of the operator $\mathcal{O}_i$.
Imposing these four constraints leaves us with two independent integration variables.
The analogy with flat space scattering amplitudes becomes more explicit if we introduce fictitious momenta $p_i$ such that
\be
\sum_{i=1}^4 p_i=0\,,\ \ \ \ \ \ \ \ \ \ \ \ \delta_{ij}=p_i\cdot p_j\,.
\ee
Notice that due to momentum conservation the constraints (\ref{deltaconstraints}) are automatically satisfied.
It is natural to introduce the analogue of the Mandelstam invariants  
\footnote{The shift in the definition of $s$ is convenient to simplify the formulas for the Mack polynomials given below. In any case, this shift is irrelevant in the Regge limit of large $s$.}
\be
t=-(p_1+p_2)^2=\Delta_1+\Delta_2-2\delta_{12}\,,
\ \ \ \ \ \ \ 
s=-(p_1+p_3)^2-\D_1-\D_4=
\Delta_{3}-\Delta_{4}-2\delta_{13}\,,
\ee
which we shall use as  independent Mellin variables. 
The reduced correlator $\mathcal{A}$ is defined by
\be
A(x_i) =
\frac{1}{(x_{12}^2)^{\frac{\D_1 + \D_2}{2}} (x_{34}^2)^{\frac{\D_3 + \D_4}{2}}}
\left( \frac{x_{24}^2}{x_{14}^2}\right)^{\frac{\D_{1}-\D_2}{2}} 
\left(\frac{x_{14}^2}{x_{13}^2}\right)^{\frac{\D_{3}-\D_4}{2}}
\mathcal{A}(u,v) \,,
\ee
such that it  only depends on the conformal invariant cross ratios
\be
u= \frac{x_{12}^2 x_{34}^2}{x_{13}^2 x_{24}^2}\ ,\ \ \ \ \ \ \ 
v= \frac{x_{14}^2 x_{23}^2}{x_{13}^2 x_{24}^2}\ .
\label{CrossRatios}
\ee
Then, the reduced correlator has the following Mellin representation
\begin{align}
\mathcal{A}(u,v)  =&
 \int_{-i\infty}^{i\infty} \frac{dt ds}{(4\pi i)^2} \,M(s,t)\,
u^{t/2} v^{-(s+t)/2}\,
\Gamma\!\left( \frac{\Delta_1 +\Delta_{2} -t}{2}\right)
\Gamma\!\left( \frac{\Delta_3 +\Delta_{4} -t}{2}\right)
  \label{reducedMellin}\\&
  \Gamma\!\left( \frac{ \Delta_{34} -s}{2}\right)
\Gamma\!\left( \frac{-\Delta_{12}  -s}{2}\right)
\Gamma\!\left( \frac{t+s }{2}\right)
\Gamma\!\left( \frac{t+s +\Delta_{12} -\Delta_{34} }{2}\right)\,,
  \nonumber
\end{align}
where $\D_{ij}=\D_i-\D_j$.
The integration contours run parallel to the imaginary axis 
and should be placed such that the infinite series of poles produced by each $\Gamma$-function stays entirely to one side of the contour.
The same requirement applies to the poles of the Mellin amplitude $M(s,t)$ itself, which are described in equation  
(\ref{Mellinpoles}) below.

%

\subsection{Operator product expansion \label{secOPE} }

The structure of the OPE implies a very simple analytic structure 
for the Mellin amplitude if the CFT has a discrete spectrum of operator dimensions \cite{Mack}. 
In appendix \ref{apMellin}, we explain how this works in detail. 

The OPE of two scalar primary operators
only contains totally symmetric and traceless tensors. 
It reads
\be
\mathcal{O}_1(x)\mathcal{O}_2(0) =
\sum_k \frac{C_{12k}}{(x^2)^{\frac{1}{2}(\Delta_1+\Delta_2-\Delta)}}
\left[ \frac{x_{\mu_1}\dots x_{\mu_J}}{(x^2)^{\frac{J}{2}} }\,
\mathcal{O}_k^{\mu_1\dots \mu_J} (0) + {\rm descendants} \right] ,
\ee
where $\D$ and $J$ are respectively the dimension and spin of the 
operator $\mathcal{O}_k$, and all operators are normalized to have two-point function
\be
\left\langle \mathcal{O}_{\mu_1\dots \mu_J} (x) 
\mathcal{O}_{\nu_1\dots \nu_J} (0)  \right\rangle =
\frac{1}{J!}\sum_{{\rm perm}\ \sigma} 
 \frac{I_{\mu_1\nu_{\sigma(1)}}I_{\mu_J\nu_{\sigma(J)}} 
 }{(x^2)^{\Delta}} -{\rm traces}\,,
\label{normalized2pf} 
\ee
with
\be
I_{\mu\nu}= \eta_{\mu\nu}-2\,\frac{x_\mu x_\nu}{x^2}\,.
\ee
This implies the following conformal block expansion of the reduced correlator
\be
\mathcal{A}(u,v)= \sum_k C_{12k}C_{34k} \, G_{\D_k,J_k}(u,v)\,,
\label{CBE}
\ee
where, in the limit $ u\to 0,\ v\to 1$ with $(v-1)/\sqrt{u}$ fixed, 
the conformal block $G_{\D,J}(u,v) $ satisfies
\be
G_{\D,J}(u,v)  \approx \frac{J!}{2^J \left(h-1\right)_J} \,
u^\frac{\D}{2} C_J^{h-1} \!\left( \frac{v-1}{2\sqrt{u}}\right) .
\ee
In this expression $C_J^{h-1}$ is the Gegenbauer polynomial and 
we shall use throughout this paper
\be
h=\frac{d}{2}\,.
\ee
 
In order to reproduce the power law behavior of $\mathcal{A}$ at small cross ratio $u$ predicted by the OPE, the Mellin amplitude must have poles in the variable $t$. More precisely, 
\be
M(s,t)\approx \frac{C_{12k}C_{34k} \,\mathcal{Q}_{J,m}(s)}{t-\Delta+J-2m}\,,\ \ \ \ \ \ \ \
m=0,1,2,\dots \,, \label{Mellinpoles}
\ee
where, as before, $\Delta$ and $J$  are 
the dimension and spin of an operator $\mathcal{O}_k$ that appears in both OPEs $\mathcal{O}_1\mathcal{O}_2$ and $\mathcal{O}_3\mathcal{O}_4$.
This shows that the $m>0$ poles correspond to conformal descendant operators with twist greater than $\D-J$.
The residues of the poles are kinematical polynomials $\mathcal{Q}_{J,m}(s)$ of degree $J$ in the Mellin variable $s$.
It is convenient to write $\mathcal{Q}_{J,m}(s)$ in terms of new polynomials $Q_{J,m}( s)$ defined by
\begin{align}
\mathcal{Q}_{J,m}( s) =&-
\frac{2\Gamma(\Delta+J) (\D-1)_J }{ 4^J
\Gamma\!\left( \frac{\Delta +J+\Delta_{1 2}}{2}\right)
\Gamma\!\left( \frac{\Delta +J-\Delta_{1 2}}{2}\right) 
\Gamma\!\left( \frac{\Delta +J+\Delta_{3 4}}{2}\right)
\Gamma\!\left( \frac{\Delta +J-\Delta_{3 4}}{2}\right) }
\label{Q}
\\
&
\frac{  Q_{J,m}( s) }{m!(\Delta-h+1)_m \,
\Gamma\!\left( \frac{\Delta_1 +\Delta_{2} -\D+J}{2}-m\right)
\Gamma\!\left( \frac{\Delta_3 +\Delta_{4} -\D+J}{2}-m\right)}\,,
\nonumber
\end{align}
where we used the Pochhammer symbol
\be
(a)_m =\frac{\Gamma(a+m)}{\Gamma(a)}=a(a+1)\dots(a+m-1)\,.
\ee
In appendix \ref{apMellin} we study these kinematical polynomials in detail and show, in particular, that with the above normalization
\be
Q_{J,m}( s)=s^J +O(s^{J-1})\,.
\label{Qasymp}
\ee
  
In order to obtain the conformal block $G_{\D,J}(u,v)$ we only kept the contribution from the series of poles (\ref{Mellinpoles}) in the integral (\ref{reducedMellin}). However, the integrand in  (\ref{reducedMellin}) has more poles in the variable $t$. These poles occur at $t=\D_1+\D_2+2m$ and $t=\D_3+\D_4+2m$, which is the twist of the composite operators 
$\mathcal{O}_1\partial_{\mu_1} \dots \partial_{\mu_J} \partial^{2m}\mathcal{O}_2$ and $\mathcal{O}_3\partial_{\mu_1} \dots \partial_{\mu_J} \partial^{2m}\mathcal{O}_4$, in the limit where the external operators $\mathcal{O}_i$ interact weakly. 
In this paper, we shall focus on the planar part of the four-point function of single-trace operators in large $N$ gauge theories. In this case, the poles of the $\Gamma$-functions in the integrand of  (\ref{reducedMellin}) automatically account for the contribution of double-trace operators in the OPE and the Mellin amplitude only has poles associated to single-trace contributions to the OPE.

\subsection{Conformal partial waves}

The first step to study Regge theory is to write down a partial wave expansion. For our purposes, the best starting point is the partial 
wave expansion described in \cite{Mack}, which is the Mellin space version of \cite{Sofia}.
We write
\be
M(s,t)= \sum_{J=0}^\infty \int_{-\infty}^\infty  d\nu  \, b_J(\nu^2) \,M_{\nu,J}(s,t)\,,
\label{CPWE}
\ee
with the partial waves $M_{\nu,J}(s,t)=M_{-\nu,J}(s,t)$  given by
\be
M_{ \nu,J}(s,t)=  \omega_{\nu,J}(t) \, P_{\nu,J}(s,t)\,,
\label{MellinPartialWave}
\ee
where 
\begin{align}
 \omega_{\nu,J}(t)=\ &
 \frac{ \Gamma\!\left( \frac{\Delta_1 +\Delta_{2} +J+i\nu-h}{2}\right)
\Gamma\!\left( \frac{\Delta_3 +\Delta_{4}  +J+i\nu-h}{2}\right)
 \Gamma\!\left( \frac{\Delta_1 +\Delta_{2} +J-i\nu-h}{2}\right)
\Gamma\!\left( \frac{\Delta_3 +\Delta_{4}  +J-i\nu-h}{2}\right) 
 }{ 8\pi \Gamma(i\nu) \Gamma(-i\nu) }
 \nonumber
\\&
 \frac{\Gamma\!\left( \frac{h+i\nu-J-t}{2}\right) 
 \Gamma\!\left( \frac{h-i\nu-J-t}{2}\right)
 }{ \Gamma\!\left( \frac{\Delta_1 +\Delta_{2} -t}{2}\right)
\Gamma\!\left( \frac{\Delta_3 +\Delta_{4} -t}{2}\right)}\,,
\label{omeganuJt}
\end{align}
and $P_{\nu,J}(s,t)$ is a Mack polynomial of degree $J$ in both variables $s$ and $t$.
We normalized these polynomials such that they obey $P_{\nu,J}(s,t) = s^J+O(s^{J-1})$. The precise definition is given in appendix \ref{AppendixMack}.

The conformal partial wave expansion  (\ref{CPWE}) is closely related to the conformal block decomposition  (\ref{CBE}).
\footnote{These two terminologies are often used as synonymous in the literature. In this paper, we shall call conformal partial wave expansion to (\ref{CPWE}) and (\ref{CPWEp}), and conformal block decomposition to (\ref{CBE}).} 
This is more easily seen if we transform  (\ref{CPWE}) to position space,
\be
\mathcal{A}(u,v)= \sum_{J=0}^\infty \int_{-\infty}^\infty  d\nu  \, b_J(\nu^2) \,F_{\nu,J}(u,v)\,,
\label{CPWEp}
\ee
where $F_{\nu,J}(u,v)$ is the transform (\ref{reducedMellin}) of a single partial wave $M_{ \nu,J}(s,t)$,
\begin{align}
F_{\nu,J}(u,v)  =&
 \int_{-i\infty}^{i\infty} \frac{dt ds}{(4\pi i)^2} \,M_{\nu,J}(s,t)\,
u^{t/2} v^{-(s+t)/2}\,
\Gamma\!\left( \frac{\Delta_1 +\Delta_{2} -t}{2}\right)
\Gamma\!\left( \frac{\Delta_3 +\Delta_{4} -t}{2}\right)\\&
  \Gamma\!\left( \frac{ \Delta_{34} -s}{2}\right)
\Gamma\!\left( \frac{-\Delta_{12}  -s}{2}\right)
\Gamma\!\left( \frac{t+s }{2}\right)
\Gamma\!\left( \frac{t+s +\Delta_{12} -\Delta_{34} }{2}\right).
  \nonumber
\end{align}
In fact, $F_{\nu,J}(u,v)$ is given by the sum of two conformal blocks with dimensions $h+i\nu$ and
$h-i\nu$.
More precisely, one can write
\footnote{
A similar equation can be found in \cite{DOMellin} where the function $F_{\nu,J}(u,v)$ was defined by the integral of the product of the 3-point function of the operators $\mathcal{O}_1$, $\mathcal{O}_2$ and an operator of spin $J$ and dimension $h+i\nu$, times  the 3-point function of the operators $\mathcal{O}_3$, $\mathcal{O}_4$ and an operator of spin $J$ and dimension $h-i\nu$.
}
\be
F_{\nu,J}(u,v)= \kappa_{\nu,J}\, G_{h+i\nu,J}(u,v)+
\kappa_{-\nu,J}\,G_{h-i\nu,J}(u,v)\,,
\label{cb+shadow}
\ee
where the normalization constant
\be
\kappa_{\nu,J} =\frac{i \nu}{2\pi K_{h+i\nu,J}}
\ee
can be fixed by comparing the residues of $M_{\nu,J}(s,t)$ at $t=h\pm i \nu -J+2m$ with the general expression (\ref{Mellinpoles}), and  
\begin{align}
K_{\D,J}=\ & \frac{\Gamma(\Delta+J) \,\Gamma(\Delta-h+1) \, (\Delta-1)_J   }{ 
4^{J-1} 
\Gamma\!\left( \frac{\Delta +J+\Delta_{1 2}}{2}\right)
\Gamma\!\left( \frac{\Delta +J-\Delta_{1 2}}{2}\right) 
\Gamma\!\left( \frac{\Delta +J+\Delta_{3 4}}{2}\right)
\Gamma\!\left( \frac{\Delta +J-\Delta_{3 4}}{2}\right)}
\label{KDeltaJ}\\&
\frac{1}{
 \Gamma\!\left( \frac{\Delta_1 +\Delta_{2} -\D+J}{2}\right)
\Gamma\!\left( \frac{\Delta_3 +\Delta_{4} -\D+J}{2}\right)
\Gamma\!\left( \frac{\Delta_1 +\Delta_{2} +\D+J-d}{2}\right)
\Gamma\!\left( \frac{\Delta_3 +\Delta_{4} +\D+J-d}{2}\right)}\,.
 \nonumber
\end{align}
Notice that the two conformal blocks in (\ref{cb+shadow}) satisfy the same differential equation (\ref{CasimirPDE}) because they have the same Casimir $C_{h+i\nu,J}=C_{h-i\nu,J}$.
The second conformal block is usually called the shadow of the first (see for example \cite{SimmonsDuffin:2012uy} for details).
Inserting (\ref{cb+shadow}) in
(\ref{CPWEp}), one can write 
\be
\mathcal{A}(u,v)= 2\sum_{J=0}^\infty \int_{-\infty}^\infty  d\nu  \, b_J(\nu^2) \,
\kappa_{\nu,J}\, G_{h+i\nu,J}(u,v)\,,
\ee
which can be easily converted into the usual conformal block decomposition  (\ref{CBE}) by deforming the $\nu$-contour into the lower-half plane and picking the contribution from all poles of the integrand with negative imaginary part of $\nu$. Notice that the contribution from infinity vanishes because the conformal block $G_{h+i\nu,J}(u,v)$ decays exponentially for $\Im (\nu) \to -\infty$.
Thus, in order to reproduce the contribution of a single-trace operator $\mathcal{O}_k$ of dimension $\D$ and spin $J$ that appears in both OPEs 
$\mathcal{O}_1\mathcal{O}_2$  and
$\mathcal{O}_3\mathcal{O}_4$,
the partial amplitude $b_J(\nu^2)$ must have  poles of the form
\be
b_J(\nu^2)\approx C_{12k}C_{34k}\,\frac{ K_{\D,J}}{\nu^2+(\Delta-h)^2}\,.
\label{polesfromCB}
\ee
We  rederive this result  in  appendix  \ref{analyticstructure} where we discuss the analytic structure of the partial amplitude $b_J(\nu^2)$  more systematically.

\section{Conformal Regge theory \label{secReggeMellin}}

We are  ready to generalize Regge theory to Mellin amplitudes. To avoid cluttering of the formulae 
we shall restrict to the case $\D_{12}=\D_{34}=0$.
It is convenient to change from the variable $s$ to the variable $z=1+\frac{2s}{t}$.
We define
\be
\mathcal{P}_{\nu,J}(z,t) = P_{\nu,J}\!\left(\frac{t(z-1)}{2},t\right).
\ee
Then, for integer spin $J$, we have the symmetry
\be
\mathcal{P}_{\nu,J}(-z,t)=(-1)^J \mathcal{P}_{\nu,J}(z,t)\,.
\ee
The starting point for Regge theory is the conformal partial wave expansion (\ref{CPWE}). The construction is now analogous to that reviewed in section \ref{ReviewRegge}.
Firstly, we analytically continue the partial amplitudes $b_J(\nu^2)$ to complex values of $J$. 
Even and odd spins give rise to different analytic continuations 
$b_J^+(\nu^2)$ and $b_J^-(\nu^2)$, respectively.
Secondly, we perform a Sommerfeld-Watson transform in  (\ref{CPWE}),
\be
M(s,t)=M^+(s,t)+M^-(s,t)\,,
\ee
with
\be
M^\pm(s,t)=  \int_{-\infty}^\infty  d\nu  \int_{\rm C} \frac{dJ}{2\pi i}
\frac{\pi  }{2\sin (\pi J)} \, b_J^\pm(\nu^2)\,\omega_{\nu,J}(t)
\,\big( \mathcal{P}_{\nu,J}(-z,t)\pm\mathcal{P}_{\nu,J}(z,t)\big)\,.
\ee
The next step in Regge theory is to deform the $J$-contour and pick up the pole with maximal real part of $J$, i.e. the leading Regge pole.
Before doing this we need to consider the poles (\ref{partialwavepole})
of the partial amplitude $b_J(\nu^2)$.
We will be mostly interested in the poles associated to the leading Regge trajectory
$\D(J)$ for $J=2,4,6,\dots$. These are the operators of lowest dimension for each even spin.
This means that 
\be
b_J^+(\nu^2)\approx \frac{r(J)}{\nu^2+\big(\Delta(J)-h\big)^2}\,,
\ee
where the residue
\be
r(J)= C_{12J}C_{34J}\,   K_{\D(J),J}\,,
\label{residuerofJ}
\ee
is determined by the OPE coefficients of the operators in the leading Regge trajectory that appear in the
OPEs of the external operators (see equation (\ref{polesfromCB})).
After analytic continuation in $J$ this pole becomes a pole in $J$, more precisely
\be
b_J^+(\nu^2)\approx -\frac{j'(\nu)\,r(j(\nu))}{2\nu\big(J-j(\nu)\big)}\,,
\ee
where $j(\nu)$ is essentially the inverse function of $\D(J)$ defined by
\be
\nu^2+\big(\Delta(j(\nu))-h\big)^2=0\,.
\label{j}
\ee
The contribution of this Regge pole is then
\be
  \int_{-\infty}^\infty  d\nu  \,
\frac{\pi  }{2\sin (\pi j(\nu))} \,\frac{j'(\nu)\,r(j(\nu))}{2\nu }\,\omega_{\nu,j(\nu)}(t)
\left( \mathcal{P}_{\nu,j(\nu)}(-z,t)+\mathcal{P}_{\nu,j(\nu)}(z,t)\right).
\ee
In the Regge limit ($s\to \infty$) this pole dominates and we obtain the result
\be
M(s,t)\approx \int  d\nu \,\beta(\nu)
\,  \omega_{\nu, j(\nu)}(t) \,
\frac{s^{ j(\nu)}+(-s)^{ j(\nu)}}{\sin\! \big(\pi j(\nu)\big) }\,,
\label{MainEquationMellin}
\ee 
where
\be
\beta(\nu)= \frac{\pi}{4\nu }\,  j'(\nu)\,r\big(j(\nu)\big)=
\frac{\pi}{4\nu }\,  j'(\nu)\,   
K_{h\pm i\nu,j(\nu)}\,
C_{12j(\nu)}C_{34j(\nu)}\,.
\label{betaofnu}
\ee
Equation (\ref{MainEquationMellin}) is our main result. It encodes the contribution of a Regge trajectory to the Mellin amplitude, which is fixed by conformal symmetry up to the dynamical observables $j(\nu)$ and $\beta(\nu)$. The Reggeon spin $j(\nu)$ is determined by the dimensions $\Delta(J)$ of the physical operators in the leading Regge trajectory through (\ref{j}). The residue $\beta(\nu)$ is controlled by the OPE coefficients of the leading twist operators in the OPE of the external operators.
This result should be valid in any CFT that has a large $N$ expansion. At leading order in $1/N$ the CFT is described by the dual AdS theory at tree-level, which implies the absence of
poles in the complex $J$-plane from
multi-particle states. This justifies our assumption of single Regge pole dominance.

The definition of the Regge limit of the Mellin amplitude (large $s$ and fixed $t$) corresponds to the Regge limit defined in position space in \cite{CornalbaRegge,ourBFKL}. 
This is shown in detail in appendix \ref{secReggepos}. For the sake of clarity, here we just state how to relate the result (\ref{MainEquationMellin}) to
the Regge limit of the correlator in position space, leaving the details to the appendix. First one needs to consider a specific Lorentzian kinematical limit
where all the points are taken to null infinity. In such limit, it is convenient to introduce the  variables $\sigma$ and $\rho$ that are
related to the  cross ratios $u$ and $v$ defined in (\ref{CrossRatios}) by
\be
u=\sigma^2\,,\ \ \ \ \ \ \ 
v=(1-\sigma e^\rho)(1-\sigma e^{-\rho})\approx 1-2\sigma \cosh \rho\,.
\ee
The  Regge limit corresponds to $\sigma \to 0$ with fixed $\rho$. The  position space version of equation (\ref{MainEquationMellin}) is then
\footnote{We remark that the definition of $\alpha(\nu)$ in this paper differs from that in  \cite{CornalbaRegge,ourBFKL} by a factor of $4^{1-j(\nu)} e^{i\pi j(\nu)}$.}
\begin{align}
\mathcal{A}(\sigma,\rho) \approx 2\pi i    
\int  d\nu \,\alpha(\nu)\,
\sigma^{1-j(\nu)}   \Omega_{i\nu}(\rho)\,,
 \label{ReggeA}
\end{align}
where $ \Omega_{i\nu}(\rho)$ is a harmonic function 
on $(2h-1)$-dimensional hyperbolic space. 
In appendix \ref{secReggepos}, we show that the residues  $\alpha(\nu)$ in (\ref{ReggeA})  and $\beta(\nu)$ in (\ref{MainEquationMellin}) are related by
\be
\alpha(\nu)= -\frac{  \pi^{h-1} 
2^{j(\nu)-1}  e^{i \pi j(\nu) /2 }}
{ \sin \!\left(\frac{\pi j(\nu)}{2}\right) }    \,\g(\nu)\gamma(-\nu)\beta(\nu)\,,
\label{alpha}
\ee
where
\be
 \gamma(\nu)=
\Gamma\!\left( \frac{2\Delta_1  +j(\nu)+i\nu-h}{2}\right)
\Gamma\!\left( \frac{2\Delta_3    +j(\nu)+i\nu-h}{2}\right).
\ee
The form (\ref{ReggeA}) was first derived in \cite{CornalbaRegge} applying Regge theory to the conformal partial wave expansion. Here, we have improved the result because we related the functions $\alpha(\nu)$ and $\beta(\nu)$ to the product of OPE coefficients (see equation (\ref{betaofnu})).


\section{Pomeron-graviton Regge trajectory in SYM \label{secSYM}}

The result (\ref{MainEquationMellin}) relates, through equation (\ref{betaofnu}), 
the four-point function of a CFT in the Regge limit  to the analytic continuation of the OPE coefficients of the operators in the leading Regge trajectory in the OPEs 
$\mathcal{O}_1\mathcal{O}_2$ and 
$\mathcal{O}_3\mathcal{O}_4$ of the external operators. 
In particular, (\ref{betaofnu}) will allow us to make non-trivial predictions about these OPE coefficients.
In this section we shall consider in detail the case of  $\mathcal{N}=4$ SYM in the planar limit and its string dual,
so we set $d=2h=4$. 

We will consider the case of correlation functions that exchange the quantum numbers of the vacuum, so that they are dominated in the 
Regge limit by the exchange of the pomeron Regge trajectory. 
Furthermore, as external operators we shall consider BPS scalars operators, such that their dimensions are protected.
At weak coupling, the operators in the  pomeron-graviton Regge trajectory are a linear combination of the following twist two operators
\be
{\rm tr} \left( F_{\mu \nu_1} D_{\nu_2} \dots D_{\nu_{J-1}} F_{\nu_J}^{\ \ \mu} \right),\ \ 
{\rm tr} \left(  \phi_{AB} D_{\nu_1} \dots D_{\nu_{J}}\phi^{AB}  \right),\ \ 
{\rm tr} \left(   \bar{\psi}_{A} D_{\nu_1} \dots D_{\nu_{J-1}} \Gamma_{\mu_J} \psi^{A} \right).
\label{zeroorderstates}
\ee
At finite 't Hooft coupling the degeneracy is lifted and there are three different Regge trajectories.
\footnote{These three trajectories are related by supersymmetry \cite{KotikovLipatov02}. 
In fact, their anomalous dimensions are simply related by $\tilde{\tilde{\g}}
(J) = \tilde{\g}
(J + 2) = 
\g(J + 4)$.} We will consider the operators in the leading Regge 
trajectory. 
In appendix \ref{Appendix3pt}, we give
the precise linear combination of the above operators that yields the spin $J$ and twist two  operators of lowest dimension, to first order in perturbation theory. 
At strong coupling these operators are dual to massive  string states, on the leading Regge trajectory of type IIB strings in AdS.

Below we shall divide the discussion in two parts, to address both weak and strong coupling expansions. We start by reviewing the consequences of conformal Regge
theory to the spin-anomalous dimension function of twist two operators, as first considered in \cite{DressingWrapping}, and also derive some new results. We then consider
the OPE coefficients.

\subsection{Weak coupling}

\begin{figure}
\begin{centering}
\includegraphics[scale=0.37]{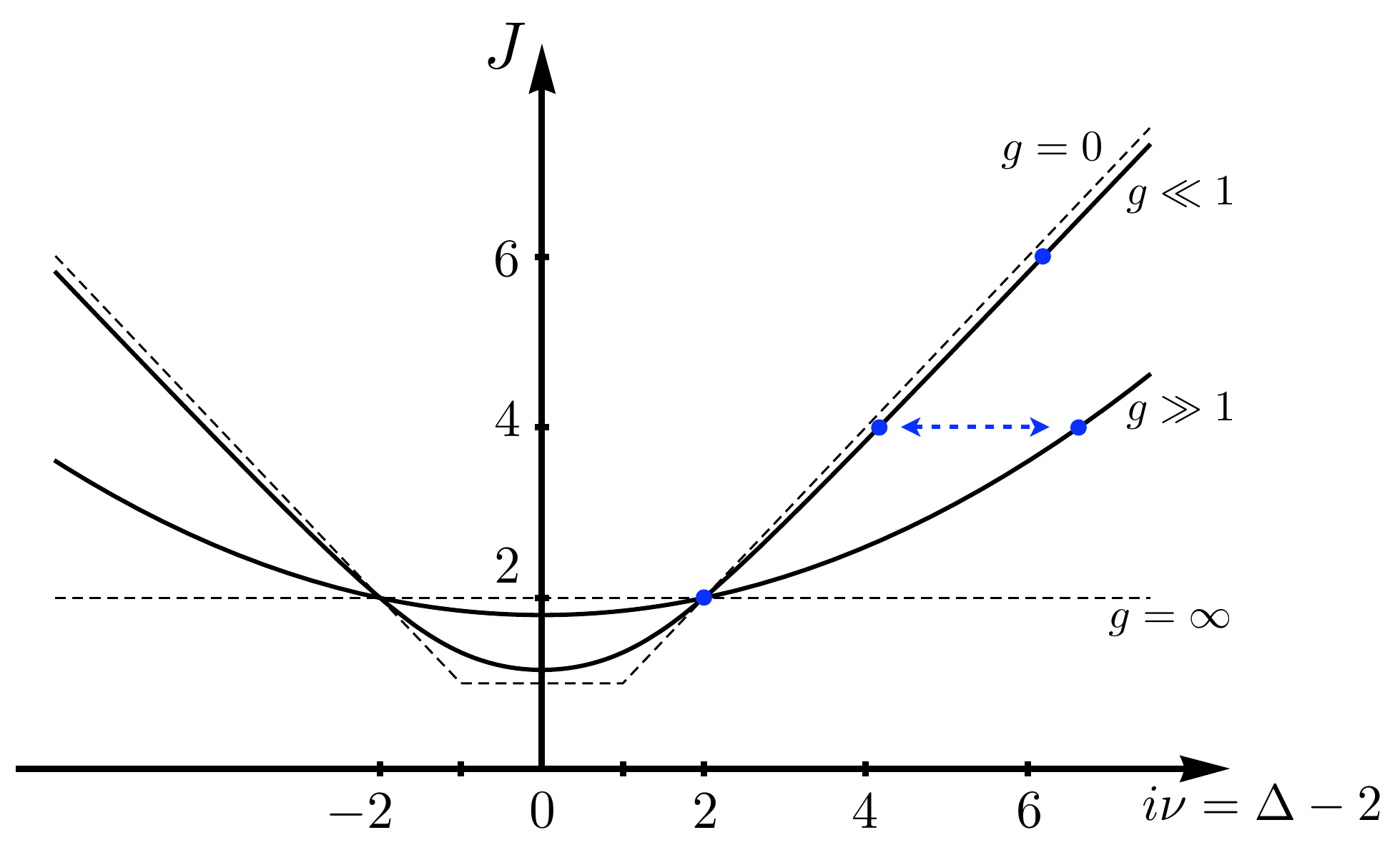}
\par\end{centering}
\caption{Leading Regge trajectory in the dimension--spin plane for various values of the 't Hooft coupling. 
The physical operators have even spin $J$ and positive dimension, and are represented by blue dots along the curves.
The horizontal dashed line $J=2$ corresponds to the strong coupling limit $g\to \infty$.
The weak coupling limit $g \to 0$, is described by the other dashed line, with 3 branches: $J=1$,
$\D-2=J$ and  $2-\D=J$.
The intercept $j(0)$ moves from $1$ to $2$, as the coupling $g$ goes from 0 to $\infty$.
\label{figJ(Delta)}}
\end{figure}

The anomalous dimension of the operators in the leading Regge trajectory is a function of the spin  and of the 't Hooft coupling.
It admits  the following weak coupling expansion 
\be
\g(J)=
\Delta(J) - J -2=   \sum_{n=1}^{\infty} g^{2n}  \g_n(J)\,.
\label{gamma(J)}
\ee
We use notation with coupling $g$  related to the 't Hooft coupling $\lambda=g^2_{YM} N$ by
\be
g^2 = \frac{\lambda}{16\pi^2}\,.
\ee
The anomalous dimensions 
$\g_n$ are known up to  five loops \cite{KotikovLipatov02, 3loopsTwist2, 4loopsTwist2, 5loopsTwist2} and obey the principle of maximal transcendentality \cite{KotikovLipatov02}.
The first two terms in this expansion are
\begin{align}
 \g_{1} &=8S_{1}(x)\,,
 \nonumber
 \\
\g_{2} &=-32\big(S_{3}(x)+S_{-3}(x)\big)+64S_{-2,1}(x)-64\big(S_{1}(x)S_{2}(x)+S_{1}(x)S_{-2}(x)\big)\,,
\label{Anomalous2ndOrder}
\end{align}
where $x=J-2$ and the functions $S$ are (nested) harmonic sums, which are recursively defined by
\begin{equation}
S_{a_1,a_2,\dots,a_n}(x)=\sum_{y=1}^{x}\frac{({\rm sign}(a_{1}))^{y}}{y^{|a_1|}} \, S_{a_2,\dots,a_n}(y)\,, \label{Hsums}
\end{equation}
starting from the trivial seed without indices, $S(y)=1$. 
Some properties of these functions are given in appendix \ref{HarmonicSums}. This weak coupling expansion of the function $\Delta(J)$ is an expansion around 
the free theory line $\Delta(J)=J +2$ (see figure \ref{figJ(Delta)}).

The function $\Delta(J)$ defines the Reggeon spin $j(\nu)$. By inverting  (\ref{j}) we have
\be
\Delta\big(j(\nu)\big)= 2 \pm i \nu\,.
\ee
However, the Reggeon spin can also be computed directly from the Regge limit of the four point correlation function \cite{ourBFKL}. 
At weak coupling, BFKL methods \cite{Fadin:1975cb,Kuraev:1977fs,Balitsky:1978ic}  give an expansion around
the free theory value $j=1$,  also shown in figure \ref{figJ(Delta)},  associated to the exchange of two free gluons in a colour singlet.
The function $j(\nu)$ is known up to next to leading order. Let us first transcribe the result presented in 
\cite{KotikovLipatov02}\footnote{Our definition of $\nu$ differs from that used in \cite{KotikovLipatov02} by a factor of 2.}
\begin{equation}
j(\nu)=1+4g^{2} \Big( \chi(\nu)+g^{2}\delta(\nu) \Big )+O(g^{6})\,,
\label{BFKLspinj}
\end{equation}
with
\begin{align}
\chi(\nu)& =2\Psi(1)-\Psi\!\left(\frac{1+i\nu}{2}\right)-\Psi\!\left(\frac{1-i\nu}{2}\right) ,
\label{BFKLspin2ndOrder}
\\
\delta(\nu)&= 4\chi''(\nu)+6\zeta(3) - 2\zeta(2) \chi(\nu)\,
-2\Phi\!\left(\frac{1+i\nu}{2}\right) -  2\Phi\!\left(\frac{1-i\nu}{2}\right),
\end{align}
where $\Psi(x)=\Gamma'(x)/\Gamma(x)$ is the Euler $\Psi$-function and the function $\Phi$ \footnote{
In the BFKL literature there are two different functions that are usually denoted by $\Phi$. The one defined above is used in \cite{KotikovLipatov02}.  
We denote the other function, used in \cite{DressingWrapping},
by $\tilde{\Phi}$ and explain the connection between the two in Appendix \ref{HarmonicSums}.  } is defined by
\begin{equation}
\Phi(x) =\frac{1}{2} \sum_{k=0}^\infty \frac{\Psi'\! \left(\frac{k+2}{2}\right) -\Psi'\!\left(\frac{k+1}{2}\right) }{k+x}\,.
\ \ \ \ \ \ \ \ \ \ \ \ \ \
\label{eq:funcao phi} \end{equation}
We remark that the  variable $\nu$ in the spin $j(\nu)$,  which 
 appears in the BFKL amplitude as an integration variable, is exactly the same as the one in our treatment 
of conformal Regge theory  \cite{ourBFKL}.
After some manipulation, $j(\nu)$ can be written in a form which makes maximal transcendentality manifest, \footnote{We thank Pedro Vieira for collaboration in this point.}
\begin{equation}
j(\nu) =1+\sum_{n=1}^{\infty}g^{2n}\left[F_{n}\!\left(\frac{i\nu-1}{2}\right)+F_{n}\!\left(\frac{-i\nu-1}{2}\right)\right] ,
\label{BFKLspin}
\end{equation}
where
\begin{align}
F_{1}(x)&=-4S_{1}(x)\,,
\\
F_{2}(x)&=4\left(-\frac{3}{2} \,\zeta(3)+\pi^{2}\ln(2)+\frac{\pi^{2}}{3}\,S_{1}(x)+2S_{3}(x)+\pi^{2}S_{-1}(x)-4S_{-2,1}(x)\right) .
\nonumber
\end{align}

It is important to realize 
that, although the functions $\Delta(J)$ and $j(\nu)$ are basically the inverse of each other, their perturbative expansions, either at weak or at strong coupling,  contain different information \cite{DressingWrapping}.
In other words, the process of inverting the functions does not commute with perturbation theory. Let us consider first the limit  $j\to1$ and $g^2\to 0$, with $(j-1)/g^2$ fixed, of the 
BFKL spin (\ref{BFKLspin}). In this limit, only  the leading order term in the expansion  (\ref{BFKLspin}) survives, and we have
\be
\frac{j(\nu)-1}{g^2} = -4S_{1}\!\left(\frac{i\nu-1}{2}\right)-4S_{1}\!\left(\frac{-i\nu-1}{2}\right) .
\ee
Now, the function in the RHS of this equation has simple poles at $i\nu=\pm 1$. If we expand this equation around one of this points, say around $i\nu= 1$,
the fixed quantity in the LHS of this equation will be very large.  This expansion has the following form
\be
\frac{j(\nu)-1}{-4g^2} = \frac{2}{i\nu-1} - 2 \sum_{k=1}^\infty   \zeta(2k+1) \, \left(\frac{i\nu -1}{2}\right)^{2k}\,,
\label{expansion1}
\ee
where the coefficients can be read from formulae presented in 
 appendix  \ref{HarmonicSums}. We can now invert this equation, solving for
$i\nu -1= \Delta(j)-3$, to obtain the behaviour of $\Delta(J)$ around $J=1$, as a power expansion in the  small quantity  $g^2/(J-1)$.
The result for the  first order terms in this expansion  is \cite{DressingWrapping}
\begin{equation}
\Delta(J) - 3 =   2 \left(\frac{-4g^2}{J-1} \right)+ 0  \left(\frac{-4g^{2}}{J-1}\right)^{2}+0 \left(\frac{-4g^{2}}{J-1}\right)^{3}
- 4\zeta(3)  \left(\frac{-4g^{2}}{J-1}\right)^{4}+ \dots \,.
\end{equation}
The remarkable thing about this expression is that, after inversion of the leading order BFKL spin, one has a prediction for the leading singularities 
of the anomalous dimension function (\ref{gamma(J)}) around $J=1$ at  all orders in perturbation theory.
This fact was explored in \cite{DressingWrapping} and served as a guide for the computation of
anomalous dimensions using integrability, most notably to check  wrapping corrections that appear at four loops \cite{Bajnok:2008qj}.
In the next section we shall follow a similar procedure to study the behaviour of  OPE coefficients.

Let us close these introductory remarks by explaining how higher order terms in the BFKL expansion can be taken into account in the above argument. 
In this case one considers the expansion (\ref{BFKLspin}) with $(j-1)/g^2$ fixed, keeping all terms in the $g^2$ expansion,
instead of only the leading term as we did in (\ref{expansion1}). Then one expands around $i\nu=1$ and inverts. The result is a
prediction for the expansion of the anomalous dimension around $J=1$ at  all orders in perturbation theory. In particular,
the next to leading BFKL spin allows one to predict the next to leading singularity around $J=1$ at  all orders in perturbation theory 
\cite{DressingWrapping}
\begin{align}
\Delta-3
=&\,\Big(  0 + (J-1) \Big) + \Big(2+0\left(J-1\right)\Big)\left(\frac{-4g^{2}}{J-1}\right) + \Big(0+0\left(J-1\right)\Big)\left(\frac{-4g^{2}}{J-1}\right)^{2}
\\
&-\Big(0+\zeta(3)(J-1)\Big)\left(\frac{-4g^{2}}{J-1}\right)^{3}-\left(4\zeta(3)+\frac{5\zeta(4)}{4} \,(J-1)\right)\left(\frac{-4g^{2}}{J-1}\right)^{4}+\dots\,.
\nonumber
\end{align}


Finally let us also note that we can twist around these arguments, and use the knowledge of the anomalous dimension function (\ref{gamma(J)}) to some fixed order in perturbation theory,
to study the behaviour of the BFKL spin 
around $i\nu=1$. Considering
the first two orders in perturbation theory for the anomalous dimension given in (\ref{Anomalous2ndOrder}), one obtains the prediction
\begin{align}
j(\nu)-1 =& \,\Big(1+0(i\nu-1)\Big)\left(\frac{8g^{2}}{i\nu-1}\right)+\left(\frac{1}{i\nu-1}+0\left(i\nu-1\right)\right)\left(\frac{8g^{2}}{i\nu-1}\right)^{2}
\\
&
+\!\left(\frac{2}{\left(i\nu-1\right)^{2}}+\frac{0}{i\nu-1}\right)\!\left(\frac{8g^{2}}{i\nu-1}\right)^{3}
\!+\left(\frac{5}{\left(i\nu-1\right)^{3}}+\frac{0}{\left(i\nu-1\right)^{2}}\right)\!\left(\frac{8g^{2}}{i\nu-1}\right)^{4} + \dots\,.
\nonumber
\end{align}
This result agrees with the known leading order and next to leading order BFKL spin.

\subsubsection{OPE coefficients - leading order prediction}

The idea is to use
the  knowledge of the Regge residue $\beta(\nu)$ to derive non-trivial predictions for OPE coefficients. 
We remind
the reader that we are considering OPE coefficients with operators normalized as in (\ref{normalized2pf}).
The product of these normalized OPE coefficients is also defined by the ratio of correlators
\be
C^2(J) \equiv C_{11J}C_{33J}\sim 
\frac{\langle \Ocal_1(x_1)\Ocal_1^*(x_2) \Ocal_J(x_5) \rangle \,
 \langle \Ocal_J(x_6)  \Ocal_3(x_3)\Ocal_3^*(x_4) \rangle}{
\langle \Ocal_1(x_1)\Ocal_1^*(x_2) \rangle \,
\langle \Ocal_J(x_5) \Ocal_J(x_6) \rangle\, 
\langle \Ocal_3(x_3)\Ocal_3^*(x_4) \rangle}\,.
\label{Cs}
\ee
The precise relation between the OPE coefficients and the ratio of correlators involves many kinematical factors that we give in
appendix \ref{Appendix3pt}. We omit these details to avoid 
dealing with all the indices in the main text.
From direct computation of the 
the four point correlator in the Regge limit we can extract 
the Regge residue that appears in the correlation function (\ref{MainEquationMellin}).
Then, using (\ref{betaofnu}), this is related to  
the analytic continuation of the product of OPE coefficients   $ C^2\!\big(j(\nu)\big) $.

We start with the weak coupling side of the story. We can compute in free theory the OPE coefficient $C(J)$ of the spin $J$ operator
of the leading Regge trajectory in the OPE of two protected scalar operators of the form $\Ocal_1={\rm tr}(\phi_{12}\phi^{12})$, where $\phi_{12}$ is a complex scalar field of SYM.
This requires lifting the degeneracy of the twist two operators and some combinatorics in doing Wick contractions. The computation
is presented in appendix \ref{Appendix3pt}, and the result is
\begin{equation}
C^2(J)= \frac{1}{N^2}\frac{2^{1+J} J (J-1)\Gamma^{2}(J+1)}{\left(4J^{2}-1\right)\Gamma(2J+1)} + O(g^2)\,.
\label{freeOPE}
\end{equation}
In particular, we can continue this result to the  region around $J=1$, with an expansion of the form
\be
C^2(J)= \frac{J-1}{N^2}\left( \frac{2}{3} +O(J-1) \right) + O(g^2)\,.
\label{JexpansionFree}
\ee 

Now let us look at the Regge residue $r(j(\nu))$ computed in perturbation theory from the four point correlation function.
In  \cite{ourBFKL} the  Regge residue in position space $\alpha(\nu)$ was shown to be
\be
\alpha(\nu)= i\,
\frac{16 \pi^5 g^4}{N^2} 
\frac{ \tanh\!\left(\frac{\pi \nu}{2} \right)}{\nu \cosh^2\!\left(\frac{\pi \nu}{2} \right)}
+ O(g^6)\,.
\ee
Using (\ref{betaofnu}) and (\ref{alpha}) this translates into a residue $r(j(\nu))$ given by
\begin{align}
r\big(j(\nu)\big)=- \frac{2^8 \pi g^2}{  N^2}
\frac{  \tanh\!\left(\frac{\pi \nu}{2} \right)}{ \chi'(\nu) (1+\nu^2)^2} + O(g^4) \,,
\label{residueBFKL}
\end{align}
where we used the leading term in the BFKL spin as written in (\ref{BFKLspinj}).
It is clear that (\ref{residueBFKL}) computes the behaviour of the function $r(J)$
around $J=1$, which starts with a power of  $J-1\sim g^2$. The same thing happens 
to the square of the OPE coefficients (\ref{JexpansionFree}) computed directly in free theory. This is not a coincidence
because both $r(J)$ and $C^2(J)$ are related by (\ref{residuerofJ}).
Thus, for $J=j(\nu)$, we can use (\ref{residuerofJ}) in the form
\be
r\big(j(\nu)\big)=C^2\!\big(j(\nu)\big)\,K_{\Delta(j({\nu})),j(\nu)}\,,
\ee
to compute the 
OPE coefficients from the Regge residue in the region $J-1\sim g^2$, i.e.
in the double limit $g^2\to 0$ and $J\to 1$ with $g^2/(J-1)$ fixed.
In particular, we will recover  the above 
free field theory result  and also make 
predictions to arbitrary high
order in perturbation theory. The analysis is entirely analogous to that of the anomalous dimension reviewed above.

From (\ref{JexpansionFree})
we conclude that the continuation of the OPE coefficients $C(J)$ to the region where $J-1\sim g^2$ admits the 
following  general perturbative expansion
\be
C^2(J) = (J-1)\, a\!\left( \frac{g^2}{J-1}\right)
 + O(g^4)\,,
\ee
for some function $a$ that will be determined bellow.
More precisely, 
for $J=j(\nu)$ given by (\ref{BFKLspin}), 
we have the Regge theory prediction
\be
C^2\!\big(j(\nu)\big)= 4 g^2 \chi(\nu) \, a\!\left( \frac{1}{4\chi(\nu)}\right) = \frac{r\big(j(\nu)\big)}{ K_{\Delta(j(\nu)),j(\nu)}} = 
 - \frac{2^8 \pi g^2}{  N^2}
\frac{  \tanh\!\left(\frac{\pi \nu}{2} \right)}{ \chi'(\nu) (1+\nu^2)^2 K_{\Delta(j(\nu)),j(\nu)}}\,,
\label{matching}
\ee
and therefore 
\be
 a\!\left( \frac{1}{4\chi(\nu)}\right) =  
 - \frac{2^6 \pi }{  N^2}
\frac{  \tanh\!\left(\frac{\pi \nu}{2} \right)}{  \chi(\nu)\chi'(\nu) (1+\nu^2)^2 K_{2\pm i\nu,j(\nu)}}\,.
\label{matching2}
\ee

We want a prediction for the OPE coefficients around $J=1$, so that the function  $a(x)$
can be expanded as
\be
a(x)=\frac{1}{N^2}\left( a_0+a_1 x+a_2 x^2+\dots \right)\,,
\label{aExpansion}
\ee
with small $x=  g^2/(J-1)$, therefore giving a prediction  to all loops for the OPE coefficients. Thus, just like in 
(\ref{expansion1}) for the anomalous dimension, we need to consider equality (\ref{matching2}) for $i\nu=\pm 1$.
Notice that the function $\chi(\nu)$ has poles at $\nu=\pm i(1+2n)$, with $n=0,1,2,\dots$, so indeed we are expanding the 
function $a(x)$ around the origin. Moreover, we started the expansion (\ref{aExpansion}) with a constant term because 
the RHS of (\ref{matching2}) is regular at the origin. Doing the computation we obtained the following prediction
for the first order terms in this expansion
\be
a_0  = \frac{2}{3} \,,\ \ \ \ \
a_1 =   \frac{64}{9} \,,\ \ \ \ \ 
a_2  =  \frac{32}{27}\left( 61-3\pi^2 \right) ,\ \ \ \ \
a_3  =  \frac{256}{81}\left( 223 - 12 \pi^2 - 27 \zeta(3) \right) ,\ \dots\,.
\ee
The first term  matches the  free theory result, the other terms give a prediction for 
the behaviour of the OPE coefficients to any order in perturbation theory.

\subsubsection{OPE coefficients - next to leading order prediction}

The next to leading order correction to the function $\a(\nu)$ was computed in \cite{Balitsky:2009yp}, with the result
\begin{equation}
\alpha(\nu)\!=\!i\frac{16\pi^{5}g^{4}\tanh\!\left(\frac{\pi \nu}{2} \right)}{N^{2}\nu\cosh^{2}\!\left(\frac{\pi\nu}{2}\right)}\!
\left(\!1+8g^{2}\!\left(\!\frac{i\pi\chi\!\left(\nu\right)}{4}+\frac{\pi^{2}}{4}-\frac{4}{1+\nu^{2}}-\Psi'\!\left(\frac{1-i\nu}{2}\right)-\Psi'\!\left(\frac{1+i\nu}{2}\right)\!\right)\!\right)\,.
\label{NLOalpha}
\end{equation}
In order to reproduce this result using Regge theory, we consider the continuation of the OPE coefficients to the region close to $J= 1$, keeping $g^2/(J-1)$ fixed, in the form
\begin{equation}
C^2(J) =\left(J-1\right) \left[ a\!\left(\frac{g^{2}}{J-1}\right)+\left(J-1\right)b\!\left(\frac{g^{2}}{J-1}\right) \right]+ O(g^6)\,.
\end{equation}
The first terms in the expansion of the function $a(x)$ were obtained in the previous section. Following the same procedure of matching expansions 
around $\nu=\pm i$, we can 
obtain the predictions for the function $b\left(x\right)=\sum_{k=0}^{\infty}b_{k}x^{k}$. In particular, from the 
result (\ref{NLOalpha}) for $\alpha(\nu)$ one obtains for the first term in the expansion of $b(x)$ the coefficient
\begin{equation}
b_{0}=\frac{2}{9}\big(-14+3\ln(2)\big)\,.
\end{equation}
The prediction can be tested against the expansion of (\ref{freeOPE}) computed in free theory around $J=1$. The results do not match, since
(\ref{freeOPE}) gives
\begin{equation}
b_{0}=\frac{2}{9}\big(-8+3\ln(2)\big)\,.
\end{equation}
The two values differ by the rational number $4/3$, though the $log$'s do agree. Given that the free theory result is by far more trivial to derive, we speculate that perhaps some term in
$\a(\nu)$, given by (\ref{NLOalpha}), is not correct. In fact, 
all the terms in  (\ref{NLOalpha}) have definite transcendentality, with the exception of the term $-8/(1+\nu^{2})$. We find that 
if this term, which violates maximal transcendentality, is not present then there is agreement. To see this we
study the dependence of $b_{0}$ on the next leading order correction to $\alpha(\nu)$. First we write, 
\begin{equation}
\alpha(\nu)= i\,\frac{16\pi^{5}g^{4}\tanh\!\left(\frac{\pi \nu}{2} \right)}{N^{2}\nu\cosh^{2}\!\left(\frac{\pi\nu}{2}\right)}\left(1+4g^{2}\left(\sum_{k=-2}^{\infty}d_{k}(\nu-i)^{k}\right)\right)\,.
\end{equation}
Matching with the expansion of $ C^2\!\big(j(\nu)\big)$ around $\nu=\pm i$, 
we find that the coefficient 
$d_{-2}$ needs to have the value 8, in exact agreement with (\ref{NLOalpha}). However, in order to match the free theory result for the coefficient $b_0$, we must have 
a vanishing coefficient $d_{-1}$.  This suggests that the term $-8/(1+\nu^{2})$ should indeed be absent from  (\ref{NLOalpha}).
Assuming this is  the case, the conformal Regge theory prediction for the first next to leading order coefficients 
$b_{k}$ is\footnote{In fact, expression (\ref{NLOalpha}) computed in \cite{Balitsky:2009yp} is very incomplete and the correct expression will be given in \cite{MSCosta}. As a consequence the predictions (\ref{eq:bprediction}) are incorrect, with the exception of coefficients $b_0$ and $b_1$.},
\[
b_0  = \frac{2}{9}\big(-8+3\ln(2)\big) \,,\ \ \ \ \
b_1 =   \frac{4}{27}\big(-244+9\pi^{2}+48\ln(2)\big) \,,\ \ \ \ \ 
\]
\be
b_2  =  \frac{16}{27}\big(-892+122\ln(2)-2\pi^{2}\big(-20+3\ln(2)\big)+225\zeta(3)\big) \,,\, \dots \,.\ \ \ \ \label{eq:bprediction}
\ee

\subsection{Strong coupling}

Let us  turn to the strong coupling expansion, starting again with the relation between spin and 
anomalous dimension of the  operators 
in the leading Regge trajectory.
\footnote{
Some of the results presented in this section
were obtained after  many discussions with Diego Bombardelli and Pedro Vieira,
who also participated in some of these computations.}

The anomalous dimensions of the leading twist operators can be computed at strong coupling from the energy of short strings
in AdS, and admit an expansion of the type 
\begin{equation}
\Delta(J) \big( \Delta(J) -4 \big) = x  \sum_{n=0}^{\infty} g^{1-n}H_{n}(x)\,,
\label{AnomalousDimStrong}
\end{equation}
where we conveniently defined $x=J-2$. The overall factor of $x$ guarantees that the energy momentum tensor has protected dimension.
The latest results for this expansion  include the one and  the two loop corrections  \cite{Gromov:2011de,Gromov:2011bz,Basso:2011rs}. It is  simpler to present these
results in terms of the quantity $\Delta(\Delta-4)$, instead of $\Delta$, because this is the combination that appears in the  Casimir of the conformal 
group and makes the symmetry $\Delta\to4-\Delta$ (or $\nu\to-\nu$) explicit. The results of \cite{Gromov:2011de,Gromov:2011bz,Basso:2011rs} then give (recall that $\sqrt{\lambda}= 4\pi g$)
\be
\Delta(J) \big( \Delta(J) -4 \big)  = x \left[   2\sqrt{\lambda} +\left( -1 +\frac{3x}{2}\right)  -\frac{3}{8}\Big( -10 + (8\zeta(3) -1)x+x^2   \Big)\frac{1}{\sqrt{\lambda}} + \cdots \right]\,.
\label{ADimStrong}
\ee
This structure suggests that  $H_n(x)$ is a polynomial of degree $n$. 

On the other hand, 
at strong coupling the Reggeon spin was computed using the dual string description \cite{BPST,CornalbaRegge}
\be
j(\nu)=2- \sum_{n=1}^\infty \frac{j_n(\nu^2)}{g^n} = 2 -  \frac{4+\nu^2}{2\sqrt{\lambda}} \left( 1 +   \sum_{n = 2}^\infty  \frac{\tilde{j}_n(\nu^2)}{\lambda^{(n-1)/2}}\right)\,,
\label{ReggeSpinStrong}
\ee
where $\tilde{j}_n(\nu^2)$, defined for $n\ge2$, is a polynomial of degree $n-2$. The $n=1$ term in this expansion was computed in  \cite{BPST} and gives the
linear Regge trajectory of strings in the flat space limit. The general form
that constrains the degree of the polynomial $\tilde{j}_n(\nu^2)$ was derived in \cite{CornalbaRegge} by requiring that such limit is well defined. We will actually see that
this polynomial can be further restricted.

\begin{figure}
\begin{centering}
\includegraphics[scale=0.3]{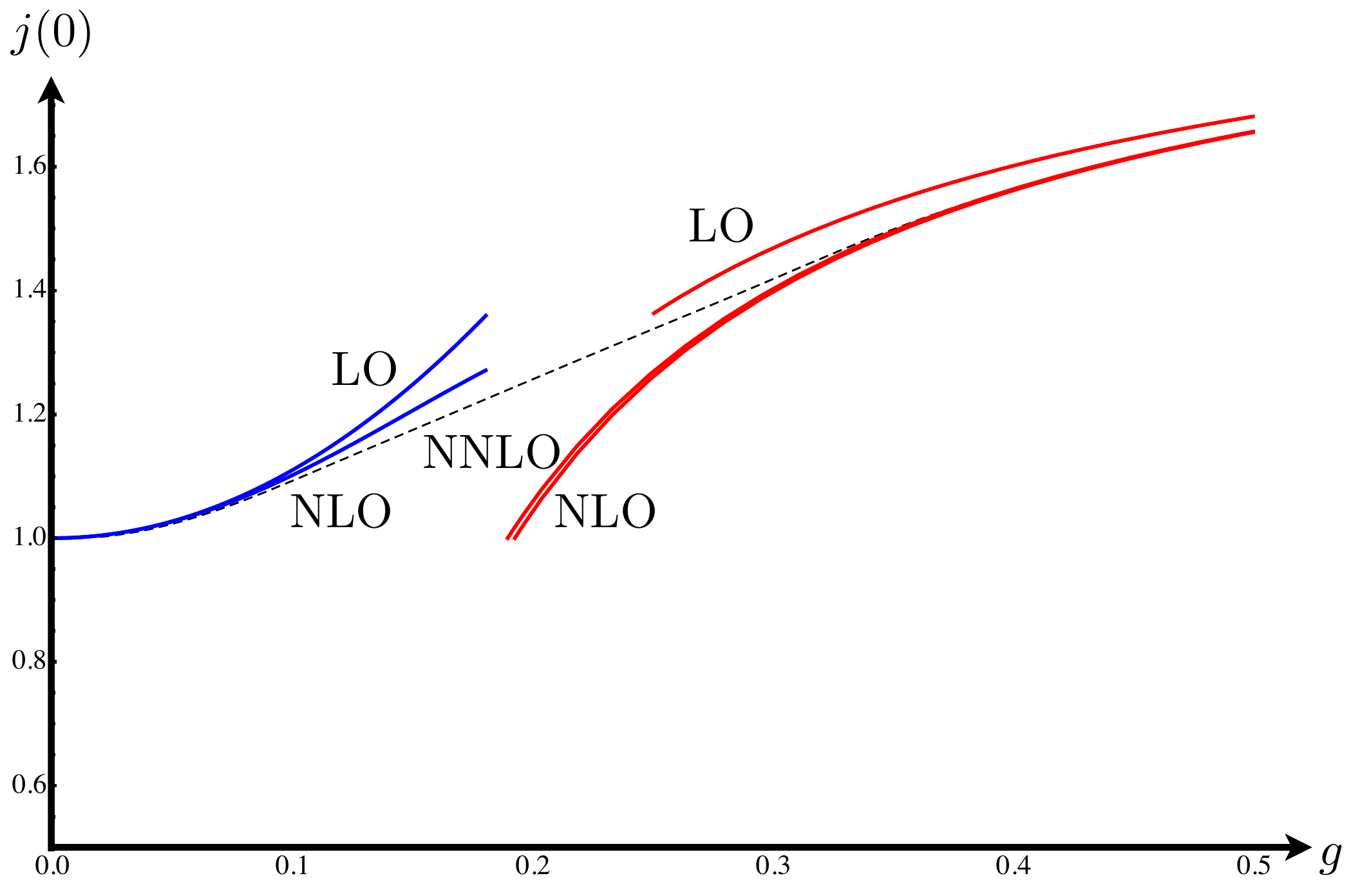}
\par\end{centering}
\caption{Weak (in blue) and strong (in red) coupling expansions of the BFKL intercept $j(0)$.
The plot suggests a smooth interpolation, like the black dashed curve, from 1 at 
$g=0$ to 2 at $g=\infty$.
\label{figIntercept}}
\end{figure}

Next we consider the limit $J\rightarrow2$ and $\lambda\rightarrow\infty$ , with $(J-2)\sqrt{\lambda}$ fixed, of the expression for the anomalous dimension 
(\ref{ADimStrong}). Noting that $ - \Delta ( \Delta -4 )= 4 +\nu^2 $, we can equate both expansion (\ref{ADimStrong}) and (\ref{ReggeSpinStrong}) to obtain new 
data for the polynomials $\tilde{j}_n(\nu^2)$, with $n\ge 2$, that characterise the AdS graviton Regge trajectory. Writing 
\be
\tilde{j}_n(\nu^2) = \sum_{k=0}^{n-2} c_{n,k}  \nu^{2k}\,,
\ee
we can fix the coefficients $c_{n,n-2}$ and $c_{n,n-3}$. More precisely, we obtained that 
\be
c_{2,0} = \frac{1}{2}\,,\ \ \ 
c_{3,0} = - \frac{1}{8} \,,\ \ \ 
c_{3,1} =  \frac{3}{8} \,,\ \ \ 
c_{4,1} = - \frac{3}{32}\big( 8\zeta(3) -7 \big)   \,,\ \ \ 
c_{5,2}= \frac{21}{64}\,,
\label{coefc}
\ee
and the remaining coefficients of this type vanish ($c_{n,n-2}=0$ for $n\ge 4$, $c_{n,n-3}=0$ for $n\ge 6$).
In particular, we derived the next and the next to next leading order correction to the intercept\footnote{In the first version of this paper, the signs of the NLO and NNLO  corrections in equation (\ref{intercepttexto}) were misprinted, although the coefficients in (\ref{coefc}) were correctly computed. We thank Chung-I Tan for pointing this out to us. The correct signs also appeared recently in \cite{KotikovLipatov2013}.}
\be
j(0) = 2 - \frac{2}{\sqrt{\lambda}} - \frac{1}{\lambda} + \frac{1}{4\lambda^{3/2}} +\dots\,.
\label{intercepttexto}
\ee
From figure \ref{figIntercept} we conclude that this strong coupling expansion works reasonably well for $g \gsim 0.3$. 
Such a strong coupling expansion has been recently used to construct  phenomenological models of high 
energy processes in QCD that are dominated by Pomeron exchange, following the proposal of \cite{BPST}. These models start from the conformal limit here studied, 
and then introduce a hard wall in AdS to cut off the IR scale. Data analysis of  deep inelastic scattering (DIS) \cite{satDIS,Levin:2010gc,Brower:2010wf} 
and deeply virtual Compton scattering (DVCS) \cite{Costa:2012fw} gives an intercept in the region $j(0)=1.2-1.3$. 
At a first glance it may seem surprising how the fits of data in a region reasonably close to $j(0)=1$ are so successful, even better than those fits that use the weak coupling 
expansion (see \cite{Kowalski:2010ue} for the latest analysis  on DIS). However, in SYM, figure \ref{figIntercept} shows that indeed the strong coupling expansion seems to 
already work reasonably well around the region of $j(0)=1.2-1.3$.

Finally, we remark that the coefficients $c_{n,k}$ can be further restricted if we assume that   $H_l(x)$  is a polynomial of degree $l$. This assumption leads to the
conclusion that for  $n\geq4$  
the coefficients satisfy
\begin{align}
c_{n,k}=0 \  \ \  \textrm{for} \ \  \  \left[\frac{n}{2}\right]\leq k\leq n-2\,.
\end{align}


\subsubsection{OPE coefficients \label{strongOPEcoe}}

Let us start with the  simple case of graviton exchange in AdS between external scalar fields dual to operators of protected dimension $\Delta_1$ and $\Delta_3$.
In this strict $\lambda \to \infty$ limit the spin $j(\nu)= 2$ and the scattering is elastic. The Regge amplitude in position space (\ref{ReggeA}) has real residue and is
given by\footnote{
In \cite{CornalbaRegge,ourBFKL} we computed $\alpha(\nu)$ for external states of dimension $\Delta_i=2$. From the results in those papers it is simple to see that,
for arbitrary dimension of the external fields, graviton exchange in AdS gives
\be
\alpha(\nu)= -\frac{\pi}{4N^2}  V_1(\nu) \,\frac{1}{4+\nu^2} \,V_3(\nu)\,,
\ \ \ \ \ \ \ \ \ \ \ 
V_k(\nu)  =   4\, \frac{\Gamma\!\left(\Delta_k+\frac{i\nu}{2}\right)\Gamma\!\left(\Delta_k -\frac{i\nu}{2}\right) }{\Gamma\!\left(\Delta_k\right) \Gamma\!\left(\Delta_k-1\right)}\,.
\nonumber
\ee
}
\be
\alpha(\nu)= -\frac{4\pi}{N^2}  
\,\frac{1}{4+\nu^2} \,
\frac{\Gamma\!\left(\Delta_1 +\frac{i\nu}{2}\right)\Gamma\!\left(\Delta_1 -\frac{i\nu}{2}\right) \Gamma\!\left(\Delta_3 +\frac{i\nu}{2}\right)\Gamma\!\left(\Delta_3 -\frac{i\nu}{2}\right)}
{\Gamma\!\left(\Delta_1\right) \Gamma\!\left(\Delta_1-1\right)\Gamma\!\left(\Delta_3\right) \Gamma\!\left(\Delta_3-1\right)}\,.
\ee
Using (\ref{alpha}) and (\ref{betaofnu}) we can relate the function $\alpha(\nu)$ to the  residue $r(j(\nu))$, and therefore to the product of OPE coefficients. 
In the particular case of  graviton exchange, we can use the first term in the expansion (\ref{ReggeSpinStrong}), 
$j(\nu)= 2 - (4+\nu^2)/(2\sqrt{\lambda}) + O(\lambda^{-1})$, to  obtain the $\lambda \to \infty$ result
\be
\alpha(\nu)=  -r(2) \,\frac{16\pi^2}{4+\nu^2} \,
\Gamma\!\left(\Delta_1 +\frac{i\nu}{2}\right)\Gamma\!\left(\Delta_1 -\frac{i\nu}{2}\right) \Gamma\!\left(\Delta_3 +\frac{i\nu}{2}\right)\Gamma\!\left(\Delta_3 -\frac{i\nu}{2}\right).
\ee
Equating the previous two equations, we can determine $r(2)$ and therefore, using (\ref{residuerofJ}), the product of OPE coefficients
\be
C^2(2) = \frac{2 \Delta_1 \Delta_3}{45N^2}\,.
\label{r(2)}
\ee
This is actually  independent of $\lambda$ because the three point function with the stress-energy tensor is determined by a Ward identity 
\cite{OsbornCFTgeneraldim, Dolan:2000ut}
\be
C^2(2)=\frac{16 \D_1 \D_3}{9 C_T}\,,
\ee
for $\D_2=\D_1$ and $\D_3=\D_4$. Note that the central charge $C_T$ appears in the denominator because, as explained in  (\ref{Cs}), we are considering normalized operators.
The central charge is known from gravity in AdS \cite{Liu:1998bu,Kovtun}
\be
C_T= 20\, \frac{\pi R^{3}}{ G_N^{(5)}} \,.
\ee
Using $G_N^{(5)} R^{-3}=\pi/(2 N^2)$, we obtain $C_T=40N^2$, and reproduce exactly the result (\ref{r(2)}).

The above was just a warm up exercise to make sure numerical factors are in place. 
We can actually compute the leading term in the strong coupling expansion of 
the function $C^2(J)$ for arbitrary $J$, therefore computing the 
OPE coefficients between the leading twist operators in 
the pomeron-graviton Regge trajectory and two external scalar operators. 
This can be done by considering the flat space limit of the CFT amplitude in the Regge limit, and then
equating it to the flat space string theory S-matrix element with external scalar fields, also in the Regge limit, 
to read the function $C^2(J)$.
As a specific example we shall  consider the Virasoro-Shapiro amplitude reviewed in section \ref{ReviewRegge}
with external  dilaton fields, which are dual to the Lagrangian operator of protected dimension $\Delta_i=4$.

The string theory S-matrix for four external scalars in the Regge limit can be recovered from the  flat space limit 
introduced in \cite{JPMellin}, \footnote{From now on, we shall denote the usual flat space Mandelstam invariants 
with capital letters $S$ and $T$, to distinguish them from the Mellin variables $s$ and $t$.}
\be
\mathcal{T}\big(S,T\big) = \frac{1}{{\cal N}} \lim_{R\to \infty}  V(S^5) \,R \int_{-i\infty}^{i\infty}
\frac{d\alpha}{2\pi i}\, \alpha^{2-\frac{\sum_i \Delta_i}{2}} e^\alpha   \,M\!\left(\frac{R^2S}{2\alpha},\frac{R^2T}{2\alpha}\right)\,,
\label{FlatSpace}
\ee
with the Mellin amplitude given by the Regge theory form (\ref{MainEquationMellin}), the volume of the 5-sphere $V(S^5)= \pi^3 R^5$ and the constant $\mathcal{N}$ given by (\ref{eq:NMellin}). 
The computation now is entirely similar to the one of appendix  \ref{FlatSpaceSubSec}
for the flat space limit of a single conformal partial wave, so we will not be so detailed here  (see equation (\ref{flatlimitCPWE})). 
The integration over $\alpha$ produces a delta function in $\nu^2$ with a characteristic width $L$, so in this case we have
\begin{align}
\mathcal{T}\big(S,T\big)\approx&\  \frac{1}{{\cal N}} \lim_{R\to \infty}  V(S^5) \,R
\int_{-\infty}^{\infty}  d\nu \,\beta(\nu)
\, \left( \frac{R^2S}{2}\right)^{ j(\nu)}
\frac{e^{i\frac{\pi}{2}j(\nu)}}{\sin\!\left( \frac{\pi j(\nu)}{2}\right) }
\label{IntermediateFlatLimit}
\\&
\hspace{3.3cm}\frac{1}{\sqrt{\nu^2}}
\left( \frac{-\nu^2}{R^2T}\right)^{\frac{1}{2}\sum\D_i +j(\nu)-2}
\left(-R^2T\right) \delta_{L}\!\left(\nu^2 + R^2T\right).
 \nonumber
\end{align}

The function $\delta_L$ should be understood as a delta function when integrated against functions that vary in a scale $\d\nu^2\gg L$. On the other
hand, for functions with characteristic scale $\d\nu^2\ll L$ one should take the average. For the above integral this gives
\be
\mathcal{T}\big(S,T\big) \approx \  \frac{ \pi R^6}{{\cal N}} \,
\langle \beta\rangle_T
\left( \frac{R^2S}{2}\right)^{J(T)}
\frac{e^{i\frac{\pi}{2}J(T)}}{\sin\!\left( \frac{\pi J(T)}{2}\right) }\,,
\label{FlatLimitRegge}
\ee
where for $R\to\infty$ the graviton Regge trajectory becomes the usual linear trajectory
\be
J(T)= 2 +\frac{\alpha'}{2}\,T\,.
\ee
As explained in section  \ref{FlatSpaceSubSec}, the integration of the function $\beta(\nu)$  
against the delta function $\delta_L$ in (\ref{IntermediateFlatLimit})
produces the average $\langle \beta\rangle_T$. We shall see bellow that this is important because
$\beta(\nu)$  contains a rapidly varying function of $\nu$. Note that in this case we can take the strong coupling 
limit of $\beta(\nu)$ given in  (\ref{betaofnu}), so that
\be
\beta(\nu) = - \frac{\pi}{4}\,\frac{\alpha'}{R^2}\,K_{\D\!\left( j(\nu) \right), j(\nu)} \,C^2\!\big(j(\nu)\big)\,.
\label{betaofnustrong}
\ee

Before we analyse in more detail the implications  of (\ref{FlatLimitRegge}), let us check that in the simplest case of $T=0$ (i.e. $J=2$) we can derive
again (\ref{r(2)}). We consider the case of scattering of four dilaton fields (dual to the Lagragian operator of dimension $\D_i=4$), so that in
this case 
 $K_{\D\!\left( 2 \right),2} = 5/256$ is constant. Thus there is no issue with averaging. It is then a simple exercise to
equate (\ref{FlatLimitRegge})  near $T=0$, to the  S-matrix for graviton exchange between four dilatons, $T(S,T)=-8\pi G_N S^2/T $,
checking again (\ref{r(2)}).

The S-matrix element (\ref{FlatLimitRegge}) can be equated to a type IIB string theory S-matrix element in the Regge limit. 
Let us consider again the case of external  fields given by the dilaton. Then we can equate (\ref{IntermediateFlatLimit}) to 
the Virasoro Shapiro S-matrix element (\ref{VS}). Although
the S-matrix element (\ref{FlatLimitRegge}) was derived in the physical scattering region of $T<0$, we can analytically continue
this expression to positive $T$. In particular this means that we can consider $J$ a positive even integer, and compute $C^2(J)$ at
strong coupling. In this kinematical region the dimension 
of the exchanged leading twist operators is real. Since we work in the strong limit we have 
\be
\Delta(J) \approx i\nu \approx \lambda^{1/4} \sqrt{2(J-2)}\,.
\ee
To compute the function $C^2(J)$ we need to be careful with the average $\langle \beta\rangle_T$ of $\beta(\nu)$ in (\ref{betaofnustrong}).
Let us first  look at the expansion of the function $K_{\Delta(J),J} $  at large $\Delta$ and for external operators of dimension four,
\be
K_{\Delta(J),J} \approx \frac{2^{9+2 J +2 \Delta(J)} \big(\Delta(J)\big)^{-10 -2 J}}{\pi^3} 
\sin^2\!\left(\frac{ \pi\D(J)}{2} \right).
\ee
Thus, when integrating $\beta(\nu)$ given by (\ref{betaofnustrong})
against the function $\delta_L$,  this $sin^2$ piece is rapidly varying and averages to $1/2$. 
In addition, we shall assume that the remaining dependence on $\beta(\nu)$ in $\nu$ is power law, so its
integration with the function $\delta_L$ works like with a delta function. 
Thus, after
some straightforward algebra, we can write the flat space limit of this  CFT amplitude  in the following form
\begin{align}
\mathcal{T}(S,T) \approx&\  \frac{1}{{\cal N}}\, \frac{N^2G_N}{2 \pi \alpha'} \, \left\langle K_{\D\!\left( j(\nu) \right),j(\nu)} \,C^2\!\big(j(\nu)\big)  \right\rangle_T
 \label{ResultFlatSpace}
\\
&
\left(  \frac{R^4}{\alpha'^2}\right)^{ \frac{\alpha'T}{4}}
e^{ \mp i \pi \frac{\alpha' T}{4}  }\,\Gamma\!\left(1 +   \frac{\alpha' T}{4} \right)\Gamma\!\left( -   \frac{\alpha'T}{4} \right)
  \left(\frac{\alpha'S}{2}  \right)^{J(T)},
    \nonumber
 \end{align}
where, for the external operators under consideration, ${\cal N}= 1/(1152 \pi^2)$. Finally equating to
the Virasoro-Shapiro amplitude in the Regge limit (\ref{eq:amplitude V-S regge}), we 
obtain the following strong coupling prediction for the OPE coefficient involving two Lagrangians  and a spin $J$ operator in the 
leading Regge trajectory,
\be
C_{{\cal L}{\cal L}J}\approx
\frac{\pi^{\frac{3}{2}}}{3N}\,
\frac{ (J-2)^{\frac{5+J}{2}}}{ 2^{1+J} \Gamma\!\left( \frac{J}{2}\right)}\,
\lambda^{\frac{7}{4}} \,2^{- \lambda^{1/4}\sqrt{2(J-2)}} \,.
\ee
The exponential dependence  on the coupling comes precisely from the dimension of the spin $J$ operator. This is expected since the AdS
computation of the three point function should be dominated by the saddle point of the dual heavy short string. It would be very interesting 
to derive this result using the methods of \cite{Klose:2011rm,Buchbinder:2011jr,Minahan:2012fh}.

\section{Conclusion \label{secConclusion}}

The first lesson of this work is that the analogy between Mellin amplitudes and scattering amplitudes can be very fruitful to guide the exploration of AdS/CFT at finite or strong coupling.
In this paper, we studied the regime of high energy scattering and successfully developed conformal Regge theory following the analogy summarized in table \ref{tabela}.
Then, 
in section \ref{secSYM}, we applied the formalism to SYM and obtained non-trivial predictions for
the dimension $\Delta(J)$ of the spin $J$ leading twist operator and its OPE coefficient $C(J)$ in the OPE of two protected scalar operators, both at weak and strong coupling. 

We conclude by discussing some questions for the future. In the context of SYM, there are two obvious directions to pursue. Firstly, we can study other Regge trajectories with different quantum numbers. In particular, it would be interesting to study the Regge limit of the four-point function 
\be
\left\langle 
\tr\! \left( XZ \right)\! (x_1)\,
\tr\! \left( \bar{X}Z \right) \!(x_2)\,
\tr \!\left( Y\bar{Z} \right)\! (x_3)\,
\tr \!\left( \bar{Y}\bar{Z} \right)\! (x_4)  \right\rangle ,
\ee
where $X$, $Y$ and $Z$ are complex scalar fields of SYM. 
This would give information about the (simpler) three-point functions 
\be
\left\langle
\tr\! \left( Y\bar{Z} \right) \!(x_1)\,
\tr\! \left( \bar{Y}\bar{Z} \right) \!(x_2) \,
\tr\! \left( Z D^J Z \right)\! (x_3) \right\rangle ,
\ee
that describe the coupling of the external operators to the leading Regge trajectory in this charged sector. These were recently computed to three-loop order in \cite{Eden:2012rr},
and were also studied in \cite{Plefka:2012rd}.
Secondly, we can apply the conformal Regge theory formalism to higher orders in perturbation theory, exploring the abundance of available data for four-point functions in SYM   \cite{Korchemsky4pt2011,Korchemsky4pt2012}.
Notice that from the four-point function at order $g^{2l}$, it is possible to extract the pomeron spin at order $g^{2l-4}$. This means that from the six loops ($l=6$) integral representation of the four-point function given in \cite{Korchemsky4pt2012}, one can, in principle, obtain the BFKL pomeron spin at order $g^{8}$ or \emph{next to next to next to leading order}!

More generally, one can pose the question: \emph{Are all planar n-point functions of SYM determined by the planar two and three point functions of single-trace operators?}
It is clear that single-trace data is not sufficient information from the OPE point of view.
However, it is sufficient information to fix all poles and residues of the Mellin amplitudes. Therefore, as speculated in \cite{JPMellin}, it should be possible to fix all Mellin amplitudes if we understand their asymptotic behaviour.
The main result (\ref{MainEquationMellin}) of this paper, can be thought as a first step in this direction. Indeed, we were able to determine the Regge limit of the four-point function solely in terms of dimensions and OPE coefficients of single-trace operators.
We believe there is a refined OPE formalism for planar (conformal) gauge theories, that distinguishes single-trace from multi-trace operators, waiting to be discovered. 

In this paper, we discussed an analogy between standard Regge theory for scattering amplitudes and conformal Regge theory. However, in some theories, one can interpolate from one to the other.
Consider a conformal gauge theory deformed by a relevant operator that leads to confinement in the infrared.
As an example, one can think of weakly coupled SYM with a large mass $M$ for the matter fields, such that there is a large hierarchy between $M$ and the lightest glueball mass $M_g$. 
Glueball scattering in this theory will be described by standard Regge theory for $M_g^2,T \ll S \ll M^2$, and by conformal Regge theory in the extreme high energy regime $S \gg T \gg M^2 \gg M_g^2$.
It would be interesting to understand this transition in detail. The basic mechanism is that the continuous variable $\nu$ of conformal Regge theory, becomes a discrete variable labelling several Regge trajectories in the confining theory. In other words, one conformal Regge trajectory breaks up into many standard Regge trajectories \cite{Lipatov:1985uk,Kowalski:2012ur,BPST}.

\section*{Acknowledgements}

We wish to thank Diego Bombardelli, Jo\~ao Caetano, Liam FitzPatrick, Jared Kaplan, Gregory Korchemsky, Hugh Osborn, Suvrat Raju, Balt van Rees and Pedro Vieira for useful discussions and 
comments on this manuscript.
We also wish to thank Perimeter Institute for the great hospitality during our visit in the summer of 2012
when a significant part of this work was done.
J.P. is grateful for the hospitality of the Kavli Institute for Theoretical Physics, UCSB, where part of this work was developed.
This work was partially funded by grants PTDC/FIS/099293/2008, CERN/FP/123599/2011.
The research leading to these results has received funding from the [European Union] Seventh Framework Programme [FP7-People-2010-IRSES] under grant agreement No 269217.
\emph{Centro de F\'{i}sica do Porto} is partially funded by the Foundation for 
Science and Technology of Portugal (FCT). 
The work of V.G. is supported  
by the FCT fellowship SFRH/BD/68313/2010. 
The research leading to these results has received funding from the
European Union Seventh Framework Programme (FP7/2007-2013) under
grant agreement No PCOFUND-GA-2009-246542 and from FCT.

\appendix

\section{Mellin amplitudes in more detail \label{apMellin}}

In this appendix, we collect several results that complement the description of Mellin amplitudes of section \ref{secMellin}.

\subsection{Mellin Poles \label{apMellinPoles} }

As stated in the main text, the structure of the conformal OPE implies a very simple analytic structure 
for the Mellin amplitude if the CFT has a discrete spectrum of operator dimensions \cite{Mack}. 
Here we shall explain how this works in more detail.
Some of the results of this section were already discussed in \cite{Mack,JPMellin, DOMellin,JLAnalyticMellin}, however we shall take the risk of repetition with the hope of making more transparent key features that are needed in Regge theory.

The OPE implies the conformal block expansion (\ref{CBE}) of the reduced correlator, which we rewrite here for convenience
\be
\mathcal{A}(u,v)= \sum_k C_{12k}C_{34k} \, G_{\D_k,J_k}(u,v)\,.
\ee
As explained in  \cite{Dolan:2000ut}, the conformal blocks have a series expansion of the form  
\be
G_{\D,J}(u,v)=u^{\frac{\D-J}{2}} \sum_{m=0}^\infty u^m g_m(v)\,,
\label{useries}
\ee
where the first term reads
\be
g_0(v)  = \left(\frac{v-1}{2}\right)^J
 \ _2F_1\! \left(\frac{\D+J-\D_{12} }{2} ,\frac{\D+J+\D_{34} }{2} , \D+J,1-v\right)
 \,.
 \label{asymptoticsuto0}
\ee

In order to reproduce the power law behavior of $\mathcal{A}$ at small cross ratio $u$ predicted by the OPE, the Mellin amplitude must have poles in the variable $t$. More precisely, 
\be
M(s,t)\approx \frac{C_{12k}C_{34k} \,\mathcal{Q}_{J,m}(s)}{t-\Delta+J-2m}\,,\ \ \ \ \ \ \ \
m=0,1,2,\dots \,, \label{MellinpolesAp}
\ee
where, as before, $\Delta$ and $J$  are 
the dimension and spin of an operator $\mathcal{O}_k$ that appears in both OPEs $\mathcal{O}_1\times\mathcal{O}_2$ and $\mathcal{O}_3\times\mathcal{O}_4$.
The integer $m$ that labels the poles in (\ref{MellinpolesAp}) corresponds precisely to the label $m$ in (\ref{useries}).
This shows that the $m>0$ poles correspond to conformal descendant operators with twist greater than $\D-J$.
The residues of the poles are the kinematical polynomials $\mathcal{Q}_{J,m}(s)$ given in (\ref{Q}).
To determine these polynomials we require that the contribution of the series of poles (\ref{MellinpolesAp}) reproduces the conformal block 
$G_{\D,J}(u,v)$. 
Picking the poles (\ref{MellinpolesAp}) in the integral (\ref{reducedMellin}) one obtains a series of the form (\ref{useries}) with
\begin{align}
&g_m(v)  =
 \frac{2\Gamma(\Delta+J) (\D-1)_J }{ 4^J 
 m!(\Delta-h+1)_m\,
\Gamma\!\left( \frac{\Delta +J+\Delta_{1 2}}{2}\right)
\Gamma\!\left( \frac{\Delta +J-\Delta_{1 2}}{2}\right) 
\Gamma\!\left( \frac{\Delta +J+\Delta_{3 4}}{2}\right)
\Gamma\!\left( \frac{\Delta +J-\Delta_{3 4}}{2}\right) }
 \label{gmofv}
\\&
 \int_{-i\infty}^{i\infty} \frac{ds}{8\pi i} \,
 v^{-(s+\tau)/2}
Q_{J,m}( s)\,
  \Gamma\!\left( \frac{ \Delta_{34} -s}{2}\right)
\Gamma\!\left( \frac{-\Delta_{12}  -s}{2}\right)
\Gamma\!\left( \frac{\tau+s }{2}\right)
\Gamma\!\left( \frac{\tau+s +\Delta_{12} -\Delta_{34} }{2}\right),
  \nonumber
\end{align}
where $\tau=\Delta-J+2m$.
This explains the position of the poles (\ref{MellinpolesAp}) of the Mellin amplitude.

Let us consider first the $m=0$ case. 
Expanding (\ref{gmofv}) in powers of $(1-v)$ and using the explicit expression of $g_0(v)$ in (\ref{asymptoticsuto0}),
we obtain the following set of equations
\begin{align}
&
 \frac{(-2)^J\Gamma\!\left(\frac{\D-J-\D_{12} }{2}+n\right) 
\Gamma\!\left(\frac{\D-J+\D_{34} }{2}+n\right) 
\Gamma\!\left( \frac{\Delta +J+\Delta_{1 2}}{2}\right)
\Gamma\!\left( \frac{\Delta +J-\Delta_{3 4}}{2}\right) n! 
}{  \Gamma (\D+n) (\D-1)_J (n-J)!} =
\label{n-equations}
\\ =& 
 \int_{-i\infty}^{i\infty} \frac{ds}{4 \pi i } \,Q_{J,0}( s)\,
  \Gamma\!\left( \frac{ \Delta_{34} -s}{2}\right)
\Gamma\!\left( \frac{-\Delta_{12}  -s}{2}\right)
\Gamma\!\left( \frac{\tau+s }{2}+n\right)
\Gamma\!\left( \frac{\tau+s +\Delta_{12} -\Delta_{34} }{2}\right).
  \nonumber
\end{align} 
For $n<J$ the LHS vanishes and this equation can be written as follows
\begin{align}
0 =& 
 \int_{-i\infty}^{i\infty} \frac{ds}{4 \pi i } \,
 Q_{J,0}( s)
 \left( \frac{\tau+s }{2}\right)_n\\&
  \Gamma\!\left( \frac{ \Delta_{34} -s}{2}\right)
\Gamma\!\left( \frac{-\Delta_{12}  -s}{2}\right)
\Gamma\!\left( \frac{\tau+s }{2}\right)
\Gamma\!\left( \frac{\tau+s +\Delta_{12} -\Delta_{34} }{2}\right).
  \nonumber
\end{align} 
Taking linear combinations of this equation with $n<J$, we conclude that
it defines an inner product under which
 $ Q_{J,0}( s)$ is orthogonal
to all polynomials of $s$ with degree less than $J$.
In other words,  the polynomials $ Q_{J,0}( s)$
must satisfy 
\begin{align}
\delta_{J,J'}  \propto& 
 \int_{-i\infty}^{i\infty} \frac{ds}{4 \pi i } \,
 Q_{J,0}( s)\,
 Q_{J',0}( s)\\&
  \Gamma\!\left( \frac{ \Delta_{34} -s}{2}\right)
\Gamma\!\left( \frac{-\Delta_{12}  -s}{2}\right)
\Gamma\!\left( \frac{\tau+s }{2}\right)
\Gamma\!\left( \frac{\tau+s +\Delta_{12} -\Delta_{34} }{2}\right).
  \nonumber
\end{align}
This fixes the polynomials $Q_{J,0}( s)$ uniquely, up to normalization.
The normalization can be fixed by imposing (\ref{n-equations}) for any $n \ge J$.

The orthogonality of the polynomials $Q_{J,0}( s)$ suggests that they are the solutions of a Sturm-Liouville problem.
Indeed, the difference operator $\mathcal{D}_s$, defined by 
\be
\mathcal{D}_s Q(s)=
(s+\tau
   +\Delta_{12}-\Delta_{34} )
   \Big[(s +\tau ) Q(s +2) 
    -2 s Q(s )\Big]
   +(s+\Delta_{12})
   (s -\D_{34})   Q(s -2)\,,
   \label{differenceoperator}
\ee
is self-adjoint with respect to the inner product above.
Therefore, eigenfunctions of $\mathcal{D}_s$ with different eigenvalues are automatically orthogonal.
By construction, the action of $\mathcal{D}_s$ on a polynomial of $s$ of degree $J$ produces another polynomial of $s$ of degree $J$. 
Thus, we can look for polynomial eigenfunctions of $\mathcal{D}_s$,
\be
\mathcal{D}_s \,Q_{J,0}(s)=\lambda_J \,Q_{J,0}(s)\,.
\ee
The eigenvalue $\lambda_J$ is fixed by comparing 
the coefficient of the highest degree term $s^J$, with the result
\be
\lambda_J=4J^2+4J(\tau-1)+(\tau+\D_{12})(\tau-\D_{34})\,.
\label{lambda}
\ee
Finally, the solution can be written in terms of hypergeometric functions
\footnote{Interestingly, these polynomials already appeared in the QCD Pomeron literature \cite{Korchemsky:1994um}. We thank Gregory Korchemsky for informing us that these polynomials are known in the mathematical literature as continuous Hahn polynomials (see \url{http://aw.twi.tudelft.nl/~koekoek/askey/ch1/par4/par4.html}).}
\begin{align}
Q_{J,0}( s)=
 \frac{2^{J}\left(\frac{\Delta_{12}+\tau}{2}\right)_{J}\left(\frac{\Delta_{34}+\tau}{2}\right)_{J}}{
 \left(\tau+J-1\right)_{J}}
   \,_3F_2\!\left(-J,J+\tau
   -1,\frac{\Delta_{34}-s
   }{2};\frac{\tau+\Delta_{12}}{2},
   \frac{\tau +\Delta
   _{34}}{2};1\right) .
   \label{QJ0}
   \end{align}

Consider now the case $m>0$.
The best way to determine the functions $g_m(v)$ is to use the differential equation that the conformal block satisfies,
\be
\mathcal{D} \,G_{\D,J}(u,v) =\frac{1}{2} \,C_{\D,J}\,G_{\D,J}(u,v) \,,
\label{CasimirPDE}
\ee
where
\begin{align}
\mathcal{D}=&\,(1-u-v)\,\frac{\partial}{\partial v} \left(v\frac{\partial}{\partial v} 
+\frac{\D_{34}-\D_{12}}{2}\right)
+u\frac{\partial}{\partial u}   \left(2u\frac{\partial}{\partial u}-d\right)
\\&-(1+u-v) \left(u\frac{\partial}{\partial u} +v\frac{\partial}{\partial v} 
-\frac{\D_{12}}{2}\right)
\left(u\frac{\partial}{\partial u} +v\frac{\partial}{\partial v} 
+\frac{\D_{34}}{2}\right),
\nonumber
\end{align}
and $C_{\D,J}=\D(\D-d)+J(J+d-2)$ is the conformal quadratic Casimir.
This equation was derived in \cite{Dolan:2003hv} and it has a simple meaning: 
the conformal block is an eigenfunction of the conformal Casimir operator 
$(J^{AB}_1+J^{AB}_2)^2$ that acts on points 1 and 2.
When applied to the power series (\ref{useries}) this partial differential equation turns into the following  (differential) recursion relation for the functions $g_m(v)$,
\begin{align}
&   4 v (v-1)^2
   g_m''(v)
   -2 (v-1) g_m'(v)
   \big(2 v (-\Delta +J-2
   m-1)+\Delta _{12} (v-1)-\Delta
   _{34} (v-1)+2\big) 
   \nonumber \\&
   +g_m(v) \Big(4 m (-2 h+m v+m)+J^2
   (v-1)-2 J \big(\Delta +2 m
   (v+1)+\Delta  v-2\big)
     \nonumber\\&
     +(v-1)
   \big(\Delta _{12} \left(-\Delta
   -\Delta _{34}+J-2 m\right)+\Delta
   _{34} (\Delta -J+2 m)\big)+4
   \Delta  m (v+1)+\Delta ^2
   (v-1)\Big)  \nonumber
   \\=\ &
   4 v (v+1)
   g_{m-1}''(v)
+
   2 g_{m-1}'(v) \big(2 v (\Delta -J+2
   m-1)-\Delta _{12} (v+1)+\Delta
   _{34} (v+1)+2\big)\nonumber
\\&
   +g_{m-1}(v)
   \left(-\Delta +\Delta _{12}+J-2
   m+2\right) \left(-\Delta -\Delta
   _{34}+J-2 m+2\right)\,.
   \label{recursiong(v)}
\end{align}
This is an ugly equation which the reader should not read in detail.
Nevertheless, it is not hard to check that $g_0(v)$ given by (\ref{asymptoticsuto0}) solves the $m=0$ equation.
Most importantly, if we replace $g_m(v)$ given by (\ref{gmofv}) in  equation (\ref{recursiong(v)}),
we obtain a set of recursion relations for the polynomials $Q_{J,m}( s)$,
\begin{align}
\left(\mathcal{D}_s - \lambda_J  \right)Q_{J,m}( s) 
= 
4m(h-\D-m) \big( 2Q_{J,m}(s)-Q_{J,m-1}(s+2) - Q_{J,m-1}(s)\big)\,,
\label{eqQJm}
\end{align}
where $\mathcal{D}_s$ and  $ \lambda_J$ are respectively given by (\ref{differenceoperator}) and (\ref{lambda})
with $\tau=\Delta-J+2m$.
This equation, plus the boundary condition 
$Q_{J,0}(s)=s^J+O(s^{J-1})$ which follows from (\ref{QJ0}), determine the polynomials $Q_{J,m}( s)$ for all $m\ge 0$.
In particular, it is clear that the leading behaviour is given by (\ref{Qasymp})
for all $m$. This follows from the fact that the LHS of (\ref{eqQJm}) is automatically a polynomial 
of degree $(J-1)$, if we assume that $Q_{J,m}( s) $ is a polynomial of degree $J$. Imposing the 
same condition to the RHS implies that $Q_{J,m}(s)$ and $Q_{J,m-1}(s)$ have the same leading behaviour.

\subsection{Flat space limit}
\label{FlatSpaceSec}

In AdS/CFT, the radius $R$ of AdS in units of the string length $l_s$ is a free parameter related to the 't Hooft coupling of the gauge theory.
Therefore, it should be possible to recover bulk flat space physics by taking the limit $R/l_s \to \infty$ and keeping energies fixed in string units.
Given the similarity between Mellin amplitudes and scattering amplitudes it is not surprising that there is a simple formula that relates them.
Such a formula was proposed in \cite{JPMellin} and rederived in \cite{JLAnalyticMellin} using localized wavepackets. It reads
\be
\mathcal{T}(p_i) =\frac{1}{\mathcal{N}}
\lim_{R\to \infty} R^{2h-3}\int_{-i\infty}^{i\infty}
\frac{d\alpha}{2\pi i}\, \alpha^{h-\frac{1}{2}\sum\D_i}
\,e^\alpha \,M\!\left( \delta_{ij}= \frac{R^2 }{2\alpha}\,p_i\cdot p_j \right),
\label{flatspacelimit}
\ee
where the integration contour runs to the right of all poles of the integrand and \footnote{
This normalization differs from \cite{JPMellin} because here we are using operators normalized to have unit two-point function (\ref{normalized2pf}).}
\be
\mathcal{N}= 
 \frac{1}{8\pi^h}
\prod_{i=1}^4 \frac{1}{\sqrt{\Gamma(\D_i)\Gamma(\D_i-h+1)}}\label{eq:NMellin} \,.
\ee
In formula (\ref{flatspacelimit}), 
$M$ is the Mellin amplitude of a CFT four-point function of single-trace operators $\mathcal{O}_i$ and $\mathcal{T}$ is the scattering amplitude of the dual bulk fields $\phi_i$.

A relevant example for the present paper is the tree-level exchange of a spin $J$ and mass $m$ particle.  In flat space, this gives rise to 
\be
\mathcal{T}=g^2\, T^{J-1} f\!\left(\frac{S}{T},\frac{m^2}{T}\right) ,
\ee
where $S$ and $T$ are the Mandelstam invariants, $g$ is a dimensionful coupling constant and $f$ is a dimensionless function.
Then, formula (\ref{flatspacelimit}) tells us that the Mellin amplitude associated to the tree-level exchange of a spin $J$ and dimension $\D$ field in AdS has the following asymptotic behaviour
\be
M(s,t) =g^2R^{5-2h-2J}\, t^{J-1} \tilde{f}\!\left(\frac{s}{t},\frac{\D^2}{t}\right) +O(t^{J-2})\,,
\ee
with
\footnote{
If the particle is massless in flat space ($m=0$) then the relation between $f$ and $\tilde{f}$ is very simple
\be
 f\left(x,0\right) =\frac{2^{1-J}}{\mathcal{N}\,
 \Gamma\!\left(\frac{1}{2}\sum\D_i -h+J-1 \right) }\,
  \tilde{f}(x, 0)\,.
  \nonumber
\ee
}
\be
 f\left(x,y\right) =\frac{2^{1-J}}{\mathcal{N}}
 \int_{-i\infty}^{i\infty}
\frac{d\alpha}{2\pi i}  \,\alpha^{h-\frac{1}{2}\sum\D_i-J+1}
\,e^\alpha\,   \tilde{f}(x, 2y\alpha)\,.
\ee

\subsubsection{Flat space limit of conformal partial wave expansion}
\label{FlatSpaceSubSec}

Next we study the flat space limit (\ref{flatspacelimit}) of the conformal partial wave expansion (\ref{CPWE}),
\be
\mathcal{T}(p_i) =\frac{1}{\mathcal{N}} \sum_{J=0}^\infty
\lim_{R\to \infty} R^{2h-3}
\int_{-\infty}^\infty  d\nu  \, b_J(\nu^2)
\int_{-i\infty}^{i\infty}
\frac{d\alpha}{2\pi i} \,\alpha^{h-\frac{1}{2}\sum\D_i}
\,e^\alpha \,M_{\nu,J}\!\left(   
\frac{R^2 S}{2\alpha},\frac{R^2 T}{2\alpha} \right).
\label{flatlimitCPWE}
\ee
In order to compute the large $R$ limit of the integral it would be useful to know what is the integration region in $\nu$ and $\alpha$ that dominates the integral for large $R$.
We shall start by assuming that the integral is dominated by $\nu^2 \gg 1$ and later check that this is indeed the case.
Using the Stirling expansion of the $\Gamma$-function we find
\begin{align}
 \omega_{\nu,J}(t)\approx
 \frac{ 1}{\sqrt{\nu^2}} 
 \left( \frac{-\nu^2}{2t}\right)^{\frac{1}{2}\sum\D_i +J-h}
 \exp \left\{
\nu  \arctan\!\left(\frac{\nu }{t}\right) 
 -\frac{t}{2} \log\!
   \left(1+ \frac{\nu^2}{t^2} \right)\right\},
\end{align}
where we are assuming  
$-t=-\frac{R^2 T}{2\alpha}\gg |\nu| \gg 1$.
In appendix \ref{AppendixMack} we consider the limit 
$|t|\sim|s| \gg|\nu| \gg 1$   of the Mack polynomials, and  obtain
\be
P_{\nu,J}(s,t) \approx  \left(\frac{t}{2}\right)^J P_J(z)\,,
\ee
where $ P_J(z)$ are the partial waves in $(2h+1)$-dimensional flat spacetime
and $z=1+\frac{2s}{t}$.
Using these two approximations (\ref{flatlimitCPWE}) becomes
 \begin{align}
\mathcal{T}(p_i) =\ &\frac{1}{\mathcal{N}} \sum_{J=0}^\infty P_J(z)
\lim_{R\to \infty} R^{2h-3}
  \left(\frac{R^2T}{4}\right)^J
\int_{-\infty}^\infty  d\nu  \, b_J(\nu^2)\,
\frac{ 1}{\sqrt{\nu^2}} 
 \left( \frac{-\nu^2}{R^2T}\right)^{\frac{1}{2}\sum\D_i +J-h}
 \nonumber
\\
&\int_{-i\infty}^{i\infty}
\frac{d\alpha}{2\pi i}  
 \exp \left\{\a+
 \nu  \arctan\!\left(\frac{2\a \nu }{R^2T}\right) 
 -\frac{R^2T}{4\a} \log\!
   \left(1+ \left(\frac{2\a \nu }{R^2T}\right)^2 \right)
  \right\},
 \label{phaseint}
 \end{align}
 where now $z=\cos \theta=1+\frac{2S}{T}$ encodes the flat space scattering angle.
Let us discuss the integral (\ref{phaseint}).
If we expand the exponent at large $R$, we obtain
\be
\int_{-i\infty}^{i\infty}
\frac{d\alpha}{2\pi i}  
\exp \left\{\a\left(1+\frac{\nu^2}{R^2T}\right) -\frac{2\a^3 \nu^4}{3 R^6T^3}
+O\!\left( \frac{1}{R^{10}}  \right) \right\}.
\label{expandedint}
\ee
Keeping only the first term in  this exponential, the integral over $\a$  gives rise to a delta-function $\d\!\left(1+\frac{\nu^2}{R^2T}\right)$.
This justifies the initial assumption of large $\nu^2$.
However, one must be careful because the delta-function follows from taking the integrand to be a plane wave in $\a$, for all values of $\a$.
This is clearly wrong since  for $\a \sim R^2T/\nu^{4/3}$ the second term in the exponent becomes of order 1.
In fact, we can perform the integral over $\a$ keeping only the first two terms in the exponent, obtaining
\be
\frac{R^2T}{(2\nu^4)^{1/3}}\,{\rm Ai}\!\left(\frac{\nu^2+R^2T}{(2\nu^4)^{1/3}}\right),
\label{Airy}
\ee
where Ai is the Airy function.
This expression means that the integral over $\nu$ is dominated by the region $\nu^2\sim -R^2T \pm \nu^{4/3} \sim  -R^2T \pm (-R^2T)^{2/3}$.
At large $R$, both the mean value of $\nu^2$ and the width of the region are large, but the mean is much larger than the width.
Including higher order corrections in (\ref{expandedint}), leads to corrections to the function of $\nu^2$ (\ref{Airy}) in smaller scales than $(-R^2T)^{2/3}$ but still much larger than 1.
Therefore, we conclude that the flat space limit of the conformal partial wave expansion gives the standard partial wave expansion,
\begin{align}
\mathcal{T}= \sum_{J=0}^\infty  P_J(z) \,
a_J(T)\,,
\end{align}
with the flat space partial amplitudes given by the limit 
\be
a_J(T)=\frac{1}{ \mathcal{N}}\lim_{R\to \infty} R^{2h-3} 
   \left(\frac{ R^2T}{4 }\right)^J\,   \langle b_J \rangle_T\,,
\ee
where  
\be
  \langle b_J \rangle_T=
\int   
 dx \,
 \delta_L(x)\,
 b_J\big(-R^2T+ x\big)
 \label{averaging}
\ee
 is an averaging of the conformal partial amplitudes 
 $b_J(\nu^2)$ around 
 $\nu^2=-R^2T$ with a function $\delta_L(x)$ which is a regulated delta-function with characteristic width $L= (-R^2T)^{2/3}$. 
%

The flat space limit of conformal blocks in Mellin space was first studied in \cite{JLAnalyticMellin}. 
The main novelty of our result  is the averaging (\ref{averaging}). As we saw in the specific example 
studied in section \ref{strongOPEcoe}, the practical effect of this averaging is simply to smooth out rapid 
oscillations of  $b_J(\nu^2)$, which should not be present in the flat space partial waves $a_J(T)$.

\subsection{Example: Witten diagrams}

\begin{figure}
\begin{centering}
\includegraphics[scale=0.35]{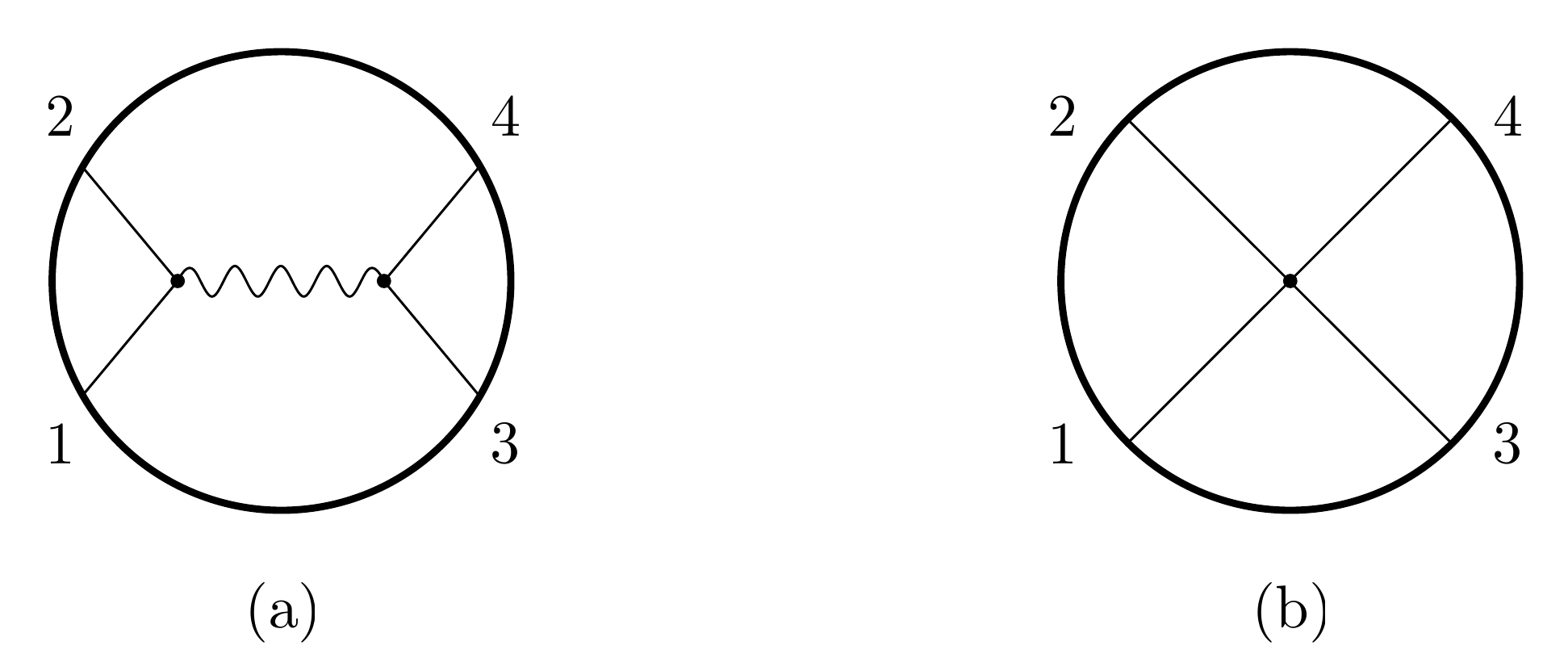}
\par\end{centering}
\caption{\label{WittenSpinJ}Witten diagrams of (a) exchange of a dimension $\D$ and spin $J$ field in AdS and (b) contact interaction.}
\end{figure}

Consider the Witten diagram in figure \ref{WittenSpinJ}a associated with the exchange of a dimension $\D$ and spin $J$ field in AdS.
The OPE expansion of the corresponding four-point function in the (12)(34) channel contains double-trace operators and the single-trace operator dual to the exchanged field in AdS \cite{Liu,D'Hoker,D'Hoker:1999jp}. The OPE expansion in the other channels only contains double-trace operators.
This means that the only poles of the associated Mellin amplitude
are given by equation (\ref{Mellinpoles}).
In addition, we know from the flat space limit analysis of the previous section
that this Mellin amplitude is polynomially bounded at large values of $s$ and $t$.
Thus, we conclude that it can be written as a sum of poles plus an analytic piece, which is a polynomial $\mathcal{R}_{J-1}$ of degree $J-1$ in both variables $s$ and $t$,
\be
M(s,t)=
C_{12k}C_{34k} \sum_{m=0}^\infty 
\frac{\mathcal{Q}_{J,m}( s ) }{t-\D+J-2m} 
 +\mathcal{R}_{J-1}(s,t)\,.
\label{MellinWitten}
\ee
Let us see if this results agrees with the expectations from the bulk point of view.
To compute the Witten diagram in figure \ref{WittenSpinJ}a we need to know what is the precise form of the cubic vertices.
However, there is a unique cubic vertex between 2 scalar fields and a spin $J$ field if we are allowed to use the equations of motion.
This is directly related to the fact that there is a unique conformal three-point function between 2 scalar operators and a spin $J$ operator (see \cite{SpinningCC} for a more complete discussion of this correspondence).
On the other hand, the internal line of the diagram \ref{WittenSpinJ}a is not on-shell 
and, therefore, the equations of motion will not give zero, but will transform the internal propagator into a delta-function.
This means that different cubic vertices will produce correlation functions
that differ by contact diagrams like the one in figure \ref{WittenSpinJ}b.
In fact, it is not hard to convince ourselves  that this contact diagrams can have at most  $2J-2$ derivatives. As explained in \cite{JPMellin}, 
this implies that the associated Mellin amplitude is a polynomial of degree $J-1$.
Thus, the result (\ref{MellinWitten}) is exactly what one expects from the bulk point of view. The Mellin amplitude contains a polynomial  $\mathcal{R}_{J-1}$ that encodes the precise choice of cubic couplings, and a sum of poles completely fixed by the OPE coefficients of the exchanged operator $\mathcal{O}_k$ in the OPEs  of
$\mathcal{O}_1\mathcal{O}_2$ and $\mathcal{O}_3\mathcal{O}_4$.
Similar arguments were recently given in \cite{JLBoundedMellin}.

\subsubsection{Regge limit}

Something nice happens in the Regge limit of large $s$ and fixed $t$.
Firstly, the non-universal part $\mathcal{R}_{J-1}$ of the Mellin amplitude 
(\ref{MellinWitten}) drops out. 
Secondly, the polynomials $Q_{J,m}(s)$, introduced in (\ref{Q}),
 can be replaced by their asymptotic
behavior (\ref{Qasymp}).
This gives
\be
M (s,t)\approx C_{12k} C_{34k}\, f (t)\, s^J \,,
\label{WittenRegge}
\ee
with
\begin{align}
f (t)=&  -\frac{2\Gamma(\Delta+J)(\Delta-1)_J   }{ 4^J
\Gamma\!\left( \frac{\Delta +J+\Delta_{1 2}}{2}\right)
\Gamma\!\left( \frac{\Delta +J-\Delta_{1 2}}{2}\right) 
\Gamma\!\left( \frac{\Delta +J+\Delta_{3 4}}{2}\right)
\Gamma\!\left( \frac{\Delta +J-\Delta_{3 4}}{2}\right)}
\\&
 \sum_{m=0}^\infty \frac{1 }{m!(\D-h+1)_m\, \Gamma\!\left( \frac{\Delta_1 +\Delta_{2} -\D+J-2m}{2}\right)
\Gamma\!\left( \frac{\Delta_3 +\Delta_{4} -\D+J-2m}{2}\right)(t-\D+J-2m)} \,.
\nonumber
\end{align}
Fortunately, this sum has a nice integral representation \cite{JPMellin} 
\be
f(t)=
 K_{\D,J} \int  d\nu  \,  \frac{ \omega_{\nu,J}(t)}{(\D-h)^2+\nu^2}\,\,,
 \label{beta(t)}
\ee
where $\omega_{\nu,J}(t)$ is given in (\ref{omeganuJt}) and  the normalization constant 
$ K_{\D,J}$ is given in (\ref{KDeltaJ}).

In the Regge limit, it is striking how similar is the behavior of the Mellin amplitude (\ref{WittenRegge}) for an exchange of a spin $J$ field in AdS, and 
the corresponding flat space scattering amplitude.
Both grow as $s^J$ (or $S^J$) times a function of $t$ (or $T$).
In the integral representation (\ref{beta(t)}) the infinite sequence of poles in $t$ is generated by a single pole in $\nu^2$.
Indeed, this pole is the best analogue to the unique pole in $T$ of the flat space scattering amplitude.

\subsection{Double trace operators \label{DoubleTraces}}

Let us briefly remark how double trace operators, that also appear in the  conformal block decomposition (\ref{CBE}), are generated in the 
conformal partial wave expansion (\ref{CPWE}). The double traces are generated by poles of $\kappa_{\nu,J}$ at 
\begin{align}
i\nu&=\D_1+\D_2+J+2m-h\,,&
m&=0,1,2,\dots\,,\\  
i\nu&=\D_3+\D_4+J+2m-h\,,&  
m&=0,1,2,\dots\,. 
\end{align}
The product of the OPE coefficients of a double-trace operator $\tilde{\mathcal{O}}_k\sim\mathcal{O}_1\partial_{\mu_1} \dots \partial_{\mu_J} \partial^{2m}\mathcal{O}_2$, of dimension 
 $\tilde{\D}_k=\D_1+\D_2+\tilde{J}+2m$ and spin $\tilde{J}$, in the OPE   $\mathcal{O}_1 \mathcal{O}_2$ and   $\mathcal{O}_3 \mathcal{O}_4$,
is given by
\begin{align}
 \tilde{C}_{12k}\tilde{C}_{34k} =\ & 
 \frac{4^J 
\Gamma\!\left(\frac{\tilde{J}+\Delta _3+\Delta _4-\tilde{\D}_k}{2}\right) 
\Gamma\!\left(\frac{-2 h+\tilde{J}+\Delta_3+\Delta _4+\tilde{\D}_k}{2}\right) 
\Gamma\!\left(\frac{\tilde{J}-\Delta _{34}+\tilde{\D}_k}{2}\right) 
\Gamma\!\left(\frac{\tilde{J}+\Delta _{34}+\tilde{\D}_k}{2}\right)}{m! \Gamma\!\left(\tilde{J}+\tilde{\D}_k\right)
   \left(h-\tilde{\D}_k+1\right)_m \left(\tilde{\D}_k-1\right)_{\tilde{J}}}
   \nonumber \\&
   \Gamma\!\left(\tilde{J}+m+\Delta _1\right) \Gamma\!\left(\tilde{J}+m+\Delta_2\right)b_{\tilde{J}}  \!\left( -\left(\tilde{\D}_k-h\right)^2 \right) \,.
   \label{dtCs}
\end{align}

To illustrate the use of this formula consider the correlator associated to the Witten diagram of figure \ref{WittenSpinJ}a describing the exchange of a spin $J$ and dimension $\D$ field in AdS.
In the previous section, we concluded that the corresponding Mellin amplitude was a polynomial of degree $J$ in the variable $s$.
This implies that the conformal partial wave expansion (\ref{CPWE}) obeys
$b_{J'}(\nu^2)=0$ for $J'>J$.
Thus, the Regge limit of (\ref{CPWE}) is simply given by
\be
M(s,t)\approx  s^J\,  \int_{-\infty}^\infty  d\nu  \, b_J(\nu^2) \,\omega_{\nu,J}(t)\,.
\ee
Comparing with the results  (\ref{WittenRegge}) and (\ref{beta(t)}) for the Regge limit of the Witten diagram of figure \ref{WittenSpinJ}a, we conclude that
\be
b_J(\nu^2)= C_{12k}C_{34k}\,\frac{ K_{\D,J}}{\nu^2+(\Delta-h)^2}\,,
\label{maximalspin}
\ee
where $C_{12k}C_{34k}$ is the product of the OPE coefficients of the operator dual to the field exchanged in AdS.
Notice that in this case, the partial amplitude $b_J(\nu^2)$ is exactly given by the sum of the two simple poles predicted in (\ref{polesfromCB}).
The partial amplitudes $b_{J'}(\nu^2)$ for $J'<J$ are more complicated and are not determined by the Regge limit.
It is then trivial to use  $b_J(\nu^2)$ given by (\ref{maximalspin}) in (\ref{dtCs}), to immediatelly obtain the OPE coefficients of the double trace operators of maximal spin $\tilde{J}=J$ produced by the Witten diagram 
of figure \ref{WittenSpinJ}a.

\subsection{Analytic structure of partial amplitudes \label{analyticstructure}}

\begin{figure}
\begin{centering}
\includegraphics[scale=.45]{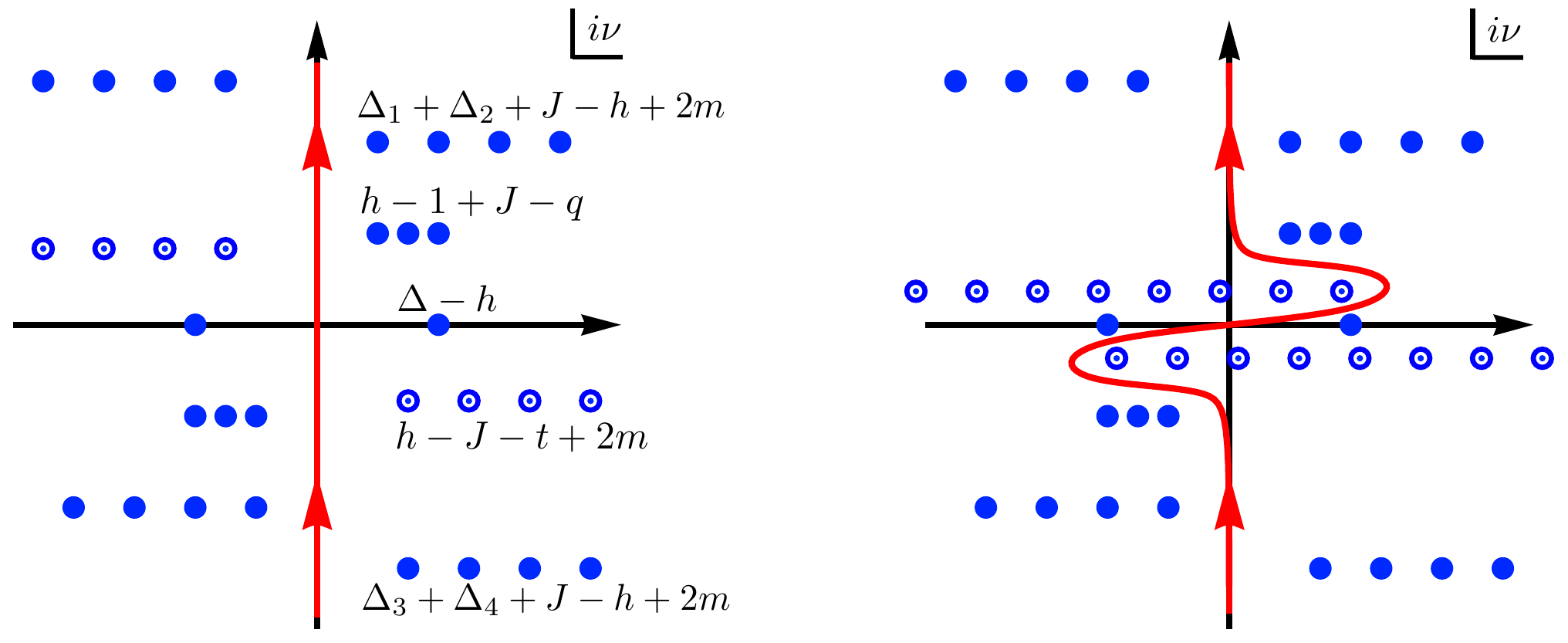}
\par\end{centering}
\caption{\label{figMellinPoles}
Integration contour in the $i\nu$ complex plane used in the conformal partial wave expansion (\ref{CPWE}). The blue dots represent poles of the integrand given by (\ref{dt12}-\ref{polypoles}) and (\ref{partialwavepole}). 
In order to make the figure readable, we have complexified several parameters to separate the poles better.
The poles (\ref{nutpoles}) that depend on $t$ are marked with a small dot enclosed by a circle. As $t$ varies these poles move and can collide with other poles pinching the integration contour, as shown on the right panel. This is the mechanism that generates the poles  (\ref{Mellinpoles}) of the Mellin amplitude.}
\end{figure}

The pole structure of the partial amplitudes  $b_J(\nu)$ is directly related to the spectrum of single-trace operators that appear in both OPEs $\mathcal{O}_1\mathcal{O}_2$ and
$\mathcal{O}_3\mathcal{O}_4$.
The mechanism is the following: the poles  (\ref{Mellinpoles}) of the Mellin amplitude arise from the integral over $\nu$ in (\ref{CPWE}) when the integration contour is pinched between two poles of the integrand as depicted in figure \ref{figMellinPoles}.
The partial wave $M_{\nu,J}(s,t)$ introduced in (\ref{MellinPartialWave})
has the following poles in the variable $\nu$,
\begin{align}
\pm i\nu&= \Delta_1 +\Delta_{2} +J -h +2m
\,,&\ \ \   m&=0,1,2,\dots\,, \label{dt12}\\
\pm i\nu&= \Delta_3 +\Delta_{4} +J -h +2m
\,,&\ \ \   m&=0,1,2,\dots\,, \label{dt34}\\
\pm i\nu&=  h-J-t +2m \label{nutpoles}
\,,&  m&=0,1,2,\dots\,,\\
\pm i\nu&=  h-1+J-q
\,,& \ \ \ q&=1,2,\dots,J\,, 
\label{polypoles}
\end{align}
where the first three sets of poles come from the function $\omega_{\nu,J}(t)$
defined in (\ref{omeganuJt}), and the last line are poles
of the polynomial $P_{\nu,J}(s,t)$ defined in appendix \ref{AppendixMack}. 
In order to obtain poles in the variable $t$   
from the integral over $\nu$ in (\ref{CPWE}), a pole from (\ref{nutpoles}) must collide with another $\nu$ pole. 
In fact, in order to reproduce the poles  (\ref{Mellinpoles}) with the correct residue one needs the partial amplitudes $b_J(\nu^2)$ to have the following pair of $\nu$ poles
\be
b_J(\nu^2)\approx C_{12k}C_{34k}\,\frac{ K_{\D,J}}{\nu^2+(\Delta-h)^2}\,,
\label{partialwavepole}
\ee
where the normalization constant $K_{\D,J}$ is given in (\ref{KDeltaJ}).
When $t$ approaches $\D-J+2m$ with $m=0,1,2,\dots$,
two poles from (\ref{nutpoles}) collide with the two poles 
(\ref{partialwavepole}), pinching the $\nu$-contour in (\ref{CPWE})  and producing a divergent integral (see figure \ref{figMellinPoles}). To check that the resulting poles in $t$ of the Mellin amplitude have the correct residues it is sufficient to keep the contribution from the poles (\ref{partialwavepole}) to the integral (\ref{CPWE}) using the Cauchy theorem,
\begin{align}
M(s,\D-J+2m+\d t)&\approx 
C_{12k}C_{34k}\,K_{\D,J}
 \,\frac{2\pi }{ \Delta-h }
\,M_{i(\D-h),J}(s,\D-J+2m+\d t)
\nonumber\\
&\approx 
C_{12k}C_{34k}\,
 \frac{\mathcal{Q}_{J,m}(s)}{ \d t}\,,
\end{align}
in perfect agreement with (\ref{Mellinpoles}).
To obtain this result it was crucial to use the property (\ref{PgivesQ}) of the Mack polynomials.
We conclude that for every single-trace operator that appears in both OPEs $\mathcal{O}_1\mathcal{O}_2$ and $\mathcal{O}_3\mathcal{O}_4$, the partial amplitudes $b_J(\nu^2)$ will have a pair of poles of the form (\ref{partialwavepole}).

Unfortunately, the story is slightly more complicated and cumbersome because $b_J(\nu^2)$ has other (spurious) poles that do not correspond to any operators appearing in the OPEs. 
To explain this let us systematically analyze all possible contour pinchings in (\ref{CPWE}) that can give rise to poles in $t$.
Suppose a pole from (\ref{nutpoles}) collides with a pole from (\ref{dt12}). This would give rise to a pole at $t=\D_1+\D_2+2m$ for $m=0,1,2,\dots$.
However, this pole is cancelled by a zero of $\omega_{\nu,J}(t)$ produced by the last $\Gamma$-functions in the denominator of (\ref{omeganuJt}). 
A similar statement applies to collision with the poles (\ref{dt34}).
Another possibility is the collision of 2 poles of the form 
(\ref{nutpoles}) themselves. 
This happens when $J+t-h$ is a non-negative integer, which means that the colliding poles are located at integer values of $i\nu$. Thus, this collision also does not generate poles in $t$ because the function $\omega_{\nu,J}(t)$ has zeros at integer values of $i\nu$.
The final possibility is for the poles (\ref{nutpoles}) to collide with the poles (\ref{polypoles}) of the Mack polynomials.
Let us focus on the contribution of one the poles 
(\ref{polypoles}) for a fixed value of $J$ and $q$.
This gives rise to a series of poles in $t$ of the form (\ref{Mellinpoles}) with dimension $\D'$, spin $J'$ and OPE coefficients $C_{12k}'C_{34k}'$ given by
\begin{align}
\D'&=2h-1+J\,,\\
J'&=J-q\,,\\
 C_{12k}'C_{34k}' &=
\frac{\mathcal{Z}_{J,q}}{ K_{\D',J'}} \,b_J\big(-(h-1+J-q)^2\big)\,,
\end{align}
where $K_{\D,J}$ is given in (\ref{KDeltaJ}) and
\be
\mathcal{Z}_{J,q}=
   \frac{J!}{ (J-q)! q!}
   \frac{2(-2)^q    \left(\frac{\Delta
   _1+\Delta _2+1-2 h-q}{2}\right)_q 
   \left(\frac{\Delta
   _3+\Delta _4+1-2 h-q}{2}\right)_q 
   \left(\frac{\Delta
   _{12}+1-q}{2}\right)_q 
   \left(\frac{\Delta
   _{34}+1-q}{2}\right)_q}{
   \Gamma (q)
   (h+J-q)_{q-1}}\,.
   \ee
In order to derive this result we used the property (\ref{MackPolypoles}) of the Mack polynomials given in appendix \ref{AppendixMack}.
This result looks strange because it says that the OPE will generically contain primary operators of dimension $\D'$ (which is an integer or half-integer). This can not be the case.
In fact, what happens is that the partial amplitudes $b_{J'}(\nu^2)$, with $J'=J-q$ have other poles that cancel this effect. This requirement fixes the new residues to be
\be
b_{J-q}(\nu^2)\approx-
\frac{ \mathcal{Z}_{J,q}\,b_{J}\big(-(h-1+J-q)^2\big)}{\nu^2+(h-1+J)^2}\,,
\ \ \ \ \ \ \ \ \ \ q=1,2,3,\dots\,J.
\label{spuriouspole}
\ee
These poles were termed spurious poles in \cite{CornalbaRegge}.
We believe that the relation (\ref{spuriouspole}) is the translation to our language of the identity (2.59b) of \cite{Sofia}, which discusses a similar conformal partial wave expansion (although in position space).

\section{Mack polynomials \label{AppendixMack}}

With our normalizations, the polynomials introduced in \cite{Mack} can be written as
\begin{align}
P_{\nu,J}(s,t)=&  \,
\sum_{r=0}^{[J/2]} a_{J,r} 
\frac{2^{J+2r} \left(\frac{h+i\nu-J-t}{2} \right)_r
\left(\frac{h-i\nu-J-t}{2} \right)_r (J-2r)!  }{\left(h+i\nu-1\right)_J\left(h-i\nu-1\right)_J}
\label{MackPol}
\\&
\sum_{\sum k_{ij}=J-2r} (-1)^{k_{13}+k_{24}}
\prod_{(ij)}\frac{(\d_{ij})_{k_{ij}}}{k_{ij}!}
\prod_{n=1}^4 \left( \a_n   \right)_{J-r-\sum_j k_{jn}}\,.
\nonumber
\end{align}
In this expression the labels $(ij)$ run over the 4 possibilities (13), (14), (23) and (24). The variables $\d_{ij}$ are as before,
\be
\d_{13}= \frac{ \Delta_{34} -s}{2}\,,\ \ \ 
\d_{24}=-\frac{\Delta_{12}  +s}{2}\,,\ \ \ 
\d_{23}= \frac{t+s }{2}\,,\ \ \ 
\d_{14}= \frac{t+s +\Delta_{12} -\Delta_{34} }{2}\,.
\ee
The variables $\a_n$ are given by\footnote{
The coefficients $\alpha_i$ had a typo in the previous version. We thank Emilio Trevisani for calling this to our attention.}
\begin{align}
&\a_{1}= 1-\frac{ h+i \nu+J +\D_{12}}{2}\,,\ \ \ \ \ \ 
\a_{2}=1-\frac{ h+i \nu +J-\D_{12}}{2}\,, \label{a1a2}\\
&\a_{3}= 1-\frac{ h-i \nu +J+\D_{34}}{2}\,,\ \  \ \ \ \ 
\a_{4}=1-\frac{ h-i \nu +J-\D_{34}}{2}\,.
\nonumber
\end{align}
The coefficients $a_{J,r}$ define the flat $(2h+1)$-dimensional spacetime partial waves
\be
P_{J}(z)=
\sum_{r=0}^{[J/2]} a_{J,r} \,z^{J-2r}\,,\ \ \ \ \ \ 
a_{J,r} = (-1)^r\, \frac{J! (h+J-1)_{-r}}{2^{2r} r! (J-2r)!}\,.
\ee
It is clear from the definition (\ref{MackPol}) that $P_{\nu,J}(s,t)$ is indeed a polynomial of degree $J$ in both variables $t$ and $s$.
Let us check that the leading term is $s^J$, as stated in the main text.
This must come from the $r=0$ term in the sum  (\ref{MackPol}),
\be
P_{\nu,J}(s,t)\approx 
\frac{s^{J}  J!  }{\left(h+i\nu-1\right)_J\left(h-i\nu-1\right)_J}
\,\sum_{\sum k_{ij}=J}  \,
\prod_{(ij)}\frac{1}{k_{ij}!}\,
\prod_{n=1}^4 \left( \a_n  \right)_{J-\sum_j k_{jn}}\,,
\label{leadingtermP}
\ee
where we have kept only the leading term in $s$ in the Pochhammer symbols $(\d_{ij})_{k_{ij}}$.
To perform the last sum we change to the variables
$q_1= J-k_{13}-k_{14}$ and $q_3= J-k_{13}-k_{23}$. Then the sum over $k_{ij}$ in (\ref{leadingtermP}) can be written as
\begin{align}
&\sum_{q_1=0}^J  \,\sum_{q_3=0}^J \,
\sum_{k_{13}=0}^{J-q_1}\,
\frac{\left( \a_1  \right)_{q_1}
\left( \a_2 \right)_{J-q_1}
\left( \a_3 \right)_{q_3}
\left( \a_4\right)_{J-q_3}
}{k_{13}!\,(J-q_1-k_{13})!\,(J-q_3-k_{13})!\,(q_1+q_3-J+k_{13})!}\\
=
&\ J!\, \sum_{q_1=0}^J \, 
\frac{\left( \a_1  \right)_{q_1}
\left( \a_2 \right)_{J-q_1}
}{q_{1}!\,(J-q_1 )! }
\,\sum_{q_3=0}^J \,
\frac{
\left( \a_3 \right)_{q_3}
\left( \a_4\right)_{J-q_3}
}{( q_3 )!\,(J-q_3 )!}
=
\frac{\left( \a_1+\a_2  \right)_{J}
\left( \a_3+\a_4 \right)_{J}
}{J! }\,.
\nonumber
\end{align}
Using the definitions (\ref{a1a2})  it follows that $P_{\nu,J}(s,t) \approx s^J$.

There are several symmetry properties that follow from the formula 
(\ref{MackPol}) by relabelling the summation variables,
\begin{align}
&k_{13} \leftrightarrow k_{24}\ \ \ \Rightarrow \ \ 
P_{-\nu,J}\big(s,t,\D_{12},\D_{34}\big) = P_{\nu,J}\big(s,t,-\D_{34},-\D_{12}\big)\\
&k_{14} \leftrightarrow k_{23}\ \ \ \Rightarrow \ \ 
P_{-\nu,J}\big(s,t,\D_{12},\D_{34}\big) = P_{\nu,J}\big(s+\D_{12}-\D_{34},t,\D_{34},\D_{12}\big)
\end{align}
and
\begin{align}
&\left\{{k_{13} \leftrightarrow k_{14} \atop  k_{23} \leftrightarrow k_{24} }\right.\ \ \ \Rightarrow \ \ 
P_{\nu,J}\big(s,t,\D_{12},-\D_{34}\big) = (-1)^J 
P_{\nu,J}\big( -s-t-\D_{12},t,\D_{12},\D_{34}\big)\\
&\left\{{k_{13} \leftrightarrow k_{23} \atop  k_{14} \leftrightarrow k_{24}}\right. \ \ \ \Rightarrow \ \ 
P_{\nu,J}\big( s,t,-\D_{12},\D_{34}\big) = (-1)^J 
P_{\nu,J}\big(-s-t+\D_{34},t,\D_{12},\D_{34}\big)
\end{align}
In fact, there is a more basic invariance, $P_{-\nu,J}(s,t)=P_{\nu,J}(s,t)$, which is not obvious from the definition (\ref{MackPol}). These symmetries were first discussed in \cite{DOMellin}.

Another important property is that the Mack polynomials, at specific values of $t$,  reduce to the polynomials $Q_{J,m}(s)$ that control the OPE as explained in section \ref{secOPE},
\be
P_{i(\D-h),J}(s,\D-J+2m) = Q_{J,m}(s)\,.
\label{PgivesQ}
\ee

Consider now the limit 
$t \sim s  \gg 1$ and $ \nu^2 \gg 1$.
In equation (\ref{MackPol}), it is sufficient to replace the Pochhammer symbols  $(x)_n$ of a large quantity $x$ by the leading term $x^n$. This gives
\begin{align}
P_{\nu,J}(s,t)\approx&
\sum_{r=0}^{[J/2]} a_{J,r}\, 
4^{r-J} \left(t^2+\nu^2\right)^{r} (J-2r)! 
\sum_{\sum k_{ij}=J-2r} 
\frac{s^{k_{13}+k_{24}} (t+s)^{k_{14}+k_{23}}}{k_{13}!\,k_{24}!\,k_{14}!\,k_{23}!}
\nonumber\\
=&  
\sum_{r=0}^{[J/2]} a_{J,r}\, 
2^{-J} \left(t^2+\nu^2\right)^{r}
\sum_{q=0}^{J-2r} 
\frac{ (J-2r)! }{q!\,(J-2r-q)!} \, s^{q} (t+s)^{J-2r-q}
\\
=&  
\left(\frac{t^2+\nu^2}{4}\right)^{\frac{J}{2}}\,\sum_{r=0}^{[J/2]} a_{J,r}\, 
 \left( \frac{t+2s}{\sqrt{t^2+\nu^2}}\right)^{J-2r} = 
\left(\frac{t^2+\nu^2}{4}\right)^{\frac{J}{2}}
 P_J\!\left( \frac{t+2s}{\sqrt{t^2+\nu^2}}\right).
 \nonumber
\end{align}
If we further assume $|t|\gg|\nu|$, we obtain
\begin{align}
P_{\nu,J}(s,t)\approx 
 \left(\frac{t}{2}\right)^{J} P_J(z)\,,
\end{align}
where $z=1+2s/t$.
This limit was first studied in \cite{JLAnalyticMellin}.

The definition (\ref{MackPol}) also makes it clear that $P_{\nu,J}(s,t)$ is  polynomial in the parameters $\D_{12}$ and  $\D_{34}$. On the other hand, we see that $P_{\nu,J}(s,t)$ has poles at $\nu=\pm i (h+J-q-1)$ for $q= 1,2,\dots,J$.
We checked that the residues of these poles are described by the formula, 
\begin{align}
P_{\nu,J}(s,t)\approx  
\frac{2^{q} J! 
\left(\frac{\D_{12}-q+1}{2}\right)_{q}
\left(\frac{\D_{34}-q+1}{2}\right)_{q}
\left(\frac{2h-t-q+1}{2}\right)_{q}
}{\left( h+J-q-1\pm i \nu \right)
q!\,(J-q)!\,\Gamma(q)\,(h-1+J-q)_{q}} \,
P_{i(h+J-1),J-q}(s,t)\,,
\label{MackPolypoles}
\end{align}
for all $J\le 8$. We believe that this is an identity, but were unable to prove it.


\section{Regge limit in position space
\label{secReggepos}}

\begin{figure}
\begin{centering}
\includegraphics[scale=0.27]{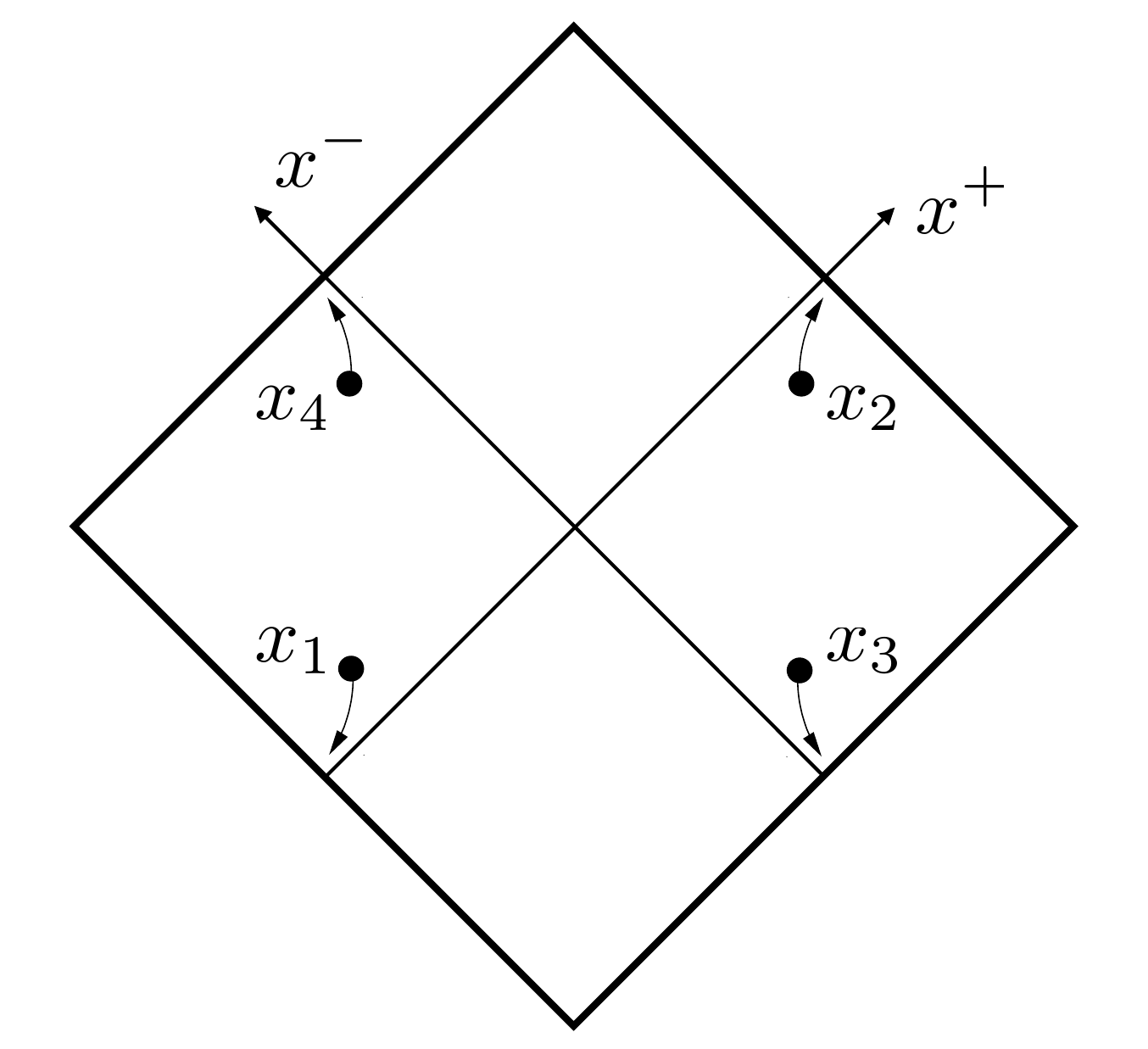}
\par\end{centering}
\caption{\label{figReggePosition}Conformal compactification of the light cone plane. In the Regge limit the positions of the operators $x_i$ go to null infinity.}
\end{figure}

This appendix has two goals. The first one is to show that our definition of the Regge limit of the Mellin amplitude (large $s$ and fixed $t$) corresponds to the Regge limit defined in position space in \cite{CornalbaRegge,ourBFKL}. This limit can be defined by 
$x_1^+ \to \lambda x_1^+ $, 
$x_2^+ \to \lambda x_2^+ $, 
$x_3^- \to \lambda x_3^- $, 
$x_4^- \to \lambda x_4^- $ and $\lambda \to \infty$, 
keeping the causal relations $x_{14}^2, x_{23}^2 <0$ and all the other $x_{ij}^2>0$.
This is depicted in figure \ref{figReggePosition}.
\footnote{Notice that in this paper we are labelling points differently from 
\cite{CornalbaRegge,ourBFKL}. 
The translation is simply the permutation $x_2 \leftrightarrow x_3$. The reason for a different notation is to follow the dominant convention in the OPE literature $\Ocal_1 \Ocal_2 \sim   \Ocal_{\Delta,J} $.}
We remark that by Fourier transforming to momentum space the position of the operators $x_i$, and defining the corresponding Mandelstam invariants, the Regge limit is just the 
usual Regge limit of large $s$ and fixed $t$.
The second goal of this appendix is to derive an expression for the position space correlator in the
Regge limit corresponding to our main equation (\ref{MainEquationMellin}).

Let us then start by the definition of the Mellin representation for the time ordered Lorentzian correlation function \cite{Mack}
\be
A(x_i) = \int [d\d] \,M(\d_{ij})\prod_{i < j} \G(\delta_{ij}) \big(x_{ij}^2+i\epsilon\big)^{- \delta_{ij}}\,.
\ee
Given the chosen causal relations for the Regge limit, we should rotate 
\be
v^{-(s+t)/2} \to v^{-(s+t)/2} e^{-i\pi (s+t)}
\ee
in the integral (\ref{reducedMellin}).
With this phase, the convergence of the integral is not obvious when $s=ix$, $x\to +\infty$.
To study this question, we approximate the second line of (\ref{reducedMellin}) using
\be
\Gamma\!\left(a+i\,\frac{x}{2}\right)\Gamma\!\left(b-i\,\frac{x}{2}\right) \approx 2\pi   \,  e^{i\frac{\pi}{2}(a-b)} \left(\frac{x}{2}\right)^{a+b-1}  e^{-\frac{\pi}{2} x}\,,
\ \ \ \ \ \ \ 
x\to +\infty\,.
\ee
This gives
\be
\mathcal{A}(u,v)  \approx
 \int_{-i\infty}^{i\infty}\frac{dt }{4 i}  \,
u^{t/2} v^{- t/2}
\,\Gamma\!\left( \frac{2\Delta_1 -t}{2}\right)
\Gamma\!\left( \frac{2\Delta_3   -t}{2}\right)
 e^{-i \pi \frac{ t}{2}  }  \int^\infty\! \!dx\, M(ix,t) \left(\frac{x}{2}\right)^{t-2} v^{-\frac{i}{2}x}\,,
\ee
where we have restricted to the case $\D_{12}=\D_{34}=0$.
Following \cite{CornalbaRegge,ourBFKL}, we introduce the variables $\sigma$ and $\rho$ via
\be
u=\sigma^2\,,\ \ \ \ \ \ \ 
v=(1-\sigma e^\rho)(1-\sigma e^{-\rho})\approx 1-2\sigma \cosh \rho\,,
\ee
such that the Regge limit corresponds to $\sigma \to 0$ with fixed $\rho$.
In this limit,
\be
\mathcal{A} \approx
 \int_{-i\infty}^{i\infty}\frac{dt }{4 i}  \,
\sigma^{t}  \,
\Gamma\!\left( \frac{2\Delta_1  -t}{2}\right)
\Gamma\!\left( \frac{2\Delta_3  -t}{2}\right)
 e^{-i \pi \frac{ t}{2}  }   \int^\infty \!\!dx \,M(ix,t) \left(\frac{x}{2}\right)^{t-2} e^{ix \sigma \cosh \rho}\,,
\ee
which shows that the small $\sigma$ behavior of $\mathcal{A}$ is controlled by the large $s$ behavior of the Mellin amplitude. 
We can now use our main result (\ref{MainEquationMellin}) to write
\be
M(ix,t)\approx \int  d\nu \,\beta(\nu)\,
\frac{x^{ j(\nu)} }
{ \sin \!\left(\frac{\pi j(\nu)}{2}\right) } \,\omega_{\nu, j(\nu)}(t)\,,
\ee 
where we have chosen the appropriate phase for $s=ix$, $x\to +\infty$.
After performing the integral over $x$, we find
\begin{align}
\mathcal{A} \approx&-\pi i   
\int  d\nu \,\beta(\nu)\,
\frac{e^{i \pi j(\nu) /2 }}{\sin \!\left(\frac{\pi j(\nu)}{2}\right)}
 \,\sigma^{1-j(\nu)}   2^{j(\nu)} \\& 
 \int_{-i\infty}^{i\infty}\frac{dt }{2\pi i}  \,
\Gamma\!\left( \frac{2\Delta_1   -t}{2}\right)
\Gamma\!\left( \frac{2\Delta_3  -t}{2}\right)
\frac{\Gamma\big(j(\nu)+t-1\big)}{(2\cosh\rho)^{j(\nu)+t-1} }\,
\omega_{\nu, j(\nu)}(t)\,.
  \nonumber
\end{align}
Finally, using the following integral representation for the harmonic functions
$\Omega_{i\nu}(\rho)$ on $(2h-1)$-dimensional hyperbolic space 
\begin{align}
\Omega_{i\nu}(\rho)&=
\int \frac{dz}{2\pi i} \frac{\Gamma(z) \,
\Gamma\!\left( \frac{h+i\nu-z-1}{2}\right) \Gamma\!\left( \frac{h-i\nu-z-1}{2}\right)}
{8\pi^{h} \Gamma(i\nu) \Gamma(-i\nu)}  \,(2\cosh \rho)^{-z} \\
&=\frac{\nu \sinh(\pi \nu) \Gamma(h-1+i\nu)
\Gamma(h-1-i\nu)
\!\ _2F_1\! \left(h-1+i\nu,h-1-i\nu,h-\frac{1}{2},
-\sinh^2\!\big(\frac{\rho}{2}\big)
\right)
}{2^{2h-1}\pi^{h+\frac{1}{2}}
\Gamma\!\left(h-\frac{1}{2} \right)}
\nonumber
\end{align}
we recover the general Regge behavior in position space 
written in (\ref{ReggeA}), with residue given by (\ref{alpha}).


\section{Harmonic Sums}
\label{HarmonicSums}

The Harmonic sums  are usefull functions to express the principle of transcendentality. They appear for example in the BFKL spin, anomalous dimensions and also in the three point functions \cite{Eden:2012rr,Vermaseren:1998uu,Blumlein:1998if}. Usually defined as a sum, they can be anallytically continued quite easily \cite{Kotikov:2005gr,Blumlein:2000hw}. %
 The simplest Harmonic sums are defined as,
\begin{align}
S_{n}\left(x\right)= &\,
\left(-1\right)^{n-1}\frac{\Psi^{(n-1)}(x+1)-\Psi^{(n-1)}(1)}{\Gamma(n)}\overset{\textrm{\textrm{integer x}}}{=}\sum_{l=1}^{x}\frac{1}{l^{n}}\,,
\\
S_{-n}(x)=&\,
(-1)^{n}\frac{\Psi^{(n-1)}\!\left(\frac{1}{2}+\frac{x}{2}\right)-\Psi^{(n-1)}\!\left(1+\frac{x}{2}\right)-\Psi^{(n-1)}\!\left(\frac{1}{2}\right)+\Psi^{(n-1)}(1)}{2^n\Gamma(n)}\overset{\textrm{even x}}{=}\sum_{m=1}^{x}\frac{(-1)^{m}}{m^{n}}\,,
\nonumber
\\
S_{-a,b}(x)=&\,
\zeta(-a,b)+\zeta(-(a+b))-\sum_{l=1}^{\infty}\frac{\left(-1\right)^{l}}{\left(l+x\right)^{a}}\,S_{b}(l+x)\,,
\nonumber
\end{align}
where $\zeta(-a)=\left(\frac{1}{2^{a-1}}-1\right)\zeta(a)$ and 
$\zeta(-a,b)$ are the Euler Zagier sums (or multivariate Zeta functions). 
The Euler Zagier $\zeta(-2,1)=\zeta(3)/8$.
Notice that the analytically continued functions $S_{a_1,\dots,a_n}(x)$ with one or more negative indices $a_i$, only match the definition (\ref{Hsums}) in terms of nested sums, for $x$ an even integer.

The expansion of the Harmonic sums around the point $x=-1$ is
\begin{align}
S_{n}(-1+\omega)=&\,
-\frac{1}{\omega^{n}}-\sum_{k=1}^{\infty}\left(-1\right)^{k}
\left(\begin{array}{c}n+k-1\\
k
\end{array}\right)\zeta(n+k)\,\omega^{k}\,,
\nonumber\\
S_{-n}(-1+\omega)=&\,
\frac{1}{\omega^{n}}+\zeta(-n)-\sum_{k=1}^{\infty}\left(-1\right)^{k}\left(\begin{array}{c}
n+k-1\\
k
\end{array}\right)\zeta(-n-k)\,\omega^{k}\,,
\\
 S_{-2,1}&\,(-1+\omega)=
\frac{\zeta(2)}{\omega}-\frac{9\zeta(3)}{4}+\frac{33\zeta(4)}{16}\omega\,.
\nonumber
\end{align}
The Harmonic sum can be related to the sine function through the equality,
\begin{equation}
S_{-1}(x)+S_{-1}(-1-x)+\frac{\pi}{\sin(\pi x)}=-2\ln2\,.
\end{equation}
In \cite{DressingWrapping} there is a function $\tilde{\Phi}(x)$ defined by
\begin{equation}
\tilde{\Phi}(x)=\sum_{k=0}^{\infty}\frac{\left(-1\right)^{k}}{\left(k+x\right)^{2}}\,\big[\Psi(k+x+1)-\Psi(1)\big]\,,
\end{equation}
that appears in the BFKL spin. This function can be related to (\ref{eq:funcao phi}) by
\begin{equation} 
\Phi(-x)+\Phi(1+x)=2\tilde{\Phi}(-x)+2\tilde{\Phi}(1+x)+\frac{\pi^{3}}{2\sin(\pi x)}\,.
\end{equation}
The function $\tilde{\Phi}$ can be written in terms of $S_{-2,1}$ through,
\begin{equation}
\tilde{\Phi}(x)=S_{-2,1}(x-1)+\frac{5}{8}\,\zeta(3)\,.
\end{equation}
\section{Leading twist two operators in SYM  \label{Appendix3pt}}

Conformal symmetry imposes constrains on the form of two and three point functions between scalar and symmetric traceless operators \cite{SpinningCC}. 
In particular, the ratio of correlators like the one in  (\ref{Cs}) contains information about the OPE coefficients.  More precisely, 
the structure of this ratio is fixed by conformal symmetry to be
\begin{align}
&\hspace{4cm}
\frac{\left\langle \Ocal_1(x_1)\Ocal_1^*(x_2) \Ocal_J(x_5) \right\rangle \,
 \left\langle \Ocal_J(x_6)  \Ocal_3(x_3)\Ocal_3^*(x_4) \right\rangle}
 { \left\langle \Ocal_1(x_1)\Ocal_1^*(x_2) \right\rangle  \left\langle \Ocal_J(x_5) \Ocal_J(x_6)\right\rangle   \left\langle \Ocal_3(x_3)\Ocal_3^*(x_4) \right\rangle  }=
\\
&C_{11J}C_{33J}
\left(\frac{x^2_{13} x^4_{56} x^2_{24} }{ x^2_{15} x^2_{35} x^2_{26} x^2_{46} }\right)^{\frac{\Delta+J}{2}}
 \frac{\big( (w\cdot x_{15}) \, x^2_{35} - (w\cdot x_{35}) \, x^2_{15}\big)^J
\big((w'\cdot x_{26}) \, x^2_{46} - (w'\cdot x_{46}) \, x^2_{26}\big)^J }
{ x^{2J}_{13} x^{2J}_{24} \big( (w\cdot w')\, x^2_{56} -2(w\cdot x_{56})\, (w'\cdot x_{56}) \big)^J }\,,
\nonumber
\end{align}
where $x_{ij} = x_i - x_j$ and $w$ and $w'$ are {\em null polarization vectors} that allow us to write the symmetric
traceless operator ${\cal O}_J$ as the polynomial ${\cal O}_J = w^{\mu_1}\dots w^{\mu_J} {\cal O}_{\mu_1\dots \mu_J}$ (see for example \cite{SpinningCC} for details).


In this appendix we compute the OPE coefficients $C_{11J}$ to leading order in perturbation theory. 
The first step in the computation is to obtain the exact linear combination of operators that makes up the leading twist operator, as already mentioned in (\ref{zeroorderstates}). This is achieved by diagonalizing the 1-loop dilatation operator and finding its eigenstates. The second step is to perform the perturbative (Wick contractions) computation of the three point functions.

\subsection{Diagonalizing the 1-loop dilatation operator }

The twist two operators   are degenerate at tree level, however at finite t'Hooft coupling the degeneracy is lifted, making explicit which operator is in the leading Regge trajectory. This is done using the dilatation operator which can be written, at first order, using harmonic oscillators. By restricting its action to the subspace generated by states of the form (\ref{zeroorderstates}) we find the eigenfunctions and eigenvalues of the dilatation operator.

In the following we review some definitions  needed for the computation, following closely \cite{Beisert:2004ry} and then apply it to our case.

\subsubsection{Definitions}

The elementary fields in SYM are $F_{\mu\nu}, \psi_{\alpha a}, \dot{\psi}_{\dot{\alpha}}^{a}$ 
and $\Phi_{m}$, which can be written using harmonic oscillators as
\begin{align}
 {\cal D}^{k}{\cal F}\equiv &\left(a^{\dagger}\right)^{k+2}\left(b^{\dagger}\right)^{k}\left(c^{\dagger}\right)^{0}\left|0\right\rangle ,
\nonumber\\
{\cal D}^{k}\psi\equiv&\left(a^{\dagger}\right)^{k+1}\left(b^{\dagger}\right)^{k}\left(c^{\dagger}\right)^{1}\left|0\right\rangle ,
\nonumber\\
{\cal D}^{k}\phi\equiv&\left(a^{\dagger}\right)^{k}\left(b^{\dagger}\right)^{k}\left(c^{\dagger}\right)^{2}\left|0\right\rangle ,
\label{eq:fields}
\\
{\cal D}^{k}\dot{\psi}\equiv&\left(a^{\dagger}\right)^{k}\left(b^{\dagger}\right)^{k+1}\left(c^{\dagger}\right)^{3}\left|0\right\rangle ,
\nonumber\\
{\cal D}^{k}\dot{{\cal F}}\equiv&\left(a^{\dagger}\right)^{k}\left(b^{\dagger}\right)^{k+2}\left(c^{\dagger}\right)^{4}\left|0\right\rangle ,
\nonumber
\end{align}
where $ F_{\mu\nu}\sim\sigma_{\mu}^{\alpha\dot{\gamma}}\epsilon_{\dot{\gamma}\dot{\delta}}\sigma_{\nu}^{\dot{\delta}\beta}{\cal F}_{\alpha\beta}+\sigma_{\mu}^{\dot{\alpha}\gamma}\epsilon_{\gamma\delta}\sigma_{\nu}^{\delta\dot{\beta}}\dot{{\cal F}}_{\dot{\alpha}\dot{\beta}}$, $\Phi_{m}\sim\sigma^{ba}_{m}\phi_{ab}$, the oscillators $a_{\alpha}^{\dagger},b_{\dot{\alpha}}^{\dagger}$ have indices corresponding to  the $\mathfrak{su}\left(2\right) \times\mathfrak{su}\left(2\right)$ Lorentz algebra and $c_{a}^{\dagger}$ has a  $\mathfrak{su}\left(4\right)$ R-charge index. 
By definition ${\cal F}_{\alpha\beta}, \dot{{\cal F}}_{\dot{\alpha}\dot{\beta}} $ are symmetric and $\phi_{ab}$ is antisymmetric in the indices.
For example, 
\[
{\cal D}_{\dot{\alpha}\beta}\phi_{ab}\sim a_{\beta}^{\dagger}b_{\dot{\alpha}}^{\dagger}c_{a}^{\dagger}c_{b}^{\dagger}\left|0\right\rangle .
\]
As expected, bosonic oscillators commute and fermionic oscillators anticommute
\begin{equation}
\left[a^{\alpha},a_{\beta}^{\dagger}\right]=\delta_{\beta}^{\alpha}\,, 
\ \ \ \ \ \ \ 
\left[b^{\dot{\alpha}},b_{\dot{\beta}}^{\dagger}\right]=\delta_{\dot{\beta}}^{\dot{\alpha}}\,,
\ \ \ \ \ \ \ 
\left\{ c^{a},c_{b}^{\dagger}\right\} =\delta_{b}^{a}\,.
\end{equation}
The state $\left|0\right\rangle $ is defined as the one annihilated
by all oscillators $a^{\alpha},b^{\dot{\alpha}}\,\textrm{and}\, c^{a}$.
Though this state is not physical state, as it gives a nonzero central
charge 

\begin{equation}
C=1-\frac{1}{2}\,a^\dagger_{\alpha}a^{\alpha} +\frac{1}{2}\,b^\dagger_{\dot{\alpha}}b^{\dot{\alpha}} -\frac{1}{2}\,c^\dagger_{a}c^{a}\,.
\end{equation}
On the other hand, the elementary fields are obtained from 
the physical state $\phi_{34}\equiv c_{3}^{\dagger}c_{4}^{\dagger}\left|0\right\rangle \equiv\left|\mathcal{Z}\right\rangle $
that is the highest weight. This leads to the redefinition of the oscillators
\begin{equation}
d_{1}^{\dagger}=c^{4}\,,
\ \ \ \ \ 
d_{2}^{\dagger}=c^{3}\,,
\ \ \ \ \ 
d^{1}=c_{4}^{\dagger}\,,
\ \ \ \ \ 
d^{2}=c_{3}^{\dagger}\,,
\end{equation}
which breaks the $\mathfrak{su}(4)$ into $\mathfrak{su}(2)\times\mathfrak{su}(2)$.
This redefinition makes the state $\left|\mathcal{Z}\right\rangle $ the natural vacuum, since it is annihilated by $a_{\alpha}$,  $b_{\dot{\alpha}}$, $c_{1}$, $c_{2}$, $d_{1}$ and $d_{2}$. 

\subsubsection{Twist two operators}

Twist two operators are defined by their classical
dimension $\Delta_{0}=1+(n_{a}+n_{b})/2=2+J$,
where $J$ is the spin. 
This implies that they are of the form
\[
\textrm{Tr}\left({\cal W}_{A}{\cal W}_{B}\right),
\]
where $\mathcal{W_{A}}\in\left\{ {\cal D}^{k}{\cal F},{\cal D}^{k}\dot{{\cal F}},{\cal D}^{k}\phi,{\cal D}^{k}\psi,{\cal D}^{k}\dot{\psi}\right\}$,
or using oscillators 
\begin{equation}
\left(a^{\dagger}\right)^{n_{a}}
\left(b^{\dagger}\right)^{n_{b}}
\left(c^{\dagger}\right)^{n_{c}}
\left(d^{\dagger}\right)^{n_{d}}
\left(a^{\dagger}\right)^{J-n_{a}}
\left(b^{\dagger}\right)^{J-n_{b}}
\left(c^{\dagger}\right)^{p-n_{c}}
\left(d^{\dagger}\right)^{p-n_{d}}\left|\mathcal{ZZ}\right\rangle .
\end{equation}
This requires some explanation; the first four types of oscillators
act on the first site and the others act on the second site; the number
$n_{a}$ of oscillators of type $a^{\dagger}$ on the first site is
arbitrary in principle
\footnote{Note that there is the restriction of $n_{a}<J$.} 
but as we want spin J, the number of $a^{\dagger}$ on the second
site has to be $J-n_{a}$; the same applies to oscillators of type
$b^{\dagger}$; on the second site, there is no loss of generality
when considering the number of oscillators of type $c^{\dagger}$
to be $p-n_{c}$, but the number of $d^{\dagger}$ on the same site
follows because of the central charge condition, which now reads
\be
C  
=\frac{1}{2}
\sum_{\rm sites} \left(
  b^\dagger_{\dot{\alpha}}b^{\dot{\alpha}}- a^\dagger_{\alpha}a^{\alpha}  
+d^\dagger_{a}d^{a} - c^\dagger_{a}c^{a}
\right) 
=0\ .
\ee
Requiring the state to be a $\mathfrak{su}(4)$ singlet, fixes $p=2$.
In the original operator basis, the $\mathfrak{su}(4)$ part of the state is $\epsilon^{abcd}c_{a}^{\dagger}c_{b}^{\dagger}c_{c}^{\dagger}c_{d}^{\dagger}\left|00\right\rangle$, which in the new basis becomes
 $ c_{1}^{\dagger}c_{2}^{\dagger}d^{1} d^{2} \left|00\right\rangle $ or $
 c_{1}^{\dagger}c_{2}^{\dagger}d_{1}^{\dagger} d_{2}^{\dagger} \left|\mathcal{ZZ}\right\rangle  $.
 In the previous sentence, we did not specify in which site each operator acts because this is not relevant for the $\mathfrak{su}(4)$ singlet constraint.
Note that, other $\mathfrak{su}(4)$ singlets, like 
$\mathcal{FF}$ or $\dot{\mathcal{F}}\dot{\mathcal{F}}$, are excluded because they do not have the required Lorentz structure.
Thus, the states can be labelled as
\begin{equation}
\Big|  \underbrace{1\dots1}_{n_a}\,
\underbrace{2\dots2}_{J-n_{a}}\,
\underbrace{1\dots 1}_{n_{b}}\,
\underbrace{2\dots 2}_{J-n_{b}}\,
i_{1}\dots i_{4};a^{\dagger} \dots a^{\dagger}b^{\dagger} \dots b^{\dagger}c_{1}^{\dagger}c_{2}^{\dagger}d_{1}^{\dagger}d_{2}^{\dagger}\Big\rangle \,,
\label{eq:state}
\end{equation}
where $n_{a}$ is the number of the first set of 1's, $J-n_{a}$ is the
number of the first set of 2's and, similarly, for $n_{b}$ and $J-n_{b}$ in  the
following set of 1's and 2's. The $i_{j}$ can be $1$ or $2$ encoding
the site where $c^{\dagger}$ and $d^{\dagger}$ act.  

In this representation the Hamiltonian can
be written in the following way
\be
\mathcal{H} \left|s_{1} \dots s_{n};A\right\rangle =\sum_{s'_{1} \dots s'_{n}}c_{n,n_{12},n_{21}}\delta_{C_{1},0}\delta_{C_{2},0}\left|s'_{1} \dots s'_{n};A\right\rangle ,
\ee
where $A$ is a list of $n$ operators, like the one
in (\ref{eq:state}) and $s_{1} \dots s_{n}$ is a list of 1's and 2's that specifies in which site each operator acts.
The variable $n_{12}$ counts the number of 1's in
 $s_{1} \dots s_{n}$ that  became 2's in $s'_{1} \dots s'_{n}$. 
In other words,
$n_{12}$ and $n_{21}$ is the number of oscillators
hopping from site 1 to 2 and from 2 to 1, respectively. 
Finally,
\begin{equation}
c_{n,n_{12},n_{21}}=\left(-1\right)^{1+n_{12}n_{21}}
\frac{\Gamma\!\left(\frac{n_{12}+n_{21}}{2}\right)\Gamma\!\left(1+\frac{n-n_{12}-n_{21}}{2}\right)}
{\Gamma\!\left(1+\frac{n}{2}\right)} \,.
\end{equation}
It is clear that the subspace generated by states of the form (\ref{eq:state}) is closed under the action of the Hamiltonian.
We have implemented a  Mathematica program to find the eigenstates and eigenvalues  for values of $J=2, \dots ,8$.%
\footnote{This was implemented in Mathematica by creating all possible states. Higher values of J were 
limited by this approach, as the number of states grows exponentially.} 
The results allow us to confirm that, all the odd
spin eigenstates  are descendants, and that the eigenvalues for any $J$ are $2S_{1}(J-2)$, $2S_{1}(J)$ and $2S_{1}(J+2)$,
where $S_1$ is a harmonic sum.
It also enabled us to confirm
that the eigenvectors are a linear combination of the form
\be
 a\left|\phi{\cal D}^{J}\phi\right\rangle +b\left|\mathcal{F}{\cal D}^{J-2}\dot{\mathcal{F}}\right\rangle +c\left|\psi{\cal D}^{J-1}\dot{\psi}\right\rangle,
\ee 
where  we use the shorthand notation \cite{Henn:2005mw},\footnote{On the right hand side of the equation we use (\ref{eq:fields}).}
\begin{align}
\left|\phi{\cal D}^{J}\phi\right\rangle =&\,
\sum_{k=0}^{J}\left(-1\right)^{k}
 \left( {J \atop k}\right)^{2}
\textrm{Tr}\!\left({\cal D}^{k}\phi{\cal D}^{J-k}\phi\right),
\nonumber
\\ 
 \left|\psi{\cal D}^{J-1}\dot{\psi}\right\rangle=&\
\sum_{k=0}^{J-1}\left(-1\right)^{k}
\left({J \atop k}\right) 
\left({J \atop k+1}\right) \textrm{Tr}\!\left({\cal D}^{k}\psi{\cal D}^{J-k-1}\dot{\psi}\right),
\label{basis}
\\ 
\left|\mathcal{F}{\cal D}^{J-2}\dot{\mathcal{F}}\right\rangle =&\,
\sum_{k=0}^{J-2}\left(-1\right)^{k}
\left({J \atop k}\right)
\left({J \atop k+2}\right)\textrm{Tr}\!\left({\cal D}^{k}F{\cal D}^{J-k-2}\dot{F}\right),
\nonumber
\end{align}
with
\be
\left({J \atop k}\right) = \frac{J!}{\left(J-k\right)!k!}\,.
\ee
This was expected as it is possible to construct twist two primary operators
at zero order restricting only to scalar, gauge or fermionic fields \cite{Henn:2005mw},
and so, at first order, the eigenvectors 
must be  a linear combination
of these zero order eigenstates.\footnote{In perturbation theory in quantum mechanics the eigenvectors lag in
 relation to the eigenvalues.}

The data collected also allowed to infer the matrix form of the Hamiltonian for general J in the (non-normalized) basis  
$\left\{   \left|1\right\rangle ,  \left|2\right\rangle ,  \left|3\right\rangle  \right\}
\equiv 
\left\{   \left|\phi{\cal D}^{J}\phi\right\rangle ,  \left|\psi{\cal D}^{J-1}\dot{\psi}\right\rangle ,  \left|\mathcal{F}{\cal D}^{J-2}\dot{\mathcal{F}}\right\rangle  \right\}$.
We found
\be
h_{ij}=\left(
\begin{array}{ccc}
 2 S_1(J) & -\frac{2}{J+1} & \frac{1}{(J+1) (J+2)}
   \\
 -\frac{6}{J} & 2 S_1(J)-\frac{4}{J (J+1)} &
   \frac{J^2+J+2}{J (J+1) (J+2)} \\
 \frac{12}{(J-1) J} & \frac{4 \left(J^2+J+2\right)}{(J-1)
   J (J+1)} & 2 S_1(J)-\frac{4
   \left(J^2+J+1\right)}{(J-1) J (J+1) (J+2)}
\end{array}
\right)
\ee
where
\be
\mathcal{H}  \left|i\right\rangle =
\sum_{j=1}^3   h_{ji} \left|j\right\rangle \ .
\ee
It is then simple to determine the eigenvectors of $\mathcal{H}$. 
From highest to lowest eigenvalue,
the three eigenvectors obtained are
\begin{align}
 \left| V_{1}\right\rangle = &\,
\left|\phi{\cal D}^{J}\phi\right\rangle
-2 \left|\psi{\cal D}^{J-1}\dot{\psi}\right\rangle
-2\left|\mathcal{F}{\cal D}^{J-2}\dot{\mathcal{F}}\right\rangle ,
\label{highesteigenstate}
\\
 \left| V_{2}\right\rangle = &\,
\left|\phi{\cal D}^{J}\phi\right\rangle
+\frac{3}{J} \left|\psi{\cal D}^{J-1}\dot{\psi}\right\rangle
+\frac{6\left(2+J\right)}{J}\left|\mathcal{F}{\cal D}^{J-2}\dot{\mathcal{F}}\right\rangle,
\\
 \left| V_{3}\right\rangle = &\,
\left|\phi{\cal D}^{J}\phi\right\rangle
+\frac{2\left(J+1\right)}{J} \left|\psi{\cal D}^{J-1}\dot{\psi}\right\rangle
-\frac{2(J+2)(J+1)}{J(J-1)}\left|\mathcal{F}{\cal D}^{J-2}\dot{\mathcal{F}}\right\rangle.
\label{lowesteigenstate}
\end{align}
 
\subsection{Three-point function }

In \cite{Henn:2005mw} two and three point between two scalars and the operator with highest eigenvalue (\ref{highesteigenstate}) were computed. 
Their results can be easily adapted to the case of the leading twist operators, that corresponds to the state (\ref{lowesteigenstate}) with lowest eigenvalue.
The only subtlety is that one needs to adapt  field normalizations
as follows $\phi \to \phi$,  $F\rightarrow iF$
and $\psi\rightarrow\sqrt{2i}\psi$.
Thus, in the conventions of \cite{Henn:2005mw},
the leading twist operator is written as
\begin{align}
\mathcal{O}_{J}=\left|\phi{\cal D}^{J}\phi\right\rangle+\frac{4i\left(J+1\right)}{J}\left|\psi{\cal D}^{J-1}\dot{\psi}\right\rangle+\frac{2(J+2)(J+1)}{J(J-1)}
\left|\mathcal{F}{\cal D}^{J-2}\dot{\mathcal{F}}\right\rangle\label{eq:leading twist}.
\end{align}
Its two and three point functions  can then be obtained from \cite{Henn:2005mw},
\begin{align}
\left\langle \mathcal{O}_{J}(x_{5})\mathcal{O}_{J}(x_{6})\right\rangle =&\,
\frac{2^{J+2}N^2}{\left(4\pi^{2}\right)^{2}}\left(12S_{p}+\left(\frac{J+1}{J}\right)^{2}S_{r}+
\left(\frac{(J+2)(J+1)}{J(J-1)}\right)^2
S_{s}\right)
\nonumber\\
&
\ \frac{\big( (w.w' ) x_{56}^{2}-2 (w.x_{56})\,(w'.x_{56})\big)^{J}}{\left(x^{2}\right)^{2J+2}}\,,
\nonumber
\\
\left\langle \mathcal{O}_{1}(x_{1})\mathcal{O}_{1}(x_{3})\right\rangle = &\,
\frac{N^2}{8\pi^{4}}\, \frac{1}{\left(x_{13}^{2}\right)^{2}}\,,
\\
\left\langle \mathcal{O}_{1}(x_{1})\mathcal{O}_{1}(x_{3})\mathcal{O}_{J}(x_{5})\right\rangle =&\,
\frac{2^{J+4} N^2 \,\Gamma(J+1)}{\left(4\pi^{2}\right)^{3}}\,
\frac{\big( (w.x_{15} ) x_{35}^{2}- (w.x_{35} )x_{15}^{2}\big)^{J}}{\left(x_{12}^{2}\right)^{3}\left(x_{23}^{2}\right)^{1+J}\left(x_{13}^{2}\right)^{1+J}}\,,
\nonumber
\end{align}
where
\be
S_{p}(J)=\Gamma(2J+1)\,,\ \ \ \ 
S_{r}(J)=16J \, \frac{\Gamma(2J+1)}{J+1}\,,\ \ \ \ 
S_{s}(J)=4\,\frac{J\left(J-1\right)\Gamma(2J+1)}{(J+1)(J+2)}\,.
\ee
Notice that the result for the two point function of $\mathcal{O}_{J}$ satisfies
the constraint that the two point function of the stress energy momentum
of the fermion is twice the value of the scalar and gauge part \cite{OsbornCFTgeneraldim}. 

Thus, we finally conclude that $C^2(J)\equiv C_{11J} C_{33J}$ is given by
\begin{equation}
C^2(J)=\frac{2^{1+J}(J-1) J\, \Gamma^{2}(J+1)}{N^2\left(4J^{2}-1\right)\Gamma(2J+1)}\,,
\end{equation}
which satisfies the requirement $C^2(2)=8/ (45N^2)$ that 
must be  satisfied independently of the t' Hooft coupling.

\bibliographystyle{./utphys}
\bibliography{./mybib}

\end{document}